\documentclass[aps,prb, amsmath, amssymb,twocolumn]{revtex4-2}

\usepackage{amsmath, amssymb, bbold}
\usepackage[usenames,dvipsnames]{xcolor}
\usepackage{subfig}
\usepackage{hyperref}
\hypersetup{
	colorlinks,
	citecolor=RoyalBlue,
	filecolor=black,
}

\usepackage{IEEEtrantools}
\usepackage{comment}

\usepackage{graphicx}
\usepackage{tikz,lipsum,lmodern}

\newcommand{\dbar}{d\hspace*{-0.08em}\bar{}\hspace*{0.1em}}

\graphicspath{{./}}

\begin{document}
	
\title{On the First Law of Thermodynamics in Time-Dependent Open Quantum Systems}
\author{Parth Kumar} \email{parthk@arizona.edu}
\author{Charles A. Stafford} \email{stafford@physics.arizona.edu}
\affiliation{Department of Physics, University of Arizona, 1118 East Fourth Street, Tucson, Arizona 85721, USA}
\date{\today}
	
\begin{abstract}
How to rigorously define thermodynamic quantities such as heat, work, and internal energy in open quantum systems driven far from equilibrium remains a significant open question in quantum thermodynamics.  Heat is a quantity whose fundamental definition $\dbar Q = T dS$ applies 
only to processes in systems infinitesimally perturbed from equilibrium, and as such, must be accounted for carefully in strongly-driven systems. 
A key insight from Mesoscopics is that infinitely far from the local driving and coupling of an open quantum system, reservoirs are indeed only infinitesimally perturbed, thereby allowing the heat dissipated to be defined.  The resulting partition of the entropy necessitates a Hilbert-space partition of the energetics,
leading to an unambiguous operator for the internal energy of an interacting time-dependent open quantum system. 
Fully general expressions for the heat current and the power delivered by various agents to the system are derived using the formalism of nonequilibrium Green's functions, establishing an experimentally meaningful and quantum mechanically consistent division of the energy of the system under consideration into Heat flowing out of and Work done on the system. The spatio-temporal distribution of internal energy in a strongly-driven open quantum system is also analyzed.
This formalism is applied to analyze the thermodynamic performance of a model quantum machine: a driven two-level quantum system strongly coupled to two metallic reservoirs, which can operate in several configurations--as a chemical pump/engine or a heat pump/engine. %
\end{abstract}

\maketitle
\tableofcontents

\section{Introduction}\label{intro_sec}

Quantum Thermodynamics has seen substantial progress in recent years in both experimental and theoretical directions \cite{binderThermodynamicsQuantumRegime2018}. These include but are not limited to quantum thermometry \cite{Shastry2020_STTh,mecklenburgNanoscaleTemperatureMapping2015,neumannHighprecisionNanoscaleTemperature2013,jeongScanningProbeMicroscopy2015,mengesTemperatureMappingOperating2016,shiThermalProbingEnergy2009,kimUltrahighVacuumScanning2012,Lee2013,Gomes2015,cuiQuantizedThermalTransport2017,Mosso2017}, quantum and thermal machines \cite{Toyabi2010_Szilard_exp,berutExperimentalVerificationLandauer2012,koskiExperimentalRealizationSzilard2014,devoretQuantumMachinesMeasurement2014a,liuPeriodicallyDrivenQuantum2021}, quantum stochastic dynamics \cite{campisiColloquiumQuantumFluctuation2011,jarzynskiEqualitiesInequalitiesIrreversibility2011,seifertStochasticThermodynamicsFluctuation2012} and quantum information and computation \cite{10.1088/2053-2571/ab21c6}. Despite these rapid advancements, however, broad consensus on several basic questions of the field has been elusive. The reason for this is not entirely mysterious. The field is, firstly, in relative infancy and, secondly, a melding of two well-established disciplines of physics. A large subset of unresolved questions in quantum thermodynamics thus centers around the need for a careful evaluation of the foundational concepts and laws that the subject builds upon. 

One of the key open questions has been the validity and the formulation of the First Law of Thermodynamics for an open quantum system (defined as a quantum system statistically and quantum mechanically open to the environment via some finite coupling) driven far from equilibrium. More concretely, since the First Law is a statement about more than mere energy conservation, we may ask the following: how does one establish an experimentally meaningful and quantum mechanically consistent division of the energy of a far-from-equilibrium quantum system into its %
Internal Energy, the Work done by the different agents acting and boundary conditions imposed on it, and the Heat flowing from it? An even more basic question that naturally precedes this is whether such a division is even possible given the fundamental uncertainty constraints that quantum mechanics places on the measurement of observables.

These questions have been previously addressed in several works \cite{talknerColloquiumStatisticalMechanics2020, ludovicoDynamicalEnergyTransfer2014,bruchQuantumThermodynamicsDriven2016,espositoEntropyProductionCorrelation2010,strasbergFirstSecondLaw2021a,lacerdaQuantumThermodynamicsFast2023,bergmannGreenfunctionPerspective2021,ochoaEnergyDistributionLocal2016}, which present highly divergent points of view on the issue. 
Much of the controversy revolves around how to treat the ``Interface'' between the system and the environment induced by the System-Reservoir coupling Hamiltonian.
An approach based on the `Hamiltonian of Mean Force' concept \cite{talknerColloquiumStatisticalMechanics2020} introduces an ambiguity in the definition of Internal Energy of the system and concludes that it might not even be possible to formulate the First Law, even on a quantum-statistical average level. Another set of analyses \cite{ludovicoDynamicalEnergyTransfer2014,bruchQuantumThermodynamicsDriven2016} proposes dividing up the Coupling Energy between the system and the environment as half of it belonging to the system and half of it belonging to the environment. Others still conclude that the Internal Energy of the nonequilibrium system should be identified by putting the entire Coupling with the System \cite{espositoEntropyProductionCorrelation2010,strasbergFirstSecondLaw2021a,lacerdaQuantumThermodynamicsFast2023} where the last two references specifically employ the Master equation approach and the so-called Mesoscopic leads approach, respectively. A recent study comparing some of these approaches has also emerged \cite{bergmannGreenfunctionPerspective2021}. 

In this work, we shed light on the problem of formulating the First Law of Thermodynamics for a time-dependent %
open quantum system using insights from Mesoscopic physics \cite{haugQuantumKineticsTransport2007,stefanucciNonequilibriumManyBodyTheory2013,dittrichQuantumTransportDissipation1998a,buttikerTimeDependentTransportMesoscopic2000}. We find that the partitioning of the energetics at the heart of the First Law is dictated by the partitioning of the underlying Hilbert space
\cite{shastryThirdLawThermodynamics2019,shastryTheoryThermodynamicMeasurements2019,webb2023}.  This leads naturally to the identification that the Internal Energy operator should be the System Hamiltonian plus {\em half} the coupling Hamiltonian.

The systems studied in this work are \emph{strongly coupled} to their environment and possess \emph{broken time-translational symmetry}. The mesoscopic modeling of the Hamiltonian governing the nonequilibirum dynamics is done such that the model is both experimentally realistic and computationally tractable while capturing all the relevant physics of the energy and particle flows. Specifically, we start by identifying the conditions under which heat can be unambiguously defined for a strongly-driven system and how one must account for it by carefully considering what role the Reservoirs (which make up the environment) coupled to the system play. 

With the proper thermodynamic identifications made, we proceed to derive general expressions for all the different possible forms of energy flows out of the system using the nonequilibrium Green's functions (NEGF) formalism. The NEGF formalism is a powerful framework for analyzing quantum transport which we utilize to investigate the thermodynamics of a fairly general structure, where the system consists of interacting electronic and phononic degrees of freedom strongly coupled to an environment of non-interacting electrons and phonons. As an application of these formal results, we simulate a strongly driven quantum machine based on Rabi oscillations of a double quantum dot system, investigating in detail its operation as an electrochemical pump and as a heat engine.

This work is organized as follows: In Section \ref{Hamiltonian_sec}, we give a detailed description of the model time-dependent Hamiltonian that forms the basis of all subsequent derivations and discussions. Section \ref{noneqbthermo_sec} identifies all the different possible forms of energy flow associated with the system, the role of reservoirs, and Hilbert-space partitioning of physical observables such as energy and entropy.
The consequent form of the Internal Energy Operator and the First Law are presented in Sec.\ \ref{firstlaw_subsec}. In Section \ref{NEGFresults_sec}, we derive all the nonequilibrium energy flows and values in terms of the System Green's functions using the NEGF formalism. In Section \ref{qmachine_sec}, we present results and discussion for the simulation of a strongly-driven quantum machine operating as an electrochemical pump and a heat engine, including an analysis of the spatio-temporal evolution of the energy. Finally, the conclusions of our work are presented in Section \ref{conclusions_sec}. 

\section{The Time-Dependent Open Quantum System}
\label{Hamiltonian_sec}

\begin{figure*}
	
	\centering
	\captionsetup{justification=raggedright, singlelinecheck=false}

	\begin{tikzpicture}[scale=1.5]%
		
		\draw[thick,fill=red!20] (8,1)--(4,1)|-(8,2);
			\node at (6,1.5) {$\mathcal{H_{R}}$};
			\draw[fill=green!20] (2,1.5) circle (1cm);
			\node at (2,1.5) {$\mathcal{H_{S}}$};
			\draw[<->](3,1.5)--(4,1.5);
			\node[below] at (3.5,1.5) {\small $H_{SR}$};
                \node[below] at (2,.5) {\small $H_{S}(t)$};
                \node[below] at (6,.5) {\small $H_{R}$};	
	\end{tikzpicture}

	\caption{Schematic representation of a general time-dependent open quantum system [Eq.\ \eqref{full_hamiltonian_eqn}]. $\mathcal{H_S}$ and $\mathcal{H_R}$ denote the System and Reservoir (single-particle) Hilbert spaces, respectively, whereas $H_S(t)$ and $H_{R}$ describe the System and Reservoir Hamiltoninans, respectively. The Coupling Hamiltonian $H_{SR}$ describes the \emph{Interface} between the System and the Environment.}
	
	\label{firstlawschematic_fig}
	
\end{figure*}
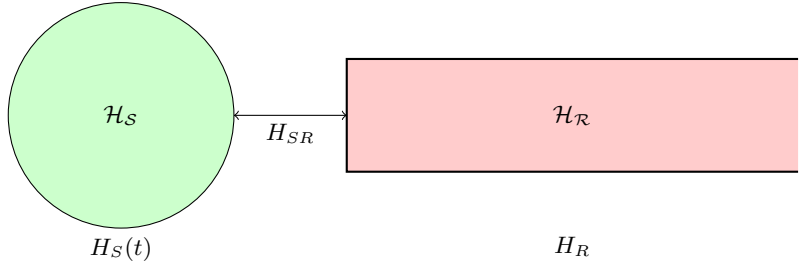	
A fairly general Hamiltonian for a time-dependent open quantum system (Fig.\ \ref{firstlawschematic_fig}) can be written as 
\begin{equation}	
H(t)=H_{S}(t)+H_{R}+H_{SR}\,.
\label{full_hamiltonian_eqn}
\end{equation}
The System Hamiltonian $H_{S}(t)$ has explicit time-dependence due to an external drive and is written as the sum of electronic, phononic and electron-phonon coupling terms
\begin{equation}
H_{S}(t)=H_{S,el}(t)+H_{S,ph}(t)+H_{S,el-ph}\,.
\label{sys_hamiltonian_eqn}
\end{equation} 
The electronic part of the system Hamiltonian $H_{S,el}(t)$ includes electron-electron interactions and has explicit time-dependence from the external electric drive  
\footnote{\label{NDOfootnote} This form of the electronic Hamiltonian is often referred to as the \textit{Complete Neglect of Differential Overlap} approximation and can also be formulated in terms of Wannier orbitals \cite{PhysRevB.86.115403}.}:
\begin{equation}
H_{S,el}(t)=\sum_{n,m}\Bigg([H^{(1)}_{S,el}(t)]_{nm}d^{\dagger}_{n}d_{m} + [H^{(2)}_{S,el}]_{nm}d^{\dagger}_{n}d^{\dagger}_{m}d_{m}d_{n}\Bigg),
\label{sys_el_hamiltonian_eqn}
\end{equation}
where $d^{\dagger}_{n}$ $(d_{n})$ creates (removes) an electron in the $n^{th}$ orbital in the system, where $n$ fully specifies both the orbital and the spin degrees of freedom of the electron, and the operators obey fermionic anti-commutation $\{d^{\dagger}_{n},d_{m}\}=\delta_{nm}$. The phononic part $H_{S,ph}(t)$ is given as the sum of a harmonic term and an anharmonic term
\begin{equation}
H_{S,ph}(t)= H_{S,ph}^{hm}(t)+H_{S,ph}^{ahm}\,,
\label{sys_ph_hamiltonian_eqn}
\end{equation}
where the time-dependent harmonic part $H_{S,ph}^{hm}(t)$ accounts for an external mechanical drive:
\begin{equation}
H_{S,ph}^{hm}(t)=\sum_{i}\Bigg(\frac{P'^{2}_{i}}{2m_i}-F'_i(t)Q'_i\Bigg)+\frac{1}{2}\sum_{i,j}K'_{ij}Q'_iQ'_j\,,
\label{sys_ph_harm_hamiltonian_0_eqn}
\end{equation}
where $Q'_i$ and $P'_i$ are the position and  momentum operators, respectively, of the $i^{th}$ atom of the system with mass $m_i$, so that we have the canonical quantization condition $[Q'_i,P'_j]=i\hbar\delta_{ij}$, and also that $P'_i=m_i\dot{Q'_i}$. The elastic coupling between any two sites $(i,j)$ is given by $K'_{ij}$ and the mechanical driving force acting on the $i^{th}$ lattice site is $F'_i(t)$. Upon rescaling to eliminate $m_i$, we get 
\begin{equation}
H_{S,ph}^{hm}(t)=\sum_{i}\Bigg(\frac{{P}_{i}^2}{2}-F_i(t)Q_i\Bigg)+\frac{1}{2}\sum_{i,j}{K}_{ij}Q_iQ_j \,,
\label{sys_ph_harm_hamiltonian_1_eqn}
\end{equation}
where $P'_{i}=\sqrt{m_i}P_i$, $Q'_{i}=\frac{Q_i}{\sqrt{m_i}}$, $K'_{ij}=\frac{{K}_{ij}}{\sqrt{m_im_j}}$, and $F'_i(t)=\sqrt{m_i}F_i(t)$, so that we now have $P_i=\dot{Q_i}$ .
Finally, upon performing an orthogonal change of basis \footnote{\label{ph_ortho_transf_footnote} 
Details of the orthogonal transformation used in Eq.\ \eqref{sys_ph_harm_hamiltonian_eqn}:
\begin{equation}
    Q_i=\sum_{r}C_{ir}Q_r\,,
\end{equation}
where from orthogonality we have $\sum_{r}C^{\mathrm{T}}_{ir}C_{rj}=\sum_{r}C_{ri}C_{rj}=\delta_{ij}$, and 
\begin{equation}
   P_i=\sum_{r}C_{ir}P_r \,,
\end{equation}%
and we can write
\begin{equation}
 \sum_{ij} C_{ir}K_{ij}C_{sj}={\omega}_r^2 \delta_{rs} \,,
\end{equation}
and 
\begin{equation}
F_{i}(t)=\sum_r C_{ir}F_r(t)\,,
\end{equation}
where $F_r$ is the force acting on the $r^{th}$ oscillator normal mode. %
}, 
we can write the Hamiltonian in diagonal form as
\begin{equation}
  H_{S,ph}^{hm}(t)= \sum_{r}\Bigg(\frac{P_{r}^{2}}{2}+\frac{\omega_{r}^{2}Q_{r}^{2}}{2} - F_r(t)Q_r \Bigg) \,,
\label{sys_ph_harm_hamiltonian_eqn}
\end{equation}
where $P_{r}$ and $Q_{r}$ are momentum and position operators, respectively, of the $r^{th}$ oscillator normal mode in the system, which has frequency $\omega_r$, and where we have the quantization condition $[Q_r,P_s]=i\hbar\delta_{rs}$. The anharmonic part $H_{S,ph}^{ahm}$ is given by 
\begin{equation}
	H_{S,ph}^{ahm}= H_{S,ph}^{(3)}+H_{S,ph}^{(4)}+... \,,
\end{equation}
where the cubic term $H_{S,ph}^{(3)}$ can be written as 
\begin{equation}
	H_{S,ph}^{(3)}=\sum_{p,r,s}[H_{S,ph}^{(3)}]_{prs} Q_p Q_r Q_s \,.
\end{equation}
The quartic term $H_{S,ph}^{(4)}$ and all higher-order anharmonic terms can be written analogously. The electron-phonon coupling is %
\begin{equation}	
H_{S,el-ph}=\sum_{r,n,m}[H_{S,el-ph}]_{rnm}Q_rd^{\dagger}_{n}d_{m}+...    	
\end{equation}
and may include higher-order terms in the lattice-electron potential expansion. 

The environment is modeled as $M$ perfectly ordered, semi-infinite Reservoirs of non-interacting electrons and phonons, each at Temperature $T_{\alpha}$ and chemical potential $\mu_{\alpha}$, with the Hamiltonian %
\begin{equation}
H_{R}=H_{R,el}+H_{R,ph}	
\label{res_hamiltonian_eqn}
\end{equation}
where the electronic sector of the $\alpha^{th}$ reservoir is 
\begin{equation}
    H_{R,el,\alpha}= %
    \sum_{k\in\alpha}\epsilon_{k}c_{k}^{\dagger}c_{k},
    \label{res_el_hamiltonian_eqn}
\end{equation}
where $c_{k}^{\dagger}$ $(c_{k})$ creates (removes) an electron in eigenmode $k$ in the reservoir and obey  $\{c_{k}^{\dagger},c_{l}\}=\delta_{kl}$, and we have  $\sum_{\alpha=1}^{M}H_{R,el,\alpha}=H_{R,el}$. The phononic sector of the $\alpha^{th}$ reservoir is given by the Hamiltonian
\begin{equation}
   H_{R,ph,\alpha}= \sum_{q\in\alpha}\Bigg(\frac{P_{q}^{2}}{2 }+\frac{\omega_{q}^{2}U_{q}^{2}}{2}\Bigg) \,,
   \label{res_ph_hamiltonian_eqn_1}
\end{equation}
where $P_{q}$ and $U_{q}$ are momentum and position operators, respectively, for the $q^{th}$ oscillator normal mode with frequency $\omega_q$ in the reservoir, satifying the canonical quantization condition $[U_p,P_q]=i\hbar\delta_{pq}$. Finally, we have  $\sum_{\alpha=1}^{M}H_{R,ph,\alpha}=H_{R,ph}$.

The coupling between the System and the Reservoirs is given by the Coupling Hamiltonian
\begin{equation}
H_{SR}= H_{SR,el} + H_{SR,ph} \,,	
\label{cpl_hamiltonian_eqn}
\end{equation}
where the electronic coupling to the $\alpha^{th}$ reservoir is given by
\begin{equation}
H_{SR,el,\alpha}=\sum_{k\in\alpha,n}(V_{kn}^{el}c_{k}^{\dagger}d_{n}+h.c.),
\label{el_cpl_hamiltonian_eqn}
\end{equation}
with 
$H_{SR,el}=\sum_{\alpha=1}^{M}H_{SR,el,\alpha}$, and the phononic coupling to the $\alpha^{th}$ reservoir is given by
\begin{equation}
H_{SR,ph,\alpha}=\frac{1}{2}\sum_{q\in\alpha,r}\left(V_{qr}^{ph}U_{q}Q_{r}+V_{rq}^{ph}Q_r U_q\right), 
\label{ph_cpl_hamiltonian_eqn}
\end{equation}
with $V_{rq}^{ph}=V_{qr}^{ph}$ and $H_{SR,ph}=\sum_{\alpha=1}^{M}H_{SR,ph,\alpha}$.
\subsection{Physical basis of the model Hamiltonian}\label{phys_basis_subsec}

We note some points here about the modeling of the Hamiltonian just presented that are important from both physical and computational points of view. 

First, the Reservoirs are modeled as non-interacting, semi-infinite, and perfectly ordered so that they faithfully represent
the external macroscopic circuit used to impose any nonequilibrium bias on the system and to measure the resultant flows of
charge and energy. As counters of charge and energy flowing out the system, it makes little sense to include two-body interactions in the Reservoirs, since that would significantly complicate the task of calculating these quantities, and many-body correlations are not relevant in such a macroscopic electric/thermal circuit in any case. The interesting case of superconducting Reservoirs \cite{arnoldSuperconductingTunnelingTunneling1987,averinAcJosephsonEffect1995,cuevasHamiltonianApproachTransport1996,scheerConductionChannelTransmissions1997}
is outside the scope of the analysis in this article. 

The coupling between the System and the Reservoir $H_{SR}$ is also taken to be quadratic.
Realistically, this is because the electrons in the metallic Reservoirs, modeled by $H_{R}$, are well screened and hence any interaction between the electrons in the System and the Reservoir can be treated with the method of image charges \cite{PhysRevB.86.115403}, the screening charges in the Reservoirs treated as slave degrees of freedom,
rather than as independent quantum modes.  This is justified because the screening charges in the reservoirs respond at the 
plasma frequency, which is much greater than typical response frequencies of a mesosopic system.  Moreover, if many-body effects in the interfacial regions of the reservoirs are important, these regions can simply be included in the System.

Second, the number of Reservoir degrees of freedom entering $H_{SR}$ is a set of measure zero \cite{bergfieldNumberTransmissionChannels2011a,bergfieldTransmissionEigenvalueDistributions2012a} compared to the full (infinite) Reservoir Hilbert space, with only a few frontier orbitals taking part in coupling the environment to the System. This is because tunnel coupling is local at the {\em Interface} between the System and Reservoir and the screening charges are also localized at the Interface \footnote{ 
The clarification about the spatial extent of the coupling is of direct relevance to the central result of this work and one that needs highlighting in view of
the claim that inclusion of Reservoir degrees of freedom in the System Internal Energy is ``problematic from an operational perspective''
 \cite{strasbergQuantumStochasticThermodynamics2022_ch3}.}.\nocite{strasbergQuantumStochasticThermodynamics2022_ch3} 
The same holds true for the phononic degrees of freedom, whose elastic coupling is short-ranged \cite{ashcroft_solid_1976_ch20}. A model calculation of the spatio-temporal dynamics of the Interface, illustrating these principles, is presented in Secs.\ \ref{int_egy_tbchain_subsubsec_1} and \ref{int_egy_tbchain_subsubsec_2}.

\section{Non-Equilibrium Thermodynamic Quantities}\label{noneqbthermo_sec}

With the open quantum system defined, we now compute the nonequilibrium thermodynamic quantities of the system.

\subsection{Work done by external forces}\label{wext_1_subsec}

Following the so-called Inclusive definition of work, standard in the Quantum Thermodynamics literature \cite{jarzynskiComparisonFarfromequilibriumWork2007} (in contrast with the Exclusive definition \cite{1977ZhETF..72..238B}), we identify the time rate of change of the expectation value of the total Hamiltonian as the rate of work done by external forces, $\dot{W}_{ext}(t)$, which for the form assumed for the Hamiltonian, we can compute as 
\begin{equation}
\frac{d}{dt}\langle H(t) \rangle \equiv \dot{W}_{ext}(t)\ = \langle\dot{H}_{S}(t)\rangle \,,	
\label{wext_def_eqn}
\end{equation}
where $\langle \, \rangle$ denotes the quantum and statistical average.

The equality can be proved \cite{ballentineChapterKinematicsDynamics2014} by noting that $\langle H(t) \rangle=\mathrm{Tr}\{\rho (t) H(t)\}$, with $\rho (t)$ denoting the full density matrix at time $t$ and $\mathrm{Tr}\{\}$ denoting the Trace operation over the full Fock Space, and using the chain rule we can write 
\begin{equation}	
\frac{d}{dt}\langle H(t) \rangle=\mathrm{Tr}\{\dot{\rho}(t)H(t)+\rho(t) \dot{H}(t)\} \,,
\end{equation}
which, with the equation of motion of the density matrix 
\begin{equation}
i\hbar \dot{\rho}=[H(t),\rho(t)] \,,
\end{equation}
and cyclicity of the Trace gives the stated result, $\dot{W}_{ext}(t)\ = \langle\dot{H}_{S}(t)\rangle$, since the time-dependence is only in the System Hamiltonian $H_{S}(t)$.	

Furthermore, since the time-dependence in $H_S(t)$ sits only in the one-body terms, we get 
\begin{equation}
 \dot{W}_{ext}(t)= \dot{W}_{ext,el}(t)+ \dot{W}_{ext,ph}(t)\,,
 \label{wext_1_eqn}
\end{equation}
where the power delivered by the electronic drive is 
\begin{equation}
 \dot{W}_{ext,el}(t)=  \sum_{n,m}\langle [\dot{H}^{(1)}_{S,el}(t)]_{nm}d^{\dagger}_{n}d_{m}\rangle\,,
 \label{wext_el_1_eqn}
\end{equation}
and that by the mechanical drive is 
\begin{equation}
 \dot{W}_{ext,ph}(t)= -\sum_{r}\dot{F}_r(t)\langle Q_r\rangle
 \label{wext_ph_1_eqn}
\end{equation}

\subsection{Particle Current and Electrochemical Power}
The electronic particle current \(I_{\alpha,el}^{N}(t)\) into reservoir $\alpha$ is given by the rate of change of the expectation value of the occupation number operator \(N_{\alpha}=\sum_{k\in\alpha}c_{k}^{\dagger}c_{k}\) of the $\alpha^{th}$ reservoir 
\begin{equation} \label{particle_el_curr_def_eqn}
  I_{\alpha,el}^{N}(t):=  \frac{d}{dt}\langle N_{\alpha}\rangle\,,
\end{equation}
which can be evaluated as
\begin{equation}
\frac{d}{dt}\langle N_{\alpha}\rangle=\frac{-i}{\hbar}\langle [N_{\alpha},H(t)]\rangle=\frac{-i}{\hbar}\langle 
[N_{\alpha},H_{SR}]\rangle \,.
\end{equation}  
Using the identities \([A,BC]=\{A,B\}C-B\{C,A\}=[A,B]C-B[C,A]\), we obtain the particle current as
\begin{equation}
I_{\alpha,el}^{N}(t)=\frac{-i}{\hbar}\sum_{k\in\alpha,n}[V_{kn}^{el}\langle c_{k}^{\dagger}d_{n}\rangle-(V^{el}_{kn})^{*}\langle d_{n}^{\dagger}c_{k}\rangle]\,.
\label{particle_curr1_eqn}
\end{equation}
The electrical current is obtained by  multiplying the above equation by the electric charge quantum.

The electrochemical power delivered to reservoir $\alpha$ is $\mu_\alpha I^N_{\alpha,el}(t)$ and the net electrochemical power delivered to the system by the reservoirs is
\begin{equation}
    \dot{W}_{elec}(t):=\sum_{\alpha}-\mu_{\alpha} I_{\alpha,el}^{N}(t).
\end{equation}

\subsection{Heat, the Role of Reservoirs, and Hilbert space partition}\label{heat_roleofres_subsec}

Heat is a quantity whose fundamental definition $\dbar Q=TdS$ is only valid for quasi-static processes %
\cite{LANDAU198034,Callen1985}, and as such, must be accounted for carefully in strongly-driven systems.
This requirement is met exactly by the non-interacting semi-infinite reservoirs to which the nonequilibrium system is coupled. The relevant properties of the reservoir model have been long established in the Mesoscopics and Quantum Transport literature \cite{buttikerRoleQuantumCoherence1986, barangerElectricalLinearresponseTheory1989,nockelAdiabaticTurnonAsymptotic1993}, where they enforce an Ordering of Limits such that one must take the limit of the Reservoir size going to infinity $L\to\infty$ \emph{before} taking limit of the adiabatic perturbation switch-on time going to infinity $t_{adiab}\to\infty$.
Physically, $t_{adiab}$ is related to the timescale of setting up and carrying out the experiment. This order of limits ensures that any electrons or phonons emitted by the System into the Reservoirs cannot be coherently backscattered into the system (causing decoherence of any quanta emitted into the Reservoirs despite the absence of 2-body interactions) and that the distributions of charge and energy in the Reservoirs cannot be depleted due to flows mediated by the System.

The entropy is given by the von Neumann formula \cite{vonneumannMathematicalFoundationsQuantum2018}
\begin{equation}
    S=- {\rm Tr}\{\rho \ln \rho\},
    \label{eq:vonNeumann}
\end{equation}
where $\rho$ is the density matrix of the universe and we have set $k_B=1$.  For small deviations from an initial density matrix $\rho_0$, the linear change in entropy is
\begin{equation}
    dS=- {\rm Tr}\{d\rho \ln \rho_0\}
\label{lineardev_entropy_eqn}
\end{equation}
(See Appendix \ref{Hspacediv_app} for a derivation.)
For an initial equilibrium state, $\rho_0$ is given by the grand canonical ensemble
\begin{equation}
    \rho_0=\frac{e^{-\beta (H-\mu N)}}{\cal Z},
\end{equation}
where ${\cal Z}={\rm Tr}\{e^{-\beta (H-\mu N)}\}$ is the grand partition function, leading to the standard result
\begin{equation}
    T dS = {\rm Tr}\{d\rho (H-\mu N)\}.
\end{equation}

In order to see how the heat $TdS$ is partitioned among the various subsystems, it is useful to first consider the case of independent fermions. Then $T dS = \sum_\gamma T dS_\gamma$, where the heat added to subsystem $\gamma$ is \cite{shastryThirdLawThermodynamics2019,shastryTheoryThermodynamicMeasurements2019}
\begin{equation}
    T dS_\gamma = \int d\varepsilon \, \mathbb {Tr}^{(1)}_\gamma \{A(\varepsilon)\}(\varepsilon-\mu) df(\varepsilon),
    \label{eq:TdS}
\end{equation}
where $A(\varepsilon)=\delta(\varepsilon-h^{(1)})$ is the spectral function, with $h^{(1)}$ the Hilbert-space operator corresponding to the Fock-space Hamiltonian $H^{(1)}$, and $df(\varepsilon)=f(\varepsilon)-f_0(\varepsilon)$ is the linear deviation of the
distribution function from the initial equilibrium Fermi-Dirac distribution. Here $\mathbb{Tr}^{(1)}_\gamma$ is the trace over the subspace $\gamma$ of the single-particle Hilbert-space. We note that $\gamma\in\{S,R\}$ where $S$ denotes the subspace/subsystem containing all the degrees of freedom belonging to the system and $R=\cup_{\alpha=1}^{M}R_{\alpha}$, where $R_{\alpha}$ denotes the subspace/subsystem containing all the degrees of freedom belonging to the $\alpha^{th}$ reservoir. Specifically,
\begin{equation}
   \mathbb{Tr}^{(1)}_\gamma \{O^{(1)}\} = 
   \mathbb{Tr}^{(1)} \{ \Pi_\gamma \,O^{(1)}\},
   \label{eq:Tr_gamma}
\end{equation}
where
\begin{equation}
    \Pi_\gamma = \sum_{x\in \gamma} |x\rangle \langle x|
\end{equation}
is the projection operator onto subspace $\gamma$ of
the single-particle Hilbert space and $O^{(1)}$ is any linear operator on the single-particle Hilbert space.

Eq.\ (\ref{eq:TdS}) 
may be integrated to obtain the equilibrium subsystem entropy
\begin{equation}
    S_\gamma = \int d\varepsilon \, \mathbb {Tr}^{(1)}_\gamma \{A(\varepsilon)\} s(\varepsilon),
    \label{subsys_entropy_eqn}
\end{equation}
which is non-singular \cite{shastryThirdLawThermodynamics2019} and respects the Third Law of Thermodynamics as $T\rightarrow 0$, where
\begin{equation}
    s(\varepsilon) = \beta (\varepsilon -\mu) f(\varepsilon) +  \ln(1+e^{-\beta(\varepsilon-\mu)}).
    \label{eq:s_eq}
\end{equation}
The connection of the entropy partition given by Eq.\ (\ref{eq:TdS}) and the required energetic partition is determined by the identity 
$\varepsilon A(\varepsilon) =  h^{(1)} A(\varepsilon)$,
which allows Eq.\ (\ref{eq:TdS}) to be rewritten as
\begin{equation}
    T dS_\gamma = d\langle \left.H^{(1)}\right|_\gamma \rangle -\langle \left.dH^{(1)}\right|_\gamma \rangle - \mu  d\langle N_\gamma \rangle,
    \label{eq:dS_gamma}
\end{equation}
where $\left.H^{(1)}\right|_\gamma$ is the Fock-space operator corresponding to the partition of $h^{(1)}$ onto subspace $\gamma$  \cite{webb2023}
\begin{equation}
    \left.h^{(1)}\right|_\gamma \equiv \frac{1}{2} \{h^{(1)}, \Pi_\gamma\},
    \label{eq:H_gamma}
\end{equation}
where $\{A,B\}=AB+BA$ is the anticommutator. (See Appendix \ref{Hspacediv_app} for details of the derivation.)
The symmetrization in the definition (\ref{eq:H_gamma}) is needed to ensure the hermiticity of the observable $\left.H^{(1)}\right|_\gamma$, which satisfies 
\begin{equation}
    H^{(1)}=\sum_\gamma \left.H^{(1)}\right|_\gamma.
\end{equation}

We point out that any energetic partition other than that dictated by Hilbert space partitioning as given by Eq.\ (\ref{eq:H_gamma}) will lead to a singular entropy partition \cite{webb2023} and a violation of the Third Law, because only the Hilbert space partition described above 
allows for the detailed cancellation of the separately divergent terms in Eq.\ (\ref{eq:s_eq}) as $\beta\rightarrow\infty$. The singular character of
the entropy for alternate energetic partitions was pointed out previously by Esposito et al.\ \cite{espositoNatureHeatStrongly2015}.

The partition of the full Hamiltonian onto reservoir $\alpha$ is
\begin{equation}
    \left.H\right|_{\gamma=R_{\alpha}} = H_{R,\alpha} + \frac{1}{2}H_{SR,\alpha},
    \label{eq:H|R}
\end{equation}
where $H_{R,\alpha}$ is diagonal on the partition, while $H_{SR,\alpha}$ is purely off-diagonal, and thus acquires a factor of 1/2 via Eq.\ (\ref{eq:H_gamma}).
The partition goes through as in the independent-particle case (\ref{eq:H_gamma}) because $H^{(2)}$ is system diagonal (see Appendix \ref{Hspacediv_app} for details).  The same holds true for phonon anharmonic terms and electron-phonon interactions, both of which vanish within the reservoirs, so that Eq.\ (\ref{eq:H|R}) is valid for the full Hamiltonian.

\subsection{Heat current}
\label{sec:heat_current}

We now compute the heat current into reservoir $\alpha$ by extending the thermodynamic identity (\ref{eq:dS_gamma}) to the time-dependent
nonequilibrium regime, where the chemical potentials and temperatures of the various reservoirs can differ by arbitrarily large amounts, 
but where
it is assumed that the separate equilibrium distributions in the macroscopic reservoirs are perturbed only linearly from their respective
equilibrium states due to energy and particle flows mediated by the system:
\begin{equation}
I_{\alpha}^{Q}(t)\equiv T_{\alpha}\frac{d}{dt} S_{\alpha}=\frac{d}{dt}\langle \left.H \right|_{\gamma=R_{\alpha}} \rangle
- \langle \left.\dot{H} \right|_{\gamma=R_{\alpha}} \rangle
-\mu_{\alpha}\frac{d}{dt}\langle N_{\alpha}\rangle \,,
\label{el_heat_curr_defn_eqn}
\end{equation}
where, to lighten notation, we write $N_{\alpha}\equiv N_{\gamma=R_{\alpha}}$, which denotes the particle number operator for reservoir $\alpha$.
We remark that the interpretation of Eq.\ (\ref{el_heat_curr_defn_eqn}) as a bona fide heat current is valid for arbitrary nonequilibrium steady-state \cite{bergfieldThermoelectricSignaturesCoherent2009} and slowly-varying flows, but that rapidly-varying transient flows 
described by Eq.\ (\ref{el_heat_curr_defn_eqn}) may simply be identified as one component of the total energy flux.

We next specialize to the case where $H_R$ and $H_{SR}$ are independent of time. (For an analysis of the general time-dependent case, see Appendix \ref{full_tdpt_generalize_app}.)
It is convenient to write the heat current as the sum of two terms 
\begin{equation}
  I_{\alpha}^{Q}(t)=  I_{\alpha}^{Q,conv}(t)+I_{\alpha}^{Q,tr}(t)\,,  
\end{equation}
where
\begin{equation}
   I_{\alpha}^{Q,conv}(t)= \frac{d}{dt}\langle H_{R,\alpha} \rangle
-\mu_{\alpha}\frac{d}{dt}\langle N_{\alpha}\rangle
\end{equation}
is the conventional component of the heat current and reduces to the Bergfield-Stafford formula \cite{bergfieldThermoelectricSignaturesCoherent2009d} in steady state, and  
\begin{equation}
    I_{\alpha}^{Q,tr}(t)= \frac{1}{2}\frac{d}{dt}\langle H_{SR,\alpha} \rangle\,.
    \label{eq:IQtr}
\end{equation}
is the transient component of the heat current which vanishes in steady state. We further define the heat currents for the electronic and phononic sectors separately by writing 
\begin{equation}
 I_{\alpha,el/ph}^{Q}(t)=  I_{\alpha,el/ph}^{Q,conv}(t)+I_{\alpha,el/ph}^{Q,tr}(t)\,, 
\end{equation}
where
\begin{equation}
    I_{\alpha,el/ph}^{Q,conv}(t)= \frac{d}{dt}\langle H_{R,el/ph,\alpha} \rangle
-\mu_{\alpha}\frac{d}{dt}\langle N_{\alpha}\rangle \,,
\label{conv_el_heatcurr_defn_eqn}
\end{equation}
with $\mu_\alpha=0$ $\forall \alpha$ for phonons, and 
\begin{equation}
    I_{\alpha,el/ph}^{Q,tr}(t)= \frac{1}{2}\frac{d}{dt}\langle H_{SR,el/ph,\alpha} \rangle\,.
    \label{tr_el_heatcurr_defn_eqn}
\end{equation}

Carrying out commutator evaluations similar to that for the particle current, we get the conventional component of electronic heat current
\cite{bergfieldThermoelectricSignaturesCoherent2009d}
\begin{equation}
I_{\alpha,el}^{Q,conv}(t)=\frac{-i}{\hbar}\sum_{k\in\alpha,n}(\epsilon_k-\mu_\alpha)[V_{kn}^{el}\langle c_{k}^{\dagger}d_{n}\rangle-(V_{kn}^{el})^{*}\langle d_{n}^{\dagger}c_{k}\rangle] \,,
\label{conv_el_heatcurr_expval_eqn}
\end{equation}
which can be cast in terms of an energy weighted particle current (for the $k^{th}$ mode) as
\begin{equation}
I_{\alpha,el}^{Q,conv}(t)=
\sum_{k\in\alpha,n}(\epsilon_k-\mu_\alpha)I_{k}^{N}(t) \,,	
\label{conv_el_heatcurr2_eqn}
\end{equation}
where $I_{k}^{N}(t)=\frac{-i}{\hbar}[V_{kn}^{el}\langle c_{k}^{\dagger}d_{n}\rangle-(V^{el}_{kn})^{*}\langle d_{n}^{\dagger}c_{k}\rangle]$. The transient component of the electronic heat current $I_{\alpha,el}^{Q,tr}(t)$ can be computed by the taking the time-derivative of 
\begin{equation}
  \frac{1}{2}\langle H_{SR,el,\alpha} \rangle=  \frac{1}{2} \sum_{k\in\alpha,n}\langle V_{kn}^{el}c_{k}^{\dagger}d_{n}+h.c.\rangle
  \label{tr_el_heatcurr_expval_eqn}
\end{equation}
The conventional component of the phononic heat current \cite{galperinHeatConductionMolecular2007,leekMathematicalDetailsApplication2012,wangNonequilibriumGreenFunction2014} can be analogously evaluated as
\begin{equation}
  I_{\alpha,ph}^{Q,conv}(t)= -\sum_{q\in\alpha,r}V^{ph}_{qr}\langle P_qQ_r\rangle\,,
\label{conv_ph_heatcurr_expval_eqn}
\end{equation}
and the transient component of the phononic heat current $I_{\alpha,ph}^{Q,tr}(t)$ can be computed by taking the time-derivative of 
\begin{equation}
  \frac{1}{2}\langle H_{SR,ph,\alpha} \rangle= \frac{1}{2} \sum_{q\in\alpha,r} V_{qr}^{ph}\langle U_{q}Q_{r}\rangle \,.
  \label{tr_ph_heatcurr_expval_eqn}
\end{equation}

The time derivative of the expectation value of the full Hamiltonian partitioned onto the Reservoir subspace $\left.H\right|_{\gamma=R}$  has a simple interpretation: it is the total energy current flowing into the Reservoirs $\sum_{\alpha}  I_{\alpha}^{E}(t)$ i.e.
\begin{equation}
\frac{d}{dt}\langle H_{R}(t) + \frac{1}{2} H_{SR} \rangle= \sum_{\alpha}  I_{\alpha}^{E}(t) \,.
\end{equation} 
\textit{At infinity}, this in turn is just the electrical (or electrochemical) power delivered to the reservoir plus the heat current flowing into it
\begin{equation}
\sum_{\alpha}  I_{\alpha}^{E}(t)=\sum_{\alpha}\mu_{\alpha}  I_{\alpha,el}^{N}(t)+\sum_{\alpha}I_{\alpha}^{Q}(t)\,,
\label{res_current_eqn}
\end{equation}
where $I_{\alpha}^{Q}(t)=I_{\alpha,el}^{Q}(t)+I_{\alpha,ph}^{Q}(t)$.  We note that while the electrochemical power and electronic heat current at infinity can be properly interpreted as such for general nonequilibrium driving of the system, according to the arguments given above, the phonon contribution to Eq.\ \eqref{res_current_eqn} may only be identified as a bona fide contribution to the heat current for steady-state or slowly varying flows, and may simply be a
contribution to the total energy flux into the reservoirs for rapid transient flows.

\section{The First Law of Quantum Thermodynamics}\label{firstlaw_subsec}

From the discussion in Sec.\ \ref{wext_1_subsec}, specifically Eq.\ \eqref{wext_def_eqn}, it follows that
\begin{equation}
	\frac{d}{dt}\langle H_{S}(t)+H_{SR}+H_{R}\rangle=\dot{W}_{ext}(t)\,.
\end{equation}	
Furthermore, with the identification of Sec.\ \ref{heat_roleofres_subsec}, specifically Eq.\ \eqref{res_current_eqn}, we get  
	\begin{eqnarray}
	    \frac{d}{dt}\langle H_{S}(t)+\frac{1}{2}H_{SR}\rangle &=& \dot{W}_{ext}(t)+(\sum_{\alpha}-\mu_{\alpha}  I_{\alpha}^{N}(t))+\nonumber\\&&(\sum_{\alpha}- I_{\alpha}^{Q}(t))\,.
	\end{eqnarray}
It is then clear that the internal energy operator for the ``System" must be recognized as sum of the System Hamiltonian and \textit{half} the Coupling Hamiltonian.
Defining $\dot{W}_{elec}(t):=\sum_{\alpha}-\mu_{\alpha} I_{\alpha,el}^{N}(t)$ as the electrochemical power delivered to the system and $\dot{Q}(t):=\sum_{\alpha}- I_{\alpha}^{Q}(t)$ as the heat current flowing into the system allows us to write the {\em First Law of Quantum Thermodynamics} as
\begin{IEEEeqnarray}{lcr}	
	\frac{d}{dt}\langle U_{sys}(t) \rangle = \dot{W}_{ext}(t)+\dot{W}_{elec}(t) +
	\dot{Q}(t) \,, 
	\label{firstlaw1_eqn}
\end{IEEEeqnarray}
where the Internal Energy operator $U_{sys}(t)$ for the open quantum system is just the Hamiltonian partitioned on the system Hilbert space
\begin{equation}
U_{sys}(t)\equiv \left.H(t)\right|_S= H_{S}(t)+\frac{1}{2}H_{SR} \,.
\label{internal_energy_op_eqn}
\end{equation}
This analysis lays to rest any ambiguity about the internal energy operator and gives a quantum mechanically consistent and experimentally meaningful division of the energetics of a fairly general open quantum system. We also note that this identification of the First Law and Internal Energy operator holds true even in the  more general case where both the coupling and reservoirs become explicitly time-dependent, as discussed in Appendix  \ref{full_tdpt_generalize_app}.%

The definition (\ref{internal_energy_op_eqn}) agrees with that proposed by Ludovico et al. \cite{ludovicoDynamicalEnergyTransfer2014} for the specific model of a single sinusoidally driven Fermionic  resonant level coupled to a single reservoir (supported in the subsequent work of Bruch et al. \cite{bruchQuantumThermodynamicsDriven2016}), as well as with that put forward by Stafford and Shastry \cite{staffordLocalEntropyNonequilibrium2017,shastryThirdLawThermodynamics2019,shastryTheoryThermodynamicMeasurements2019}
for  an independent Fermion system in steady state,
but disagrees \footnote{\label{prev_ver_footnote}We note that in the previous version of this preprint  \cite{kumarFirstLawThermodynamics2022} we presented arguments which lead to an agreement with the internal energy identification of Refs. \cite{espositoEntropyProductionCorrelation2010,strasbergFirstSecondLaw2021a,lacerdaQuantumThermodynamicsFast2023} and a disagreement with that of Refs. \cite{ludovicoDynamicalEnergyTransfer2014,bruchQuantumThermodynamicsDriven2016}--a view that has been reversed in light of the arguments presented in the preceding sections.%
} with those put forward by Esposito et al. \cite{espositoEntropyProductionCorrelation2010}, Strasberg and Winter \cite{strasbergFirstSecondLaw2021a}, and Lacerda et al. \cite{lacerdaQuantumThermodynamicsFast2023}. %

While Eq.\ \eqref{internal_energy_op_eqn} provides an explicit quantum mechanical operator for the internal energy, 
it should be noted that the other thermodynamic quantities entering the First Law cannot in general be associated with explicit quantum operators. Work, being a path (and not a state) function cannot, in general, be a Quantum Operator \cite{talknerFluctuationTheoremsWork2007,talknerAspectsQuantumWork2016}. Furthermore, Heat (again a path function) also cannot have a quantum operator associated with it since it is a statistical quantity ($\dbar Q=TdS$) defined unambiguously only under the special conditions discussed in Sec.\ \ref{heat_roleofres_subsec}. 

\subsection{Energy Density}\label{energy_density_1_sec}

To further study the spatio-temporal distribution of  energy in the driven open quantum system---for the special case where 2-body interactions are absent---we make use of a general result for the spatial density of any one-body observable \cite{webb2023,2022APS_MM_poster_densityandtopology}.
The energy density is given by
\begin{equation}
    \rho_H(x,t)=\mbox{Tr}\left\{\rho(t) \left.H^{(1)}\right|_x (t)\right\},
    \label{eq:rho_H}
\end{equation}
where $\left.H^{(1)}\right|_x (t)$ is the Fock-space operator corresponding to the Hilbert-space Hamiltonian density
\begin{equation}
    \left.h^{(1)}\right|_x (t) = \frac{1}{2} \{h^{(1)}(t),|x\rangle\langle x|\}.
\end{equation}

\section{Evaluating the Time-Dependent Energetics: NEGF Formulas}\label{NEGFresults_sec}
	
We employ the Nonequilibrium Green's function (NEGF) formalism \cite{kadanoffQuantumStatisticalMechanics1962,keldyshDiagramTechniqueNonequilibrium1964,rammerQuantumFieldTheory2007,haugQuantumKineticsTransport2007,stefanucciNonequilibriumManyBodyTheory2013} to evaluate all the terms appearing in the First Law \eqref{firstlaw1_eqn}, each  term on the LHS of which we have evaluated in terms of the expectation values in Eqs.\ \eqref{wext_1_eqn}, \eqref{particle_curr1_eqn}, \eqref{conv_el_heatcurr2_eqn}, and \eqref{conv_ph_heatcurr_expval_eqn},
and the time derivatives of Eqs.\ \eqref{tr_el_heatcurr_expval_eqn} and \eqref{tr_ph_heatcurr_expval_eqn}.
Here we present only the final results of our analysis and defer the relevant details of the derivations and methods of evaluation to Appendix \ref{NEGF_details_app}. We also note that the NEGF formalism has a well-established connection with the S-Matrix formalism \cite{gasparianPartialDensitiesStates1996}---another technique often employed in quantum transport and thermodynamics. A key advantage of NEGF is its simplicity and power in computing one-body (or even few-body) observables. This is in contrast to the Master equation approach \cite{breuerTheoryOpenQuantum2002a,dittrichQuantumTransportDissipation1998a,haugQuantumKineticsTransport2007}, which works in the full many-particle Hilbert space.

\subsection{Internal energy}
\label{sec:NEGF_U}

The expectation value of the internal energy operator defined in Eq.\ \eqref{internal_energy_op_eqn} can be evaluated in terms of the system Green's functions.  The first term in $U_{sys}(t)$ can be written as a sum of three terms
\begin{equation}\label{HS_negf_eqn}
    \langle H_{S}(t)\rangle =\langle H_{S,el}(t) \rangle + \langle H_{S,ph} (t) \rangle + \langle H_{S,el-ph} \rangle \,,
\end{equation}
where the electronic part can be written as 
\begin{eqnarray}\label{HS_el_negf_eqn}
   \langle H_{S,el}(t) \rangle &=& -i\mathbb{Tr}^{(1)}\Bigg\{H^{(1)}_{S,el}(t)G^{<}(t,t)\Bigg\}\nonumber\\ & & -i\mathbb{Tr}^{(2)}\Bigg\{H^{(2)}_{S,el}G^{(2)}(t,t)\Bigg\} \,,
\end{eqnarray}
where the system electronic Lesser Green's function is defined as 
\begin{equation}
		G^{<}_{nm}(t,t_1)= i\langle d_{m}^{\dagger}(t_1)d_{n}(t)\rangle \,,
		\label{sys_lesser_negf_eq}
\end{equation}
and the 2-body electronic Green's function is defined as
\begin{equation}
 G^{(2)}_{nnmm}(t,t)=i\langle d^{\dagger}_{n}(t)d^{\dagger}_{m}(t)d_{m}(t)d_{n}(t)\rangle    
\end{equation}(see Appendix \ref{NEGF_details_app} for the method of their evaluation). The phononic part can be written as  
\begin{eqnarray}\label{HS_ph_negf_eqn}
    \langle H_{S,ph} (t) \rangle &=& \frac{i}{2}\mathbb{Tr}^{(1)}\Bigg\{\Bigg(\frac{\partial^2}{\partial t \partial t'}+K\Bigg) \left.D^<(t,t')\right|_{t=t'}\Bigg\} \nonumber \\ && +\frac{1}{\hbar}\int_{-\infty}^{t} \,dt' \Bigg[\mathbf{F}(t)\mathcal{D}(t,t')\mathbf{F}^{\mathrm T}(t')\Bigg]\nonumber \\ & & -i\mathbb{Tr}^{(3)}\Bigg\{H_{S,ph}^{(3)}D^{ahm}(t,t,t)\Bigg\} \,,
\end{eqnarray}
 where $K$ denotes a matrix whose elements $K_{ij}$ are the harmonic couplings introduced in Eq.\ \eqref{sys_ph_harm_hamiltonian_1_eqn}, and $\mathbf{F}(t)=(...,F_r(t),...)$, is a (row) vector whose elements are the mechanical driving forces acting on the system. The system phononic Lesser Green's function is
\begin{equation} 
D^{<}_{rs}(t,t_1)=-i\langle Q_{s}(t_1)Q_{r}(t)\rangle
\end{equation}
and $\mathcal{D}(t,t')$ is the time-ordered phonon Green's function, to be evaluated in the absence of the time-dependent mechanical drive, whose matrix element is defined as 
\begin{equation}
   \mathcal{D}_{rs}(t,t')=-i\langle \mathrm{T}(Q_r(t)Q_s(t'))\rangle\,, 
\end{equation}
where $\mathrm{T}$ denotes the time-ordering operator. Finally, $D^{ahm}_{prs}(t,t,t)=i\langle Q_p(t) Q_r(t) Q_s(t)\rangle$ is the lowest-order anharmonic phonon Green's function (higher-order phononic Green's functions can be constructed analogously). The electron-phonon coupling contribution is 
\begin{equation} \label{HS_elph_negf_eqn}
	\langle H_{S,el-ph} \rangle = -i\mathbb{Tr}^{(2)}\Bigg\{
	H_{S,el-ph}G^{el-ph}(t,t,t)\Bigg\}\,,
\end{equation}
where the electron-phonon Green's function is defined as $G^{el-ph}_{rnm}(t,t,t)=i\langle Q_r(t)d^\dagger_n(t)d_m(t)\rangle$. 

The second term in the internal energy may be
evaluated as (see Appendix \ref{NEGF_details_app} for details)
\begin{equation}
    \langle H_{SR}\rangle=\sum_{\alpha=1}^{M}\Bigg(\langle H_{SR,el,\alpha}(t)\rangle+\langle H_{SR,ph,\alpha}(t)\rangle\Bigg),
    \label{HSR_negf_eqn}
\end{equation}
where the electronic part is given by 
\begin{eqnarray}
\langle H_{SR,el,\alpha}(t)\rangle &=& 2\int_{-\infty}^{+\infty}\,\frac{d\omega}{2\pi}\int_{-\infty}^{+\infty}\,dt_1\mathbb{ImTr}^{(1)}\Bigg\{e^{-i\omega(t_1-t)}\nonumber \\ &&
[G^{<}(t,t_1)\Sigma^{A}_{\alpha}(\omega)+ G^{R}(t,t_1)\Sigma^{<}_{\alpha}(\omega)]	\Bigg\}\,,\nonumber \\ 
    \label{HSR_el_negf_eqn}
\end{eqnarray}
where the system electronic Retarded Green's function is defined as 
\begin{equation}
		G^{R}_{nm}(t,t_1)= -i\theta(t-t_1)\langle \{d_{n}(t),d_{m}^{\dagger}(t_1)\}\rangle \,,
		\label{sys_ret_negf_eq}
\end{equation}
and $\Sigma^{A}(\omega)$ and $\Sigma^{<}(\omega)$
are the advanced and lesser electronic self-energy functions, respectively, given by 
\begin{equation}
 [\Sigma^{A/<}_{\alpha}(\omega)]_{nm}=\sum_{k\in\alpha}V_{kn}g^{A/<}_{kk}(\omega)V_{km}^{*}  \,,
 \label{elec_self_egy_eqn}
\end{equation} 
where $g^{A}_{kk}(t,t')=i\theta (t'-t)\langle \{c_{k}(t),c^{\dagger}_{k}(t')\}\rangle$ and $g^{<}_{kk}(t,t')=i\langle c^{\dagger}_{k}(t')c_{k}(t)\rangle$ are the  uncoupled reservoir advanced and lesser electronic Green's functions, respectively. The analogous phononic part is given by 
\begin{eqnarray}
 \langle H_{SR,ph,\alpha}(t)\rangle &=&\frac{-i}{\hbar}\int_{-\infty}^{+\infty}\,\frac{d\omega}{2\pi}\int_{-\infty}^{+\infty}\,dt_1\mathbb{Tr}^{(1)}\Bigg\{e^{-i\omega(t_1-t)}\nonumber \\ 
&\times& [D^{<}(t,t_1)\Pi^{tun,A}_{\alpha}(\omega)\nonumber \\ 
&& +D^{R}(t,t_1)\Pi^{tun,<}_{\alpha}(\omega)]\Bigg\}\,.\nonumber \\ 
\label{HSR_ph_negf_eqn}
\end{eqnarray} 
where the system phononic Retarded Green's function is defined as
\begin{equation} 
D^{R}_{rs}(t,t_1)=-i\theta(t-t_1)\langle[Q_{r}(t),Q_{s}(t_1)]\rangle	\,,
\end{equation} 
and where $\Pi^{tun,A}_{\alpha}(\omega)$ and $\Pi^{tun,<}_{\alpha}(\omega)$  are the advanced and lesser phononic tunneling self-energy functions, respectively, given by
\begin{equation}
    [\Pi^{tun,A/<}_{\alpha}(\omega)]_{rs}=\sum_{q\in\alpha}V_{qr}d^{A/<}_{qq}(\omega)V_{qs} \,,\label{phonon_self_egy_eqn}
\end{equation}
where $d^{A}_{ql}(t,t')=i\theta(t'-t)\langle[U_q(t),U_l(t')]\rangle$ and $ d^{<}_{ql}(t,t')=-i\langle U_l(t')U_q(t)\rangle$ are the uncoupled reservoir advanced and lesser phonon Green's functions, respectively. Since the reservoirs are time-independent, they can be evaluated in the energy domain as
\begin{equation}
   d^{A}_{qq}(\omega)=2\pi i \Bigg[\frac{1}{\omega-\omega_{q}+i\eta}+\frac{1}{\omega+\omega_{q}+i\eta}\Bigg]\frac{\hbar}{2\omega_q} \,,
\end{equation}
and
\begin{eqnarray}
   d^{<}_{qq}(\omega)&=&-2\pi i \Bigg[f^{Planck}_{\alpha}(\omega)\delta(\omega-\omega_{q})\nonumber\\+&&(1+f^{Planck}_{\alpha}(\omega))\delta(\omega+\omega_{q})\Bigg]\frac{\hbar}{2\omega_q}\,,
\end{eqnarray}
where 
\begin{equation}
	f^{Planck}_{\alpha}(\omega_{q})=\frac{1}{e^{\beta_{\alpha}\hbar\omega_{q}}-1}
\end{equation}
is the Planck distribution function of the $\alpha^{th}$ reservoir.
\subsection{External work}
 
The work done by external forces \eqref{wext_1_eqn} can be written in terms of the system Green's functions as (see Appendix \ref{phonon_wext_NEGF_app} for details)
\begin{equation}
	\dot{W}_{ext,el}(t)=-i\mathbb{Tr}^{(1)}\{\dot{H}^{(1)}_{S,el}(t)G^{<}(t,t)\} \,,
	\label{wext_el_negf_eqn}
\end{equation}
and
\begin{equation}
	\dot{W}_{ext,ph}(t)= \frac{1}{\hbar}\int_{-\infty}^{t} \,dt' \Bigg[\dot{\mathbf{F}}(t)\mathcal{D}(t,t')\mathbf{F}^{\mathrm T}(t')\Bigg]\,.
	\label{wext_ph_negf_eqn}
\end{equation}

\subsection{Particle currents}

The electronic particle current, given by the Jauho-Wingreen-Meir formula \cite{wingreenTimedependentTransportMesoscopic1993,jauhoTimedependentTransportInteracting1994}, is
\begin{IEEEeqnarray}{lcr}
	I_{\alpha,el}^{N}(t)&=&
	\frac{2}{\hbar}\int_{-\infty}^{\infty} \,\frac{d\omega}{2\pi} \int_{-\infty}^{t} \,dt_1\mathbb{ImTr}^{(1)}\{e^{-i\omega(t_1-t)}\Gamma_{\alpha}^{el}(\omega)\nonumber\\&&[G^{<}(t,t_1) + f_{\alpha}(\omega)G^{R}(t,t_1)]\} \,,\nonumber\\
	\label{el_particle_curr_negf_eqn}
\end{IEEEeqnarray}
where $\Gamma_{\alpha}^{el}(\omega)$, $G^{<}(t,t')$ and $G^{R}(t,t')$ are understood to be matrices acting on the 1-particle Hilbert space. 
respectively, and 
	\begin{equation}
		f_{\alpha}(\omega)=\frac{1}{e^{\beta_{\alpha}(\hbar\omega-\mu_\alpha)}+1}
	\end{equation}
is the Fermi-Dirac distribution function of reservoir $\alpha$ with $\beta_{\alpha}=(k_{B}T_{\alpha})^{-1}$, where $k_{B}$ is the Boltzmann Constant. Following \cite{meirLandauerFormulaCurrent1992}, we identify the electronic tunneling-width matrix \(\Gamma_{\alpha}^{el}(\omega)\) as
\begin{equation}
		[\Gamma_{\alpha}^{el}(\omega)]_{nm}= \sum_{k\in\alpha}2\pi\delta(\hbar\omega-\epsilon_{k})V_{n}^{el}(\omega)(V^{el}_{m})^{*}(\omega)
		\label{t_indpt_Gamma_eqn} \,,
\end{equation}
with \(V_{n}^{el}(\omega)=V_{kn}^{el}\) when \(\hbar\omega=\epsilon_{k}\). See Ref.\ \cite{lehmannTimedependentFrameworkEnergy2018} for a related treatment of particle and energy currents.  

\subsection{Heat currents}

The conventional component of the electronic heat current formula (\ref{conv_el_heatcurr_expval_eqn}) becomes (see Appendix \ref{NEGF_details_app} for a derivation)
\begin{IEEEeqnarray}{lcr}
	I_{\alpha,el}^{Q,conv}(t)&=&\frac{2}{\hbar}\int_{-\infty}^{\infty} \,\frac{d\omega}{2\pi}\int_{-\infty}^{t} \,dt_1(\hbar\omega-\mu_{\alpha})\mathbb{ImTr}^{(1)}\{e^{-i\omega(t_1-t)}\nonumber \\&&\Gamma_{\alpha}^{el}(\omega)[G^{<}(t,t_1) + f_{\alpha}(\omega)G^{R}(t,t_1)]\} \,,\nonumber\\
	\label{el_heat_curr_negf_eqn}
\end{IEEEeqnarray}
which generalizes the Bergfield-Stafford formula \cite{bergfieldThermoelectricSignaturesCoherent2009d} to the case of a time-dependent system.

The conventional component of the phononic heat current \eqref{conv_ph_heatcurr_expval_eqn} is obtained similarly in terms of the system Phononic Green's functions as (see Appendix \ref{phonon_heat_current_derv_app} for details)
\begin{eqnarray}
    I_{\alpha,ph}^{Q,conv}(t)&=& \frac{1}{\hbar}\int_{-\infty}^{+\infty}\,\frac{d\omega}{2\pi} \omega \int_{-\infty}^{+\infty}\,dt_1 \mathbb{Tr}^{(1)}\Bigg\{e^{-i\omega(t_1-t)}\nonumber\\
    &&[D^{<}(t,t_1)\Pi^{tun,A}_{\alpha}(\omega)+D^{R}(t,t_1)\Pi^{tun,<}_{\alpha}(\omega)]\Bigg\}\,.\nonumber\\ 
    \label{ph_heat_curr_negf_eqn}
\end{eqnarray} 

The transient component of the heat current is given by Eq.\ \eqref{eq:IQtr}, where the electronic and phononic contributions to 
$\langle H_{SR,\alpha}(t)\rangle$ are evaluated in 
 Eqs.\ \eqref{HSR_el_negf_eqn} and
\eqref{HSR_ph_negf_eqn}, respectively.  The total transient component of the heat transferred to reservoir $\alpha$ for a process lasting from an initial time $t_1$ to a final time $t_2$ is
\begin{equation}
    Q_\alpha^{tr}=\frac{1}{2}\left(\left.\langle H_{SR,\alpha}\rangle\right|_{t_2}-\left.\langle H_{SR,\alpha}\rangle\right|_{t_1}\right).
    \label{eq:Qtr_alpha}
\end{equation}

With this, all the terms appearing in the First Law [Eq.\ \eqref{firstlaw1_eqn}] have now been evaluated in terms of system Green's functions.

\subsection{Energy density}\label{eqy_den_negf_subsec}

The energy density, Eq.\ \eqref{eq:rho_H}, can be decomposed into contributions from the various terms in the Hamiltonian, $\rho_H(x,t)=\rho_{H_S}(x,t)+\rho_{H_{SR}}(x,t)+\rho_{H_R}(x,t)$.  Within the system, only $\rho_{H_S}$ and $\rho_{H_{SR}}$ are non-zero, and can
be evaluated in terms of the Green's functions in a similar manner to the quantities analyzed above in Sec.\ \ref{sec:NEGF_U}. For the case of noninteracting particles, the contributions from the system Hamiltonian can be further decomposed as $\rho_{H_S}=\rho_{H_S^{el}}+\rho_{H_S^{ph}}$, where 
\begin{equation}
	\rho_{H_{S}^{el}}(x,t)= \frac{1}{2}\mathbb{Im}\langle  x|\{h_{S,el}^{(1)}(t),G^{<}(t,t)\}|x \rangle
	\label{egy_den_HS_el_negf_eqn}
\end{equation} 
and 
\begin{equation}
	\rho_{H_{S}^{ph}}(x,t)= \frac{i}{2}\langle  x|\left.\left(\frac{\partial^2}{\partial t \partial t'}+K\right)D^{<}(t,t')\right|_{t=t'}|x \rangle \,.
	\label{egy_den_HS_ph_negf_eqn}
\end{equation} 
The contributions from the coupling Hamiltonian can in general be decomposed as $\rho_{H_{SR}}=\rho_{H_{SR}^{el}}+\rho_{H_{SR}^{ph}}$, where
\begin{equation}
	\rho_{H_{SR}^{el}}(x,t)= \frac{1}{2}\mathbb{Im}\langle x| \{V^{el},G^{<}_{tun}(t,t)\}|x\rangle 
		\label{egy_den_HSR_el_negf_eqn}
\end{equation} 
and
\begin{equation}
	\rho_{H_{SR}^{ph}}(x,t)= \frac{i}{2}\langle x| V^{ph}D^{<}_{tun}(t,t)|x\rangle \,,
		\label{egy_den_HSR_ph_negf_eqn}
\end{equation} 
where $V^{el}$ and $V^{ph}$ are matrices defined in Eqs.\ \eqref{el_cpl_hamiltonian_eqn} and \eqref{ph_cpl_hamiltonian_eqn}, respectively.
Here the electron tunneling Green's functions are defined as $G^{<}_{tun,kn}(t,t')=i\langle d_{n}^{\dagger}(t')c_{k}(t) \rangle$ and $G^{<}_{tun,nk}(t,t')=i\langle c_{k}^{\dagger}(t') d_{n}(t)\rangle$, and
the phonon tunneling Green's functions are defined as $D^{<}_{tun,qr}(t,t')=-i\langle Q_{r}(t')U_{q}(t) \rangle$ and
$D^{<}_{tun,rq}(t,t')=-i\langle U_q(t')Q_r(t) \rangle$. (
see Appendix \ref{egyden_app_subsec} for evaluation of the interfacial energy densities in terms of the system Green's functions.)  

\section{Proof of Concept: Application to a Driven Two-Level Quantum Machine}\label{qmachine_sec}

We have applied our formal results to perform a complete thermodynamic analysis of a strongly-driven two-level quantum system coupled to two metallic reservoirs. We first describe the basic setup of the quantum machine, followed by a detailed analysis of its operation in two different configurations. This simulation underscores the utility of our formal results, derived in the previous sections, as tools to investigate the real-time dynamics of quantum systems driven far from equilibrium.

\subsection{Machine Setup}\label{qmachine_setup_subsec}
The quantum machine consists of two adjacent quantum dots, each with a single active level, with energies $E_1$ and $E_2$, coupled to each other via inter-dot coupling $w$. The system is opened to the environment by coupling each dot to a perfectly ordered, semi-infinite Fermionic reservoir via energy-independent tunneling-width matrix elements $\Gamma_1=\Gamma_2=\Gamma$. Both reservoirs are maintained in internal equilibrium characterized by fixed Chemical Potentials and Temperatures $\mu_1\,,T_1$ and $\mu_2\,,T_2$ (see 
Ref.\ \cite{staffordResonantPhotonAssistedTunneling1996} for an analysis of a similar quantum machine). 
The time-dependent drive is applied only to one of the dots (the dot with energy $E_1$ for the results presented) and is set up as a rectangular pulse of strength $\delta$ which is active for a duration $\tau$. The pulse strength $\delta$ is set equal to the difference between the dot energies $\Delta E=E_2-E_1 $ and brings the two levels into resonance, allowing the electron density to strongly Rabi oscillate between them. The pulse duration is tuned as a $\pi$-pulse by setting  $\tau=\frac{\pi}{2w}$. This maximizes the probability of transferring the electron density from one dot to the other at the end of the pulse. We also order the parameters such that $\Delta E \gg w \gg \Gamma$ to ensure that the levels remain highly localized in the absence of a drive. 
	
The energy parameters of the system can be tuned relative to each other such that its Rabi Oscillations can be leveraged to operate it as an Electrochemical Pump, an Electrochemical Engine, a Heat Engine, or a Heat Pump. Here we present a detailed analysis for the Electrochemical Pump and Heat Engine configurations.

\subsection{Electrochemical Pump}\label{echem_pump_subsec}

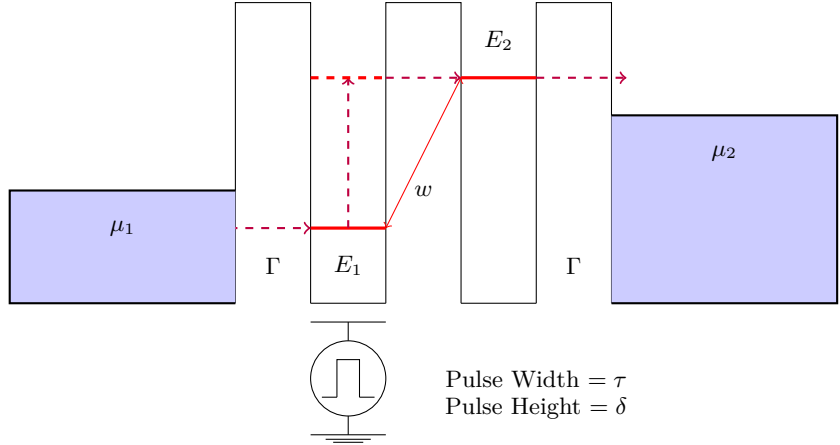
\begin{figure*}
	\centering
	\captionsetup{justification=raggedright, singlelinecheck=false}
	\begin{tikzpicture}%

		\draw (3,1)--(3,5)-|(4,1);	\draw (4,1)--(5,1); 
		\draw (5,1)--(5,5)-|(6,1); \draw (6,1)--(7,1); \draw (7,1)--(7,5)-|(8,1);
		
		\draw (4,0.75)--(5,0.75);\draw (4.5,0.75)--(4.5,0.5);\draw (4.5,0.0) circle (0.5cm);\draw (4.5,-0.5)--(4.5,-.75);\draw (4,-.75)--(5,-.75); \draw (4.2,-.81)--(4.8,-.81);\draw (4.3,-.85)--(4.7,-.85); 
		\node at (7,0) {Pulse Width $=\tau$};
		\node at (7,-.4) {Pulse Height $=\delta$};
		
		\draw (4.35,-.25)--(4.35,.25)-|(4.65,-.25); \draw(4.15,-.25)--(4.35,-.25);\draw(4.65,-.25)--(4.85,-.25);
		
		\draw [very thick,red] (4,2)--(5,2); \node at (4.5,1.5) {$E_1$};
		
		\draw [very thick,red] (6,4)--(7,4); \node at (6.5,4.5) {$E_2$};
		
		\draw [<->,red] (5,2)--(6,4);  \node at (5.5,2.5) {$w$};

		\draw(4.5,2)[->,dashed,purple,thick]--(4.5,4);\draw [very thick,red,dashed] (4,4)--(5,4);
		\draw[->,dashed,purple,thick](2.5,2)--(4,2);
		\draw[->,dashed,purple,thick](5,4)--(6,4);
		\draw[->,dashed,purple,thick](7,4)--(8.2,4);
		
		\draw[thick,fill=blue!20] (3,1)--(0,1)|-(3,2.5); \node at (1.5,2) {$\mu_1$};
		\draw[thick,fill=blue!20] (8,1)--(11,1)|-(8,3.5); \node at (9.5,3) {$\mu_2$};
		
		\node at (3.5,1.5) {$\Gamma$};
		\node at (7.5,1.5) {$\Gamma$};	
		
	\end{tikzpicture}
	\caption{Schematic representation of an Electrochemical Pump based on Rabi Oscillations between states of a double quantum dot. %
	The left and right dots in the system have on-site energies $E_1$ and $E_2$, respectively, and are coupled by a constant hopping matrix element $w$. The left dot is driven by a rectangular pulse of  width $\tau$ and height $\delta$. Each dot is coupled with the same Tunneling width matrix element $\Gamma$ to a Reservoir. The Reservoirs are modeled as non-interacting, Fermionic, semi-infinite reservoirs at Electrochemical Potentials $\mu_1$ and $\mu_2$ for the left and right reservoirs, respectively, with $T_2=T_1$. The cycle of operation of the electrochemical pump is denoted by the dashed purple arrows.}
	\label{echempump_fig}
\end{figure*}

\begin{figure*}%
	
	\centering

	\subfloat[External Work]{\label{ecehempump_wext}\includegraphics[width=.45\textwidth,height=5cm]{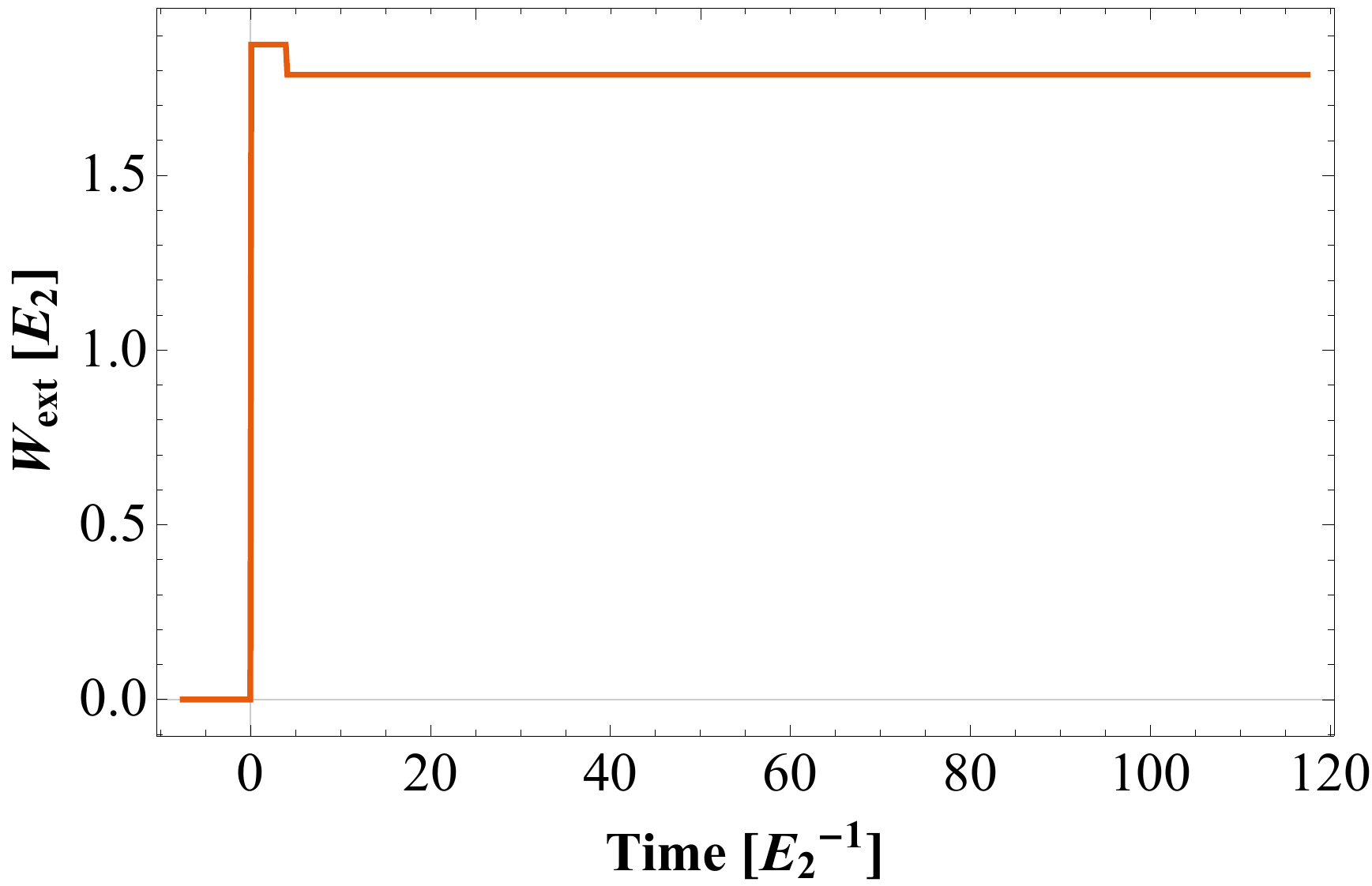}}
	\subfloat[Electrochemical Work]{\label{ecehempump_welec}\includegraphics[width=.45\textwidth,height=5cm]{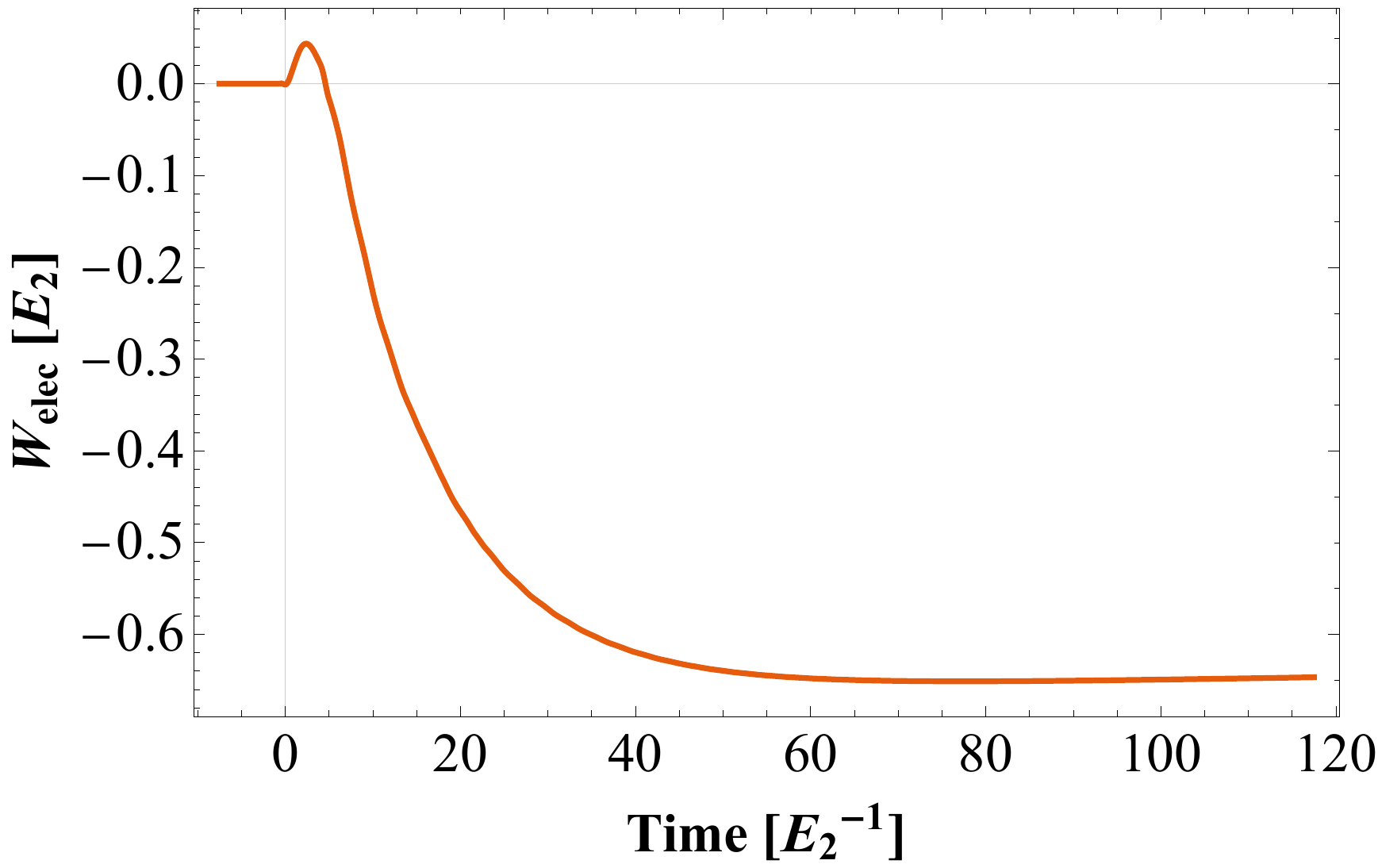}}\\
	\subfloat[Total Heat Dissipated]{\label{ecehempump_heat}\includegraphics[width=.45\textwidth,height=5cm]{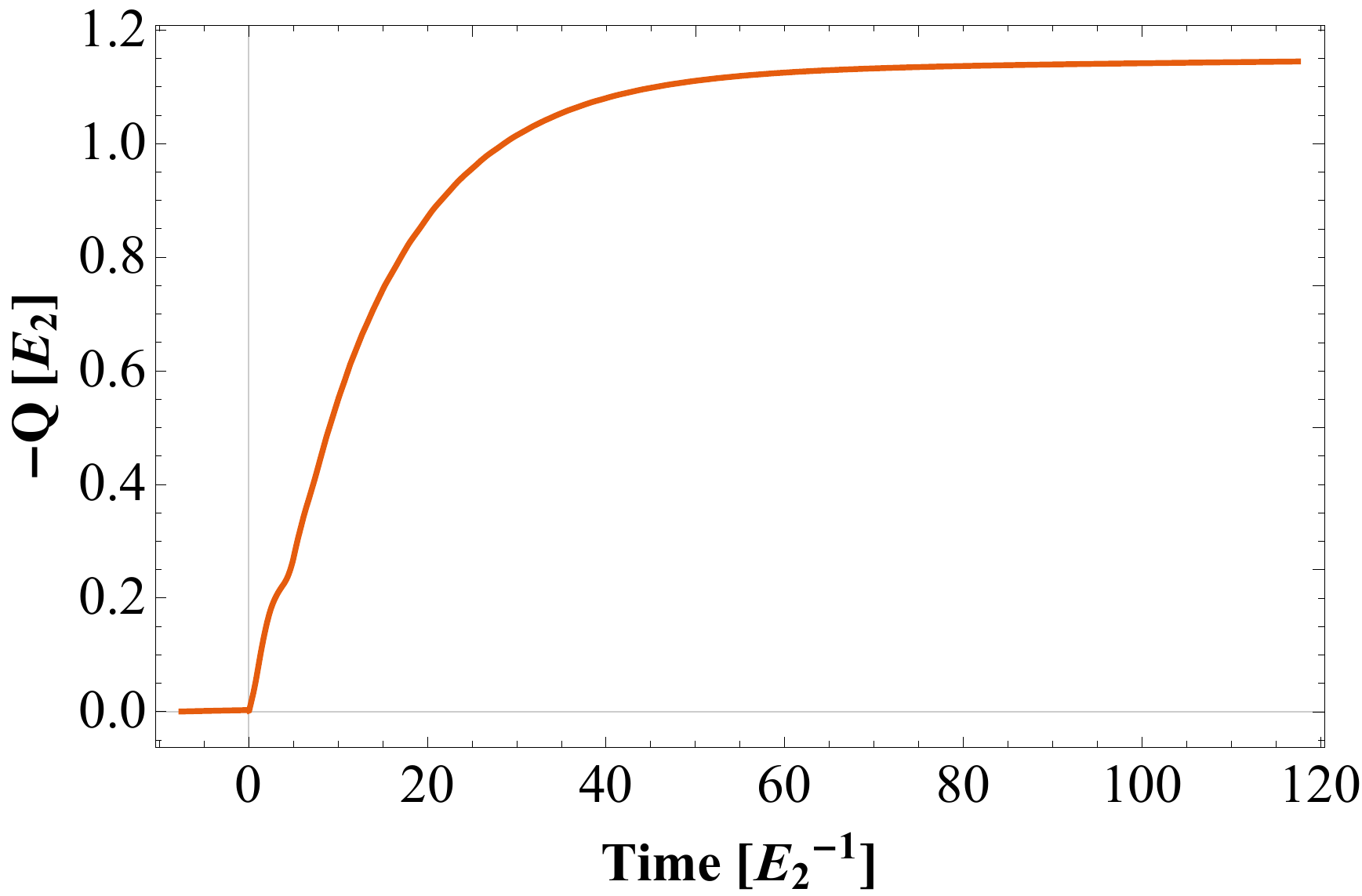}}
	\subfloat[Internal Energy (and consistency check) ]{\label{ecehempump_inten}\includegraphics[width=.45\textwidth,height=5cm]{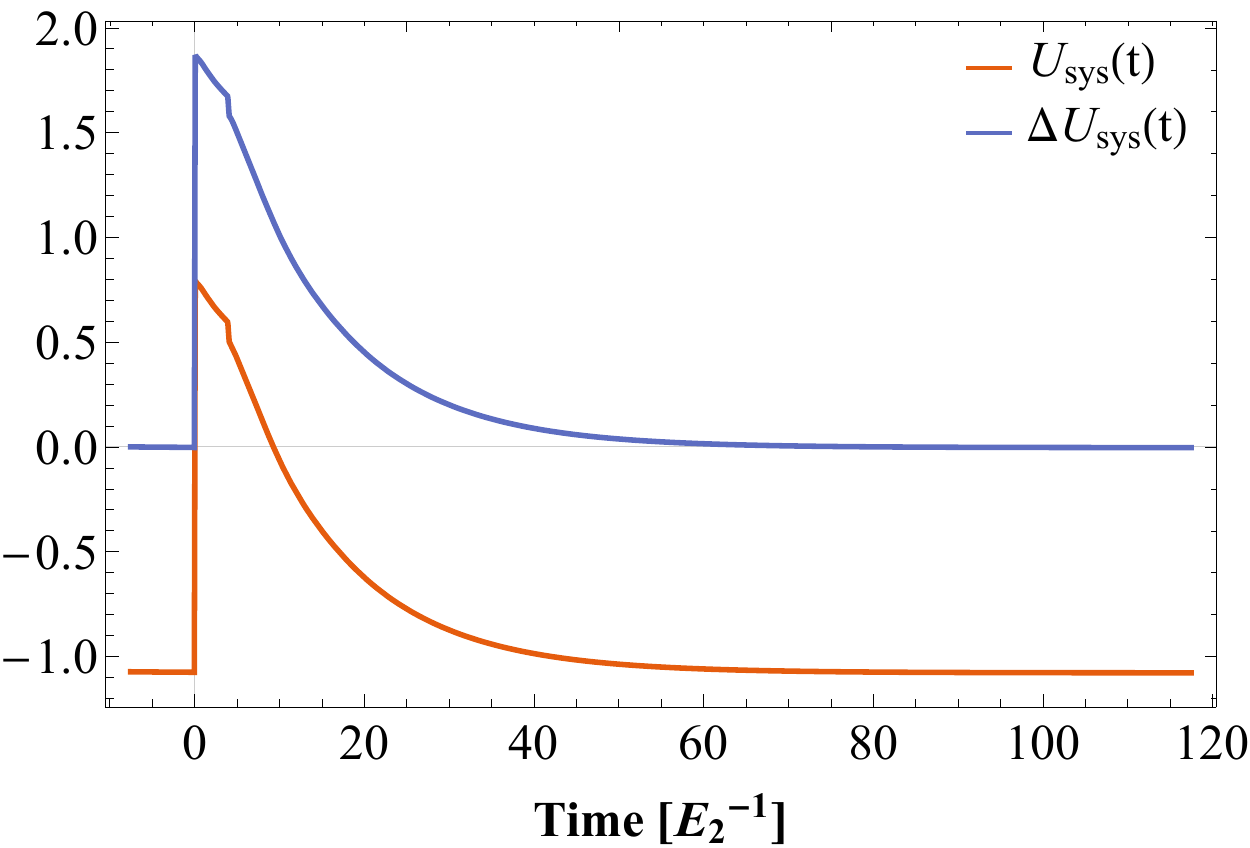}}
	
	\captionsetup{justification=raggedright, singlelinecheck=false}	
	\caption{%
 The time-dependent energetics of the Electrochemical Pump depicted schematically in Fig.\ \ref{echempump_fig}. (a) Total Work done by external drive versus time, (b) total Electrochemical Work done versus time, (c) total Heat dissipated into reservoirs versus time, and (d) Internal Energy $U_{sys}$ (in orange) and $\Delta U_{sys}$ from integration of the RHS of the First Law equation \eqref{firstlaw1_eqn} (in blue) versus time. For the results shown, we have used $\hbar=1$ and we set the on-site energy on the left and right dots $\{E_1,E_2\}=\{-1,1\}$ so that the difference in the on-site energies of the two dots is $ \Delta E=E_2-E_1=2$ in units of $E_2$. The dot hybridization is set to $w=\Delta E/5$ and the coupling to the reservoirs is  taken as $\Gamma=w/5$. The chemical potential and temperatures of the reservoirs are set $-\mu_1=\mu_2=0.5$ in units of $E_2$ and $k_{B}T_1 =k_{B}T_2 = 0.016$ in units of $E_2$, respectively. The pulse amplitude is set equal to the difference in the on-site energies $\delta=\Delta E$ and the pulse duration is set $\tau=\pi/2w$ ($\pi$-pulse).}
	\label{ecehempump_plots}
\end{figure*}

For the setup described, if the electrochemical potentials of the reservoirs are biased such that $\mu_1>E_1$ and $\mu_2<E_2$, electrons can be pumped uphill from the left to the right reservoir. A schematic of the chemical pump configuration is shown in Fig.\ \ref{echempump_fig}. In this configuration, the Temperatures $T_1\,, T_2$ of the reservoirs are set to very low values to suppress any thermal excitations from contributing to the operation. The thermodynamic cycle for the pump involves the electron tunneling from the left reservoir to dot 1 followed by the pulse raising its energy to bring it into resonance with dot 2. The $\pi$-pulse ensures a maximum probability of dot 2 capturing the electron at the end of the pulse, which is followed by the electron tunneling out into the right reservoir. The machine relaxes back to its initial state at late times, at which point we can calculate the efficiency of the electrochemical pump as
\begin{equation}
 \eta_{EP}=\frac{|W_{elec}(t\rightarrow\infty)|}{|W_{ext}(t\rightarrow\infty)|}.   
\end{equation}

The full Hamiltonian for this configuration is given by Eq.\ \eqref{full_hamiltonian_eqn}, where the System Hamiltonian is %
\begin{equation}
H_{S}(t)= E_1(t)d_{1}^{\dagger}d_{1}+E_{2}d_{2}^{\dagger}d_{2}+w(d_{1}^{\dagger}d_{2}+d_{2}^{\dagger}d_{1}) \,,
\label{echem_pump_Hsys}
\end{equation}
with the Reservoir Hamiltonian given by
\begin{equation}
	H_{R}= \sum_{i=1,2}\sum_{k\in i}\epsilon_{k}c_{k}^{\dagger}c_{k}\,,
	\label{echem_pump_HB_eqn}
\end{equation}
and the coupling is such that the $i^{th}$ dot is only coupled to the $i^{th}$ reservoir
\begin{equation}
	H_{SR}= \sum_{k\in1}(V_{k1}c_{k}^{\dagger}d_{1}+h.c.)+\sum_{k\in2} (V_{k2}c_{k}^{\dagger}d_{2}+h.c.) \,.
	\label{echem_pump_HSB_eqn}
\end{equation}
Finally, the time dependence of the system is encoded in the left dot as
\begin{equation}
E_1(t) = \begin{cases} 
	
	E_1 & t<0, \\
	E_1+\delta	& 0\leq t< \tau, \\
	E_1 & t\geq\tau.
	
\end{cases}
\label{rect_pulse_eqn}
\end{equation} 

Fig.\ \ref{ecehempump_plots} shows plots of all the thermodynamic quantities entering the First Law as functions of time. We note that we have plotted the integrated thermodynamic quantities here instead of the time-derivatives that appear in the First Law [Eq.\ \eqref{firstlaw1_eqn}], 
so that we have the total external work done on the system $W_{ext}(t)=-i\int_{-\infty}^{t}\, dt \mathbb{Tr}\{\dot{H}^{(1)}_{S}(t)G^{<}(t,t)\}$, the total electrochemical  work done on the system $W_{elec}(t)=\int_{-\infty}^{t}\, dt[\sum_{\alpha}-\mu_{\alpha} I_{\alpha}^{(0)}(t)]$, and the total heat dissipated into the reservoirs $-Q(t)=\int_{-\infty}^{t}\, dt(\sum_{\alpha} I_{\alpha}^Q(t))$. All the energies reported for this setup are in units of the on-site energy  of the right dot $E_2$ and time is in units of $E_2^{-1}$.

The total external work done by the drive $W_{ext}(t)$, (Fig.\ \ref{ecehempump_wext}) follows the expected instantaneous rise of the pulse at $t=0$ to a value of $1.87$, indicating that work is done \emph{on} the system by the drive. 
$W_{ext}(t)$ remains at this constant value for the duration of the pulse $\tau$, followed by a slight instantaneous decrease at $t=\tau$. This can be explained by noting that there is a small but finite probability for the electron being on dot 1 at the end of the $\pi$-pulse. This results in work being done \emph{by} the system on the drive as the energy of the first dot is lowered back to its initial value, appearing as a negative contribution. The external work done remains constant at a value of $1.78$ thereafter since the drive is inactive for $t>\tau$. 

As the pulse starts, the electrochemical  work $W_{elec}(t)$, (Fig. \ref{ecehempump_welec}) is done \emph{on} the system as the particle current flows mostly into the left reservoir initially. At a certain time $t>\tau$ the electrochemical work becomes negative indicating that electrochemical  work is now done \emph{by} the system as the particle current starts to flow mostly into the right reservoir. It increases in magnitude for some time and eventually attains a constant value of $-0.65$ at late times as the particle currents flowing into the reservoirs relax to $0$. For the complete cycle, electrochemical  work is thus done by the system in transferring an electron from the reservoir with the lower chemical potential $\mu_{1}$ to the one with the higher chemical potential $\mu_{2}$. With the late time values of $W_{elec}(t)$ and $W_{ext}(t)$, we can see that the efficiency of the electrochemical pump $\eta_{EP}$ is about $36.5\%$. 

The total heat dissipated into the reservoirs $-Q(t)$ (Fig.\ \ref{ecehempump_heat}) always remains positive implying that heat is only ever flowing into the reservoirs in the electrochemical pump configuration for the parameters chosen. This can be explained by noting that the (steady-state) heat flowing into the $i^{th}$ reservoir goes as $~(\omega_{i}-\mu_{i})v_{i}f_{i}(\omega)$, where $v_{i}$ is the velocity of the electron \cite{dattaElectronicTransportMesoscopic1995}. The levels in the numerics are biased such that $(\omega_{i}-\mu_{i})v_{i}$ always remains positive for both reservoirs, no matter which direction the electron is moving. The pronounced knee at $t=\tau$ is due to the opening up of another transport channel at the first dot in the form of a hole channel once the electron has been captured by the second dot at the end of the pulse and the left dot is in an empty state.
The total heat dissipated asymptotes to a constant value of $1.17$ at late times. 

We finally plot the internal energy $U_{sys}(t)$ (Fig. \ref{ecehempump_inten}) and note that it matches up well with the plot of the sum of all the terms appearing on the right hand side of the First Law i.e., $\Delta U_{sys}(t)=W_{ext}(t)+W_{elec}(t)+Q(t)$. %
The offset is expected since we have plotted $U_{sys}(t)$ directly from the Eqs.\ \eqref{HS_negf_eqn} and \eqref{HSR_negf_eqn}, which are expressions for energy, whereas the plot of $\Delta U_{sys}(t)$ was generated by integration of the current formulas where the constant of integration was set to $0$. 

\subsubsection{Spatio-Temporal Distribution of Energy---Electrochemical Pump}
\label{int_egy_tbchain_subsubsec_1}

\begin{figure*}

\centering
\includegraphics[width=\textwidth,height=5cm]{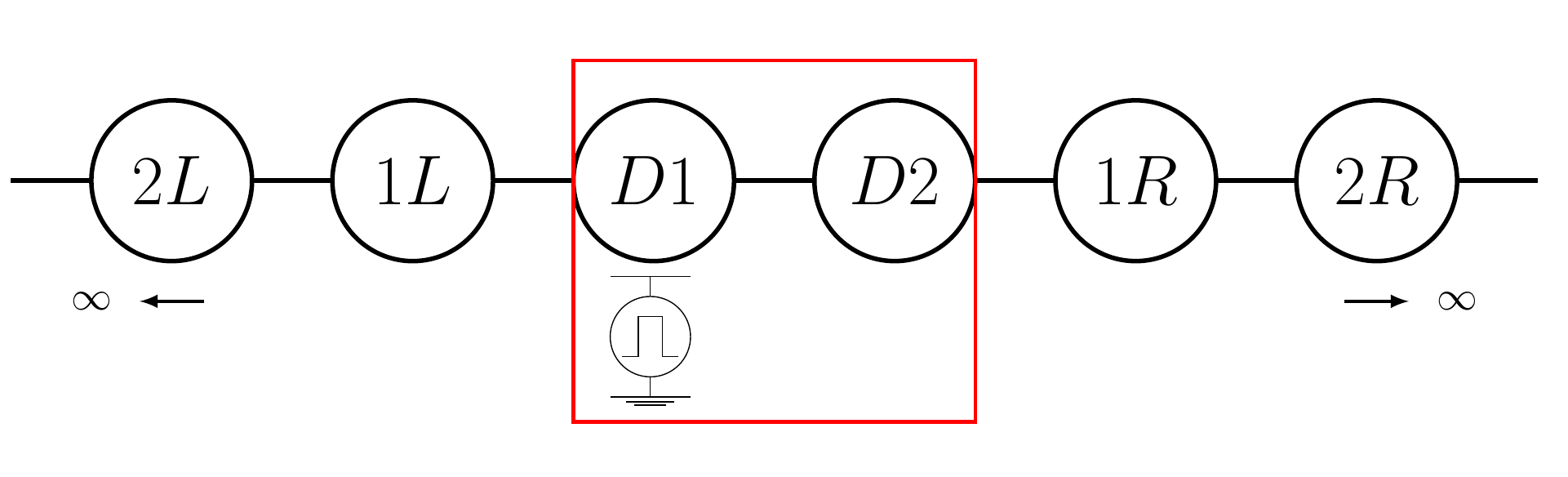}
\captionsetup{justification=raggedright,singlelinecheck=false}
\caption{Schematic representation of the driven double quantum dot system (red box) discussed in Secs.\ \ref{int_egy_tbchain_subsubsec_1} and \ref{int_egy_tbchain_subsubsec_2}, now coupled specifically to semi-infinite tight-binding chains. The left and right dots are labeled as D1 and D2, respectively, and the chain sites on the left and right reservoirs are labelled 1L, 2L,... and 1R, 2R,..., respectively. Only the site $1L$ is directly coupled to $D1$ and only the site $1R$ is directly coupled to $D2$.}
\label{chain_schematic_fig}

\end{figure*}

To develop a finer-grained understanding of the operation of this strongly-driven quantum machine, we
investigate the spatio-temporal distribution of energy using the theory of energy density developed in Sec.\ \ref{energy_density_1_sec}. We work with essentially the same setup that was used in the preceding paragraphs for the integrated thermodynamic quantities, but now with a specific model for the reservoirs: 1D tight-binding chains. The System Hamiltonian $H_S(t)$ is chosen the same as before [Eq.\ \eqref{echem_pump_Hsys}]. The Reservoir Hamiltonian is
\begin{equation}\label{tbchain_res_ham_eqn}
	H_{R}= \sum_{\alpha=L,R}\Bigg[\sum_{j=1}^\infty\epsilon_{0}c_{j\alpha}^{\dagger}c_{j\alpha}+\sum_{j=1}^\infty
	t_0(c_{j\alpha}^{\dagger}c_{j+1\alpha}+h.c.)\Bigg]\,,
\end{equation}
where $\epsilon_{0}$ is the on-site energy at any given site and $t_0$ is the hopping integral between nearest-neighbor sites on a given chain. 
The coupling Hamiltonian modelling the Interface is 
\begin{equation}
H_{SR}=	t\Bigg[(d_{1}^{\dagger}c_{1L}+h.c.)+(d_{2}^{\dagger}c_{1R}+h.c.)\Bigg]\,,
\label{tbchain_cpl_ham_eqn}
\end{equation}
where $t$ is the coupling between the first site in the left reservoir $1L$ and the left dot, and that between the first site in the right reservoir $1R$ and the right dot. A schematic of this setup is given in Fig.\ \ref{chain_schematic_fig}.

We study the energy density of the system, defined in Eq.\ \eqref{eq:rho_H}, as a function of time.
The contributions of $H_S(t)$ and $H_{SR}$ to the energy density are evaluated using Eqs.\ \eqref{egy_den_HS_el_negf_eqn} and \eqref{egy_den_HSR_el_negf_eqn}, respectively.
The tunneling-width matrices for the 1D tight-binding chains coupled to the double quantum dot system are \cite{dattaChapterLevelBroadening2005,bergfieldManyBodyTheoryElectrical2010}
\begin{equation}
[\Gamma_{\alpha}(\omega)]_{ij}= 
\begin{cases}
\frac{2t^2}{t_0}\Big(1-\Big(\frac{\epsilon_0-\hbar\omega}{2t_0}\Big)^{2}\Big)^{\frac{1}{2}}\delta_{i\alpha}\delta_{j\alpha}, & |\hbar\omega-\epsilon_0|<2t_0, \\
0, & \text{otherwise}.
\end{cases}
\label{tbchain_Gamma_eqn}
\end{equation}

\begin{figure}
\centering
\includegraphics[width=.45\textwidth,height=16cm]{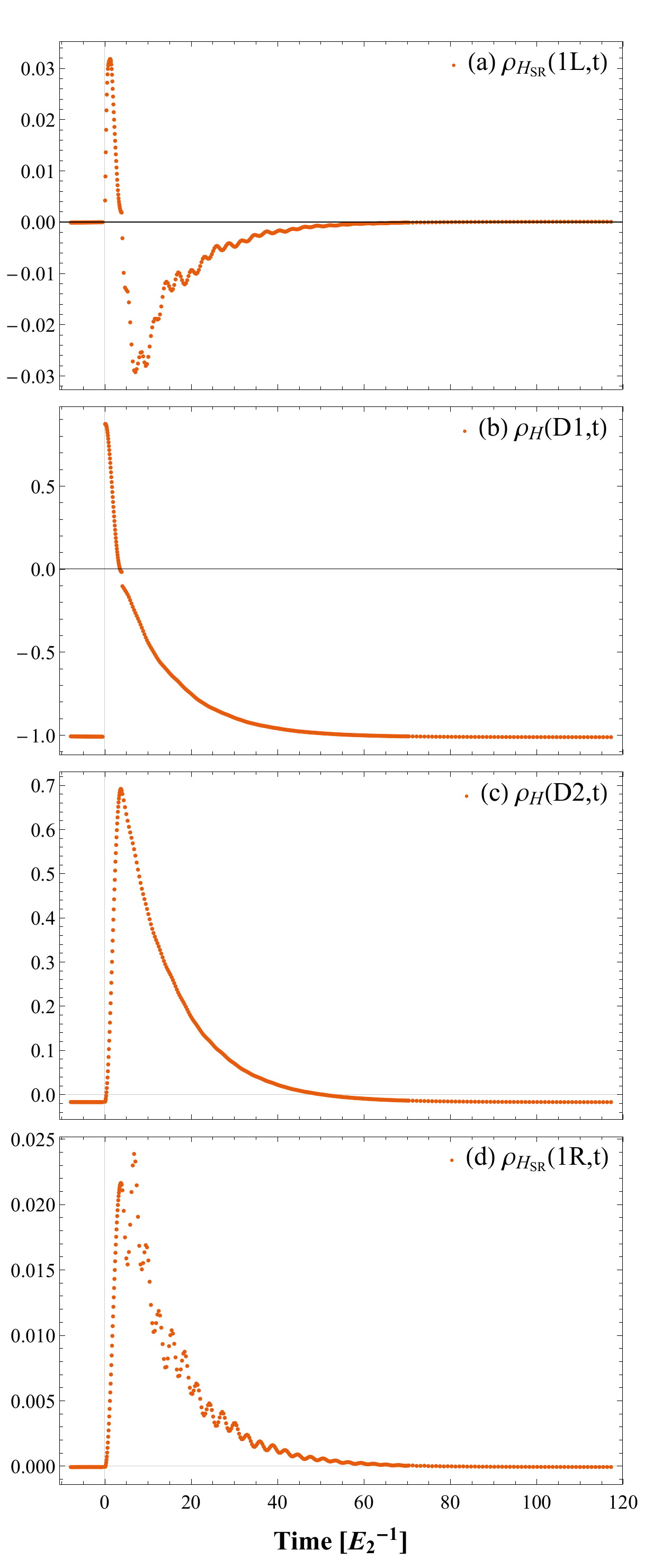}

\captionsetup{justification=raggedright,singlelinecheck=false}

\caption{
Plots of spatially resolved energy as a function of time for the \emph{Electrochemical Pump} configuration for the same Hamiltonian parameters as those for Fig.\ \ref{ecehempump_plots}. %
For the interfacial sites of both the reservoirs, we plot the contribution of the interface energy [Eq.\ \eqref{tbchain_cpl_ham_eqn}] denoted by $\rho_{H_{SR}}(1L,t)$ and $\rho_{H_{SR}}(1R,t)$ for the left and right reservoirs, respectively. For the two dots, we plot the total energy density as a function of time denoted by $\rho_{H}(D1,t)$ and $\rho_{H}(D2,t)$. The energies are in units of the on-site energy of the right dot $E_2$ and time is in units of $E_2^{-1}$in this configuration. Note that the energy scale for $\rho_{H_{SR}}(1L,t)$ and $\rho_{H_{SR}}(1R,t)$ (panels (a) and (d)) is different from that for $\rho_{H}(D1,t)$ and $\rho_{H}(D2,t)$ (panels (b) and (c)) in the figure.}
\label{echempump_energy_density_plots_fig}
\end{figure}
 
The numerical results for the energy density for the electrochemical pump are shown in Fig.\ \ref{echempump_energy_density_plots_fig} for the system 
in the electrochemical pump configuration %
\footnote{\label{numerics_bb_footnote}In the computational solution, the broad-band limit was used, which consists in taking the limit $t_0\rightarrow \infty$ while keeping $t^2/t_0$ fixed.}. We plot the total energy density for the two dots $D1$ and $D2$, and just the contribution of the coupling Hamiltonian $H_{SR}$ [Eq.\ \eqref{tbchain_cpl_ham_eqn}] to the energy density for the first sites $1L$ and $1R$ on the left and right reservoirs, respectively, which give half the 
energies of the left and right interfaces. 
Note that, by design, for the two dots we have $\rho_{H_R}(D1,t)=\rho_{H_R}(D2,t)=0$ and similarly for the sites on the reservoirs we have  $\rho_{H_S}(1L,t)=\rho_{H_S}(1R,t)=0$. 

The total energy on the left dot $\rho_{H}(D1,t)$ (Fig. \ref{echempump_energy_density_plots_fig}b) rises instantaneously to its maximum positive value at the start of the pulse as the external drive does work \emph{on} the dot. As the electron Rabi oscillates onto the right dot, the energy on the left dot begins to decrease and at the end of the pulse becomes zero and then asymptotes to its intial value as the initial conditions of the system are established again at late times. The Rabi oscillations of the system become more evident for the case of a $3\pi$-pulse, as seen in %
Fig.\ \ref{fig:Usys_den_3tau}, where we have again plotted the energy density [Eq.\ \eqref{eq:rho_H}] for both the dots and first sites of both reservoirs.

The interfacial energy at the first site in the left reservoir $\rho_{H_{SR}}(1L,t)$ (Fig.\ \ref{echempump_energy_density_plots_fig}a) rises and falls in response to changes in the energy on Dot 1. This ``in-phase with $D1$" behavior can be understood by noting that $H_{SR}$ models the covalent bond between the dot and the first site on the chain which is equally shared between the two. Negative (positive) bond energies correspond to bonding  (anti-bonding) character of the Dot-Reservoir interface. The bond energy tends to zero at long times as the system relaxes to its initial state wherein the occupancy of Dot 1 is nearly unity so that the hybridization with the left Reservoir tends to zero. The oscillatory behavior during the relaxation is a signature of the Rabi oscillations between the two dots $D1$ and $D2$, which are enhanced in frequency (but suppressed in amplitude) when the pulse is inactivated. 

The energy on the right dot $\rho_{H}(D2,t)$ (Fig.\ \ref{echempump_energy_density_plots_fig}c) rises as the electron Rabi oscillates onto it from the left dot and begins to decrease as the electron tunnels out onto the first site in the right reservoir, eventually relaxing back to its zero initial value. The interfacial coupling energy on the first site in the right reservoir $\rho_{H_{SR}}(1R,t)$ (Fig. \ref{echempump_energy_density_plots_fig}d) can be understood similarly to that on site $1L$ within the bonding picture discussed in the previous paragraph. As the electron tunnels onto $1R$ from the right dot the energy there increases followed by an asymptotic decrease to its initial value again almost mimicking the temporal behavior at $D2$.

\subsection{Heat Engine}\label{heat_engine_subsec}

\begin{figure*}
	\centering
	
	\captionsetup{justification=raggedright, singlelinecheck=false}
	
	\begin{tikzpicture}%

		\draw (3,1)--(3,5)-|(4,1);	\draw (4,1)--(5,1); 
		\draw (5,1)--(5,5)-|(6,1); \draw (6,1)--(7,1); \draw (7,1)--(7,5)-|(8,1);
		
		\draw (4,0.75)--(5,0.75);\draw (4.5,0.75)--(4.5,0.5);\draw (4.5,0.0) circle (0.5cm);\draw (4.5,-0.5)--(4.5,-.75);\draw (4,-.75)--(5,-.75); \draw (4.2,-.81)--(4.8,-.81);\draw (4.3,-.85)--(4.7,-.85); 
		\node at (7,0) {Pulse Width $=\tau$};
		\node at (7,-.4) {Pulse Height $=\delta$};
		\draw (4.35,-.25)--(4.35,.25)-|(4.65,-.25); \draw(4.15,-.25)--(4.35,-.25);\draw(4.65,-.25)--(4.85,-.25);
		
		\draw [very thick,red] (4,2)--(5,2); \node at (4.5,1.5) {$E_1$};
		
		\draw [very thick,red] (6,4)--(7,4); \node at (6.5,4.5) {$E_2$};
		
		\draw [<->,red] (5,2)--(6,4);  \node at (5.5,2) {\small $w(t)$};
		
		\draw(4.5,2)[<-,dashed,purple,thick]--(4.5,4);
		\draw [very thick,red,dashed] (4,4)--(5,4);
		\draw[<-,dashed,purple,thick](2.5,2)--(4,2);
		\draw[<-,dashed,purple,thick](5,4)--(6,4);
		\draw[<-,dashed,purple,thick](7,4)--(8.5,4);
		
		\draw[thick,fill=blue!20] (3,1)--(0,1)|-(3,1.5); \node at (1.5,2) {$T_1 $};
		\draw[thick,fill=red!20] (8,1)--(11,1)|-(8,4.5); \node at (9.5,3) {$T_2 $};	
		
		\node at (3.5,1.5) {$\Gamma$};
		\node at (7.5,1.5) {$\Gamma$};	
		
	\end{tikzpicture}
	
	\caption{Schematic representation of a Heat Engine based on Rabi Oscillations between states of a double quantum dot. The left and right dots in the system have on-site energies $E_1$ and $E_2$, respectively, and are dynamically coupled by a time-dependent hopping term $w(t)$. The left dot is driven by a rectangular pulse of  width $\tau$ and strength $\delta$. Each dot is coupled with the same Tunneling width matrix element $\Gamma$ to a Reservoir. The Reservoirs are modeled as non-interacting, Fermionic, semi-infinite systems at Temperatures $T_1$ and $T_2$ for the left and right reservoirs, respectively, with $T_2>T_1$. The cycle of operation of the heat engine is denoted by the dashed purple arrows.}
	\label{heatengine_fig}
\end{figure*}
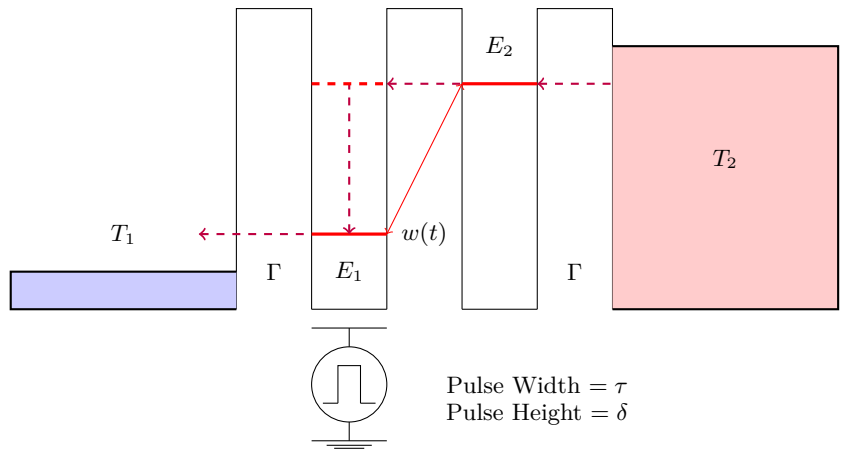

\begin{figure*}
	
	\centering

	\subfloat[External Work]{\label{heatengine_wext}\includegraphics[width=.45\textwidth,height=5cm]{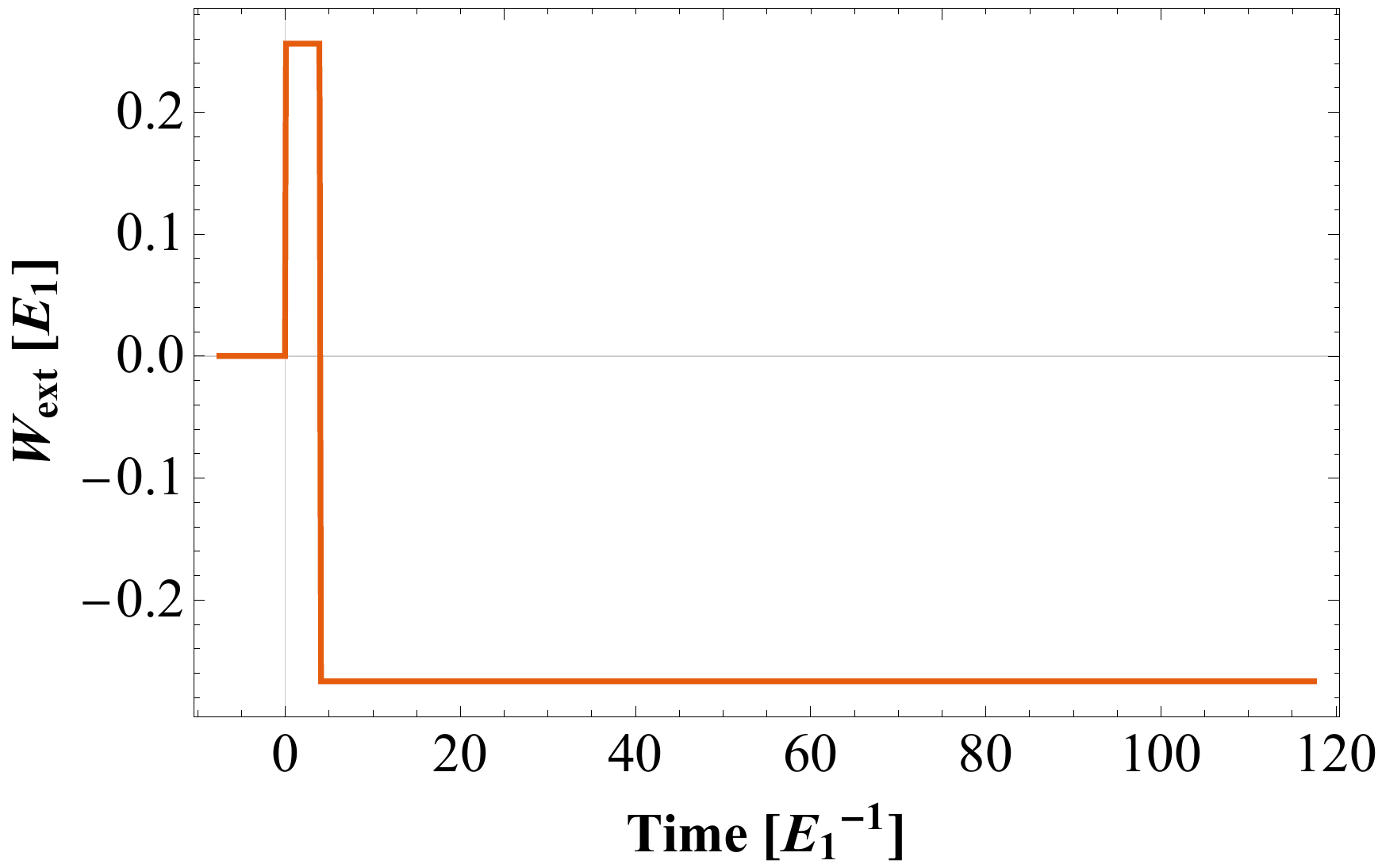}}
	\subfloat[Heat extracted from Right Reservoir]{\label{heatengine_rightheat}\includegraphics[width=.45\textwidth,height=5cm]{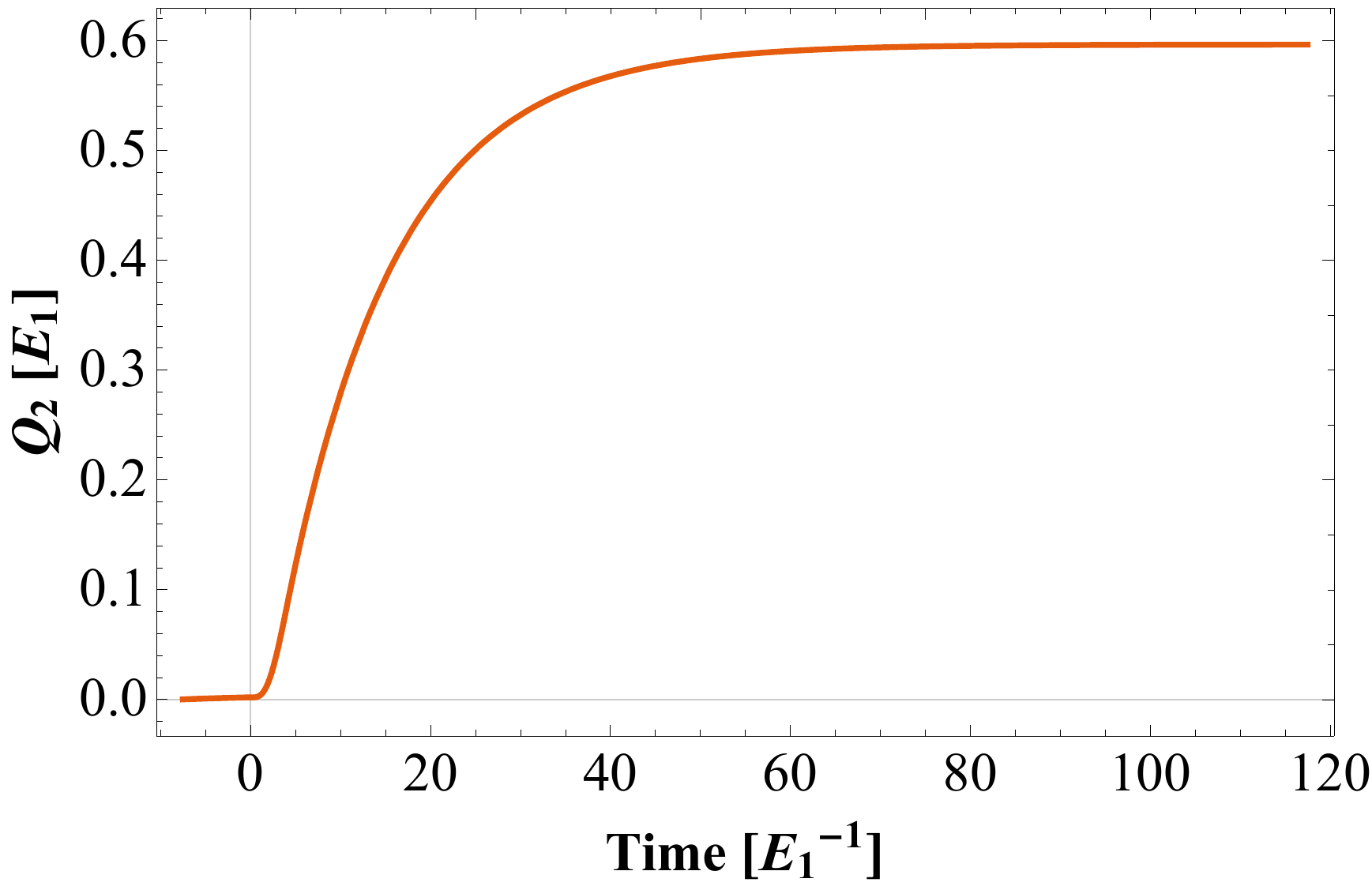}}\\
	\subfloat[Electrochemical Work]{\label{heatengine_welec}\includegraphics[width=.45\textwidth,height=5cm]{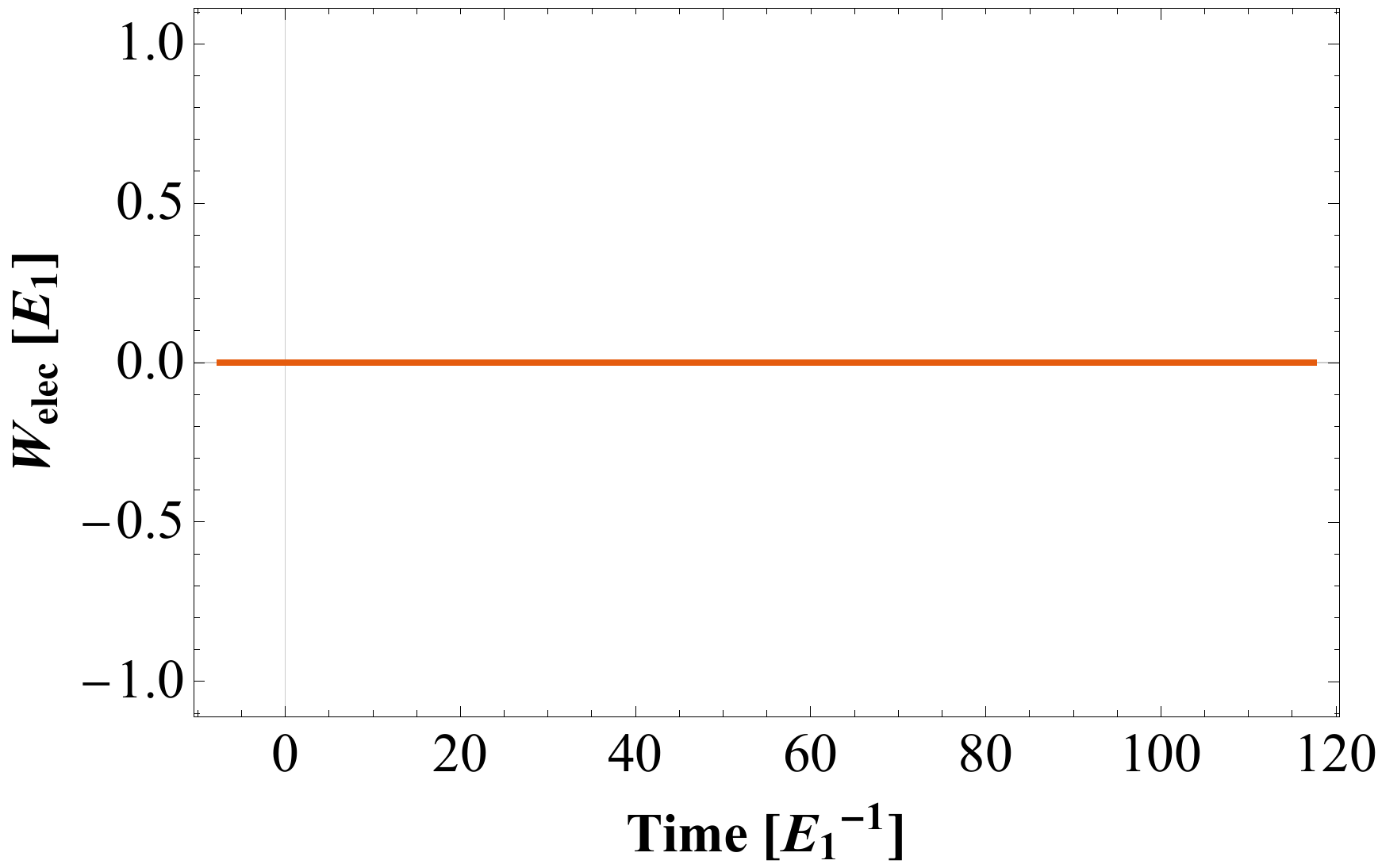}}
	\subfloat[Internal Energy (and consistency check)]{\label{heatengine_inten}\includegraphics[width=.45\textwidth,height=5cm]{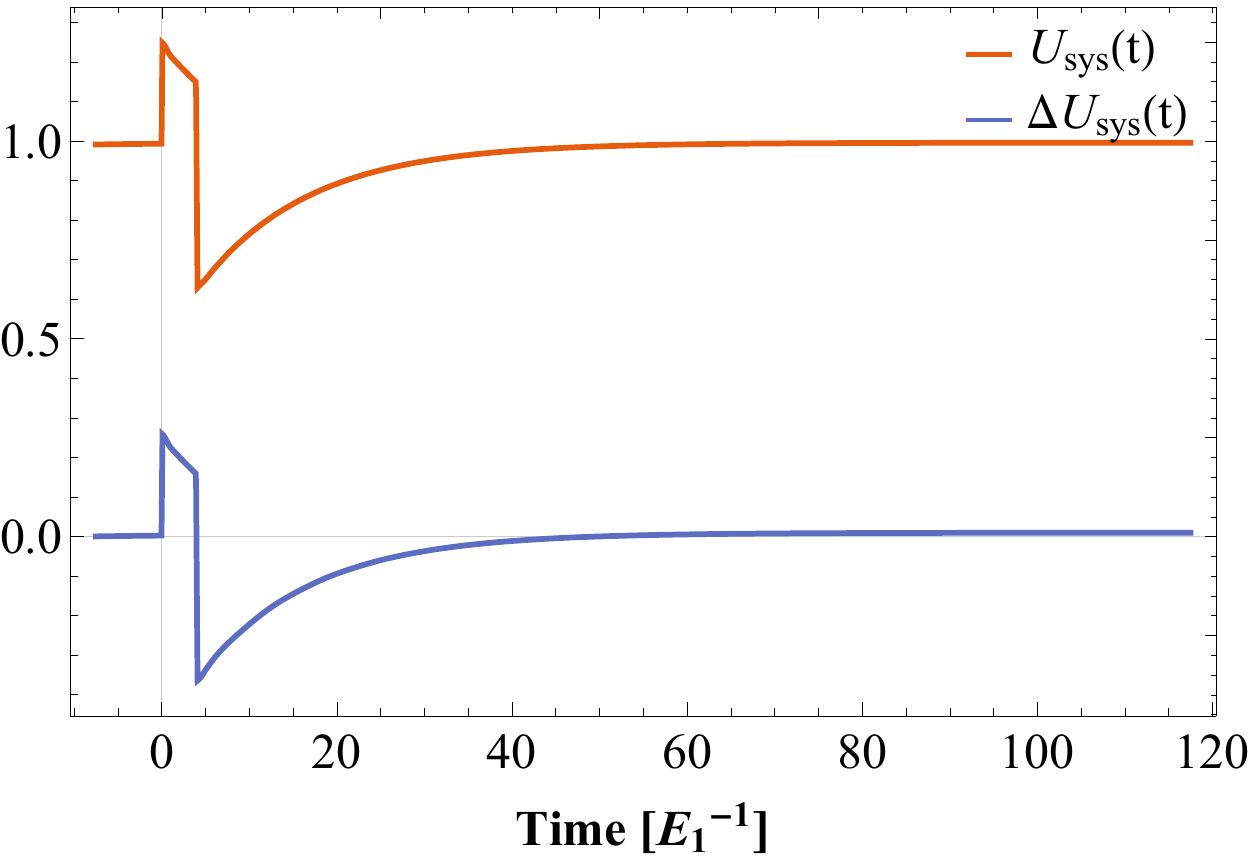}}
	
	\captionsetup{justification=raggedright, singlelinecheck=false}
	
	\caption{The Time-dependent energetics of the Heat Engine depicted schematically in Fig.\ \ref{heatengine_fig}. (a) Total Work done by external drive versus time, (b) Heat extracted from the right (hot) Reservoir, (c) total Electrochemical Work done versus time, and (d) Internal Energy (in orange) and RHS of the First Law equation \eqref{firstlaw1_eqn} (in blue) versus time. For the results shown, we have used $\hbar=1$ and we set the on-site energy on the left and right dots to $\{E_1,E_2\}=\{1,3\}$ so that the difference in the on-site energies of the two dots is $\Delta E=E_2-E_1=2$ in units of $E_1$. The dot hybridization (when non-zero during the pulse) is set $w=\Delta E/5$ and the coupling to the reservoirs is  taken as $\Gamma=w/5$. The chemical potentials and temperatures of the reservoirs are set to $\mu_1=\mu_2=0$ and $\{k_{B}T_1,k_{B}T_2\} = \{0.5,3.6\}$ in units of $E_1$, respectively. The pulse amplitude is set equal to the difference in the on-site energies $\delta=\Delta E$ and the pulse duration is set $\tau=\pi/2w$ ($\pi$-pulse).}

	\label{heatengine_plots}
	
\end{figure*}

For the heat engine configuration, the temperatures are raised to significant unequal values $(T_2>T_1)$ comparable in magnitude to the system energy levels, and the chemical potentials of both the reservoirs are set to zero, $\mu_{1}=\mu_{2}=0$. Furthermore, in addition to the left dot energy, the inter-dot coupling is now a function of time $w \rightarrow w(t)$. Specifically, it is activated to a constant value $w$ only during the pulse and is zero before and after it. This is done to suppress heat flow into the reservoirs in the quiescent state of the machine.  The efficiency of the cycle is
\begin{equation}
    \eta_{HE}=\frac{|W_{ext}(t\rightarrow\infty)|}{|Q_2(t\rightarrow\infty)|}, 
\end{equation}
where $Q_2(t\rightarrow\infty)$ is the heat extracted from the right reservoir at late times. 

Again, the full Hamiltonian for this configuration is given by Eq.\ \eqref{full_hamiltonian_eqn} and the System Hamiltonian is %
\begin{equation}\label{heateng_HS_eqn}
	H_{S}(t)= E_1(t)d_{1}^{\dagger}d_{1}+E_{2}d_{2}^{\dagger}d_{2}+w(t)(d_{1}^{\dagger}d_{2}+d_{2}^{\dagger}d_{1}) ,
\end{equation}
with the Reservoir and Coupling Hamiltonians exactly the same as in the electrochemical pump case given by Eqs.\ \eqref{echem_pump_HB_eqn} and \eqref{echem_pump_HSB_eqn}, respectively. 
As before, the rectangular pulse acts only on the left dot with $E_1(t)$ given by Eq.\ \eqref{rect_pulse_eqn}, and the time-dependent inter-dot coupling is 
\begin{equation} w(t) = \begin{cases} 
	
	0 & t<0, \\
	w	& 0\leq t< \tau, \\
	0 & t\geq\tau.
	
\end{cases}
\end{equation}

In the heat engine configuration, the electron transport path is essentially reversed from the pump configuration. The electron tunnels in from the right reservoir onto the right dot, Rabi Oscillates onto the left dot, is lowered to energy $E_1$ at the end of the pulse, and tunnels out into the left reservoir, accomplishing engine operation. A schematic for the engine is given in Fig.\ \ref{heatengine_fig}.

The time-dependent energetics shown in Fig.\ \ref{heatengine_plots}, where all the energies are reported in units of the on-site energy  of the left dot $E_1$ and time is in units of $E_1^{-1}$, demonstrate that external work $W_{ext}(t)$ (Fig. \ref{heatengine_wext}) is now done by the engine on the drive as the left dot is lowered at the end of the pulse. For the parameters chosen $W_{ext}(t)$ at late times attains a constant value of $-0.18$. This work is done by extracting heat $Q_2(t)$ (Fig. \ref{heatengine_rightheat}) from the hot reservoir at $T_2$, and eventually attains a constant value of $-0.60$ at late times after the pulse and inter-dot coupling are inactivated. %

By design, no net electrochemical work is done (Fig.\ \ref{heatengine_welec}) since we had set the chemical potentials of both the reservoirs to $0$ in this configuration. Finally, the left and right hand sides of the First Law are again in good agreement 
(Fig.\ \ref{heatengine_inten}), up to a constant of integration offset. For the parameters chosen, the heat engine operates at an efficiency $\eta_{HE}$ of about $46.5\%$ which is close to $53\%$ of Carnot efficiency.

\subsubsection{Spatio-Temporal Distribution of Energy---Heat Engine}\label{int_egy_tbchain_subsubsec_2}

\begin{figure}
\centering
\includegraphics[width=.45\textwidth,height=16cm]{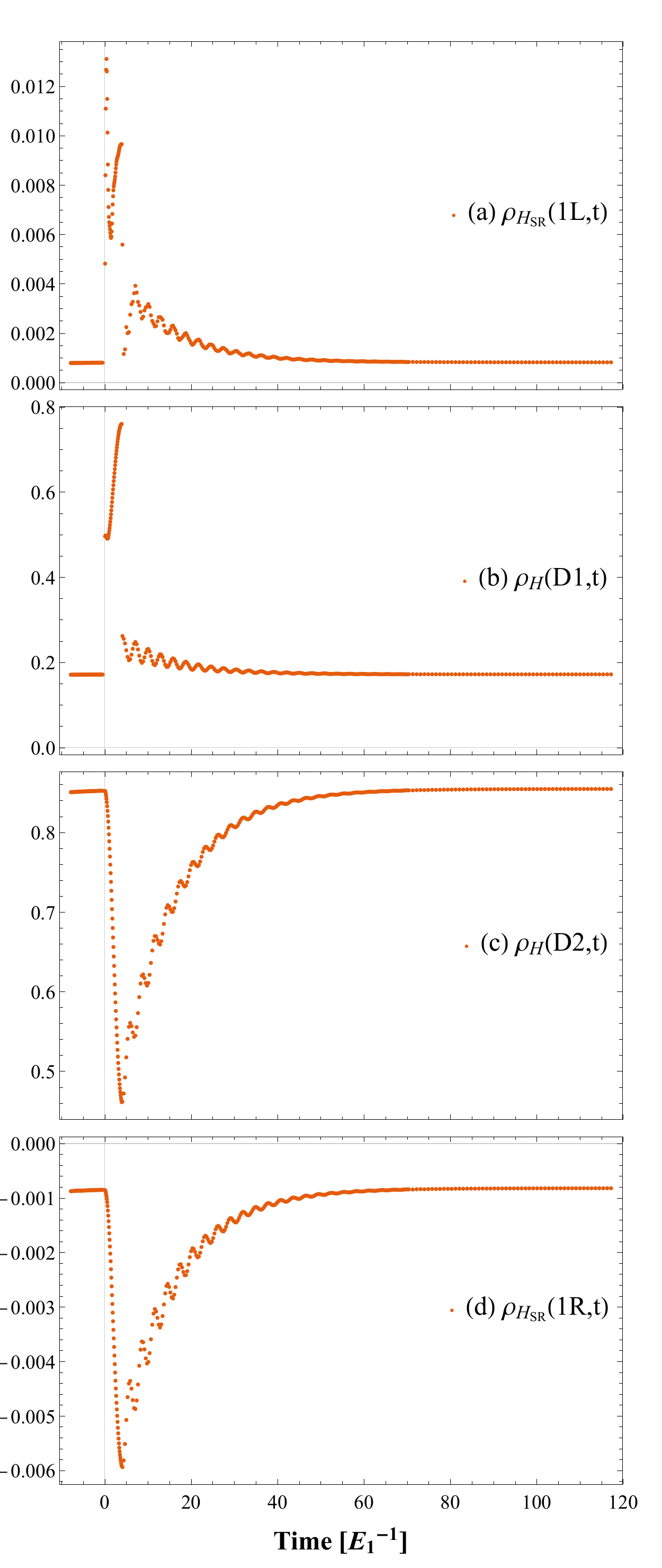}

\captionsetup{justification=raggedright,singlelinecheck=false}

\caption{ %
Plots of spatially resolved energy as a function of time for the \emph{Heat Engine} configuration for the same Hamiltonian parameters as those for Fig.\ \ref{heatengine_plots}. For the interfacial sites on both the reservoirs, we plot the contribution of the interface energy [Eq.\ \eqref{tbchain_cpl_ham_eqn}] denoted by $\rho_{H_{SR}}(1L,t)$ and $\rho_{H_{SR}}(1R,t)$ for the left and right reservoirs, respectively. For the two dots, we plot the total energy density as a function of time denoted by $\rho_{H}(D1,t)$ and $\rho_{H}(D2,t)$. The energies are in units of the on-site energy of the right dot $E_1$ and time is in units of $E_1^{-1}$in this configuration. Note that the energy scale for $\rho_{H_{SR}}(1L,t)$ and $\rho_{H_{SR}}(1R,t)$ (panels (a) and (d)) is different from that for $\rho_{H}(D1,t)$ and $\rho_{H}(D2,t)$ (panels (b) and (c)) in the figure.}
\label{heateng_energy_density_plots_fig}

\end{figure}

Fig.\ \ref{heateng_energy_density_plots_fig} gives the results for the spatiotemporal distribution of the energy of the machine in the Heat Engine configuration. For these results, the system Hamiltonian is the same as Eq.\ \eqref{heateng_HS_eqn} used for the integrated thermodynamic quantities in the preceding paragraphs, and the functional form of the reservoir and coupling Hamiltonians is kept the same as that in the electrochemical pump case, Eqs.\ \eqref{tbchain_res_ham_eqn} and \eqref{tbchain_cpl_ham_eqn}. %

We again see the total energy for the left dot $\rho_{H}(D1,t)$ (Fig.\ \ref{heateng_energy_density_plots_fig}b) rise instantaneously as the pulse is activated at $t=0$. It increases further to its maximum value as the electron Rabi oscillates from the right dot to the left dot. The energy on the left dot decreases as the electron tunnels out into the left reservoir, eventually asymptoting to its initial value at late times. 

On the first site on the left reservoir, the interfacial energy $\rho_{H_{SR}}(1L,t)$  (Fig.\ \ref{heateng_energy_density_plots_fig}a) rises first at $t=0$ in response to the activation of the pulse at the left dot, reflecting again the fact that $H_{SR}$ models the bond between the system and reservoir. The energy rises for a second time in response to the Rabi oscillation of the electron onto the left dot from the right dot, and yet again for a third and final time as the electron tunnels onto the first site on the left reservoir itself. As initial conditions are re-established, the coupling energy again assumes its initial value. 

On the right dot, the energy $\rho_{H}(D2,t)$  (Fig.\ \ref{heateng_energy_density_plots_fig}c) begins to decrease upon the activation of the pulse as the electron tunnels onto the left dot. It asymptotes back to its initial value as another electron tunnels in from the right reservoir and initial conditions are re-established. Again, the interfacial contribution to the energy $\rho_{H_{SR}}(1R,t)$ (Fig.\ \ref{heateng_energy_density_plots_fig}d) on the first site on right reservoir almost mimics the behaviour at the right dot.

\section{Conclusions}\label{conclusions_sec}

Formulation of the First Law of Thermodynamics requires proper definitions of Work, Heat, and Internal Energy.  While the definition of External Work was established early on in Quantum Thermodynamics \cite{jarzynskiComparisonFarfromequilibriumWork2007,talknerColloquiumStatisticalMechanics2020}, the latter two concepts have proven problematic for open quantum systems driven out of equilibrium \cite{talknerColloquiumStatisticalMechanics2020, ludovicoDynamicalEnergyTransfer2014,bruchQuantumThermodynamicsDriven2016,espositoEntropyProductionCorrelation2010,strasbergFirstSecondLaw2021a,lacerdaQuantumThermodynamicsFast2023,bergmannGreenfunctionPerspective2021}.  In this article, we have shown that the proper definition of the Internal Energy operator $U_{sys}(t)$ for an open quantum system is the partition of the Hamiltonian operator onto the system Hilbert space, confirming the validity of the so-called
half-and-half partition \cite{ludovicoDynamicalEnergyTransfer2014,bruchQuantumThermodynamicsDriven2016}, $U_{sys}(t)=H_{S}(t)+\frac{1}{2}H_{SR}$, wherein the internal energy of the system is the sum of the system Hamiltonian and half of the coupling Hamiltonian describing the System-Reservoir interface.
Furthermore, we have been able to identify the Electrochemical Work done on the system and the Heat flowing out of the system into the environment by employing a key insight from Mesoscopics---that far away from the local driving and coupling of the system, the reservoirs remain arbitrarily close to equilibrium.  This allows the Electrochemical Work and Heat %
to be properly defined for open quantum systems driven far from equilibrium, although there is a transient component of the energy flux, carried in particular by phonons, that is not identifiable as heat per se except for steady-state or slowly-varying driving conditions.

Our definition (\ref{internal_energy_op_eqn}) of Internal Energy agrees with that proposed by Ludovico et al.\ \cite{ludovicoDynamicalEnergyTransfer2014} for the specific model of a single sinusoidally driven Fermionic  resonant level coupled to a single reservoir (supported in the subsequent work of Bruch et al.\ \cite{bruchQuantumThermodynamicsDriven2016}), as well as with that put forward by Stafford and Shastry \cite{staffordLocalEntropyNonequilibrium2017,shastryThirdLawThermodynamics2019,shastryTheoryThermodynamicMeasurements2019}
for  an independent Fermion system in steady state,
but disagrees 
with those put forward by Esposito et al. \cite{espositoEntropyProductionCorrelation2010}, Strasberg and Winter \cite{strasbergFirstSecondLaw2021a}, and Lacerda et al.\ \cite{lacerdaQuantumThermodynamicsFast2023}.
We remark that our definition of $U_{sys}(t)$ based on Hilbert-space partition appears to coincide with the definition of Internal Energy
advocated in Ref.\ \cite{bruchQuantumThermodynamicsDriven2016}---based on the $H_S$-dependent part of the total energy---in the broad-band limit only, but our definition remains valid outside the broad-band limit.

We have derived fully general expressions for all the terms appearing in the First Law using the formalism of non-equilibrium Green's functions (NEGF). In our analysis, the system-reservoir coupling can be arbitrarily strong, and our analysis can be readily extended to the general case where the system, coupling, and reservoir are all explicitly time-dependent (see Appendix \ref{full_tdpt_generalize_app}). Furthermore, our formal results incorporate electronic and phononic degrees of freedom in both the system and the reservoir as well as electron-electron,  electron-phonon, and phonon-phonon interactions in the system. In our analysis, the external electromagnetic field used to drive the electron-phonon system is treated classically, which allows for absorption and stimulated emission of energy quanta, but does not allow spontaneous emission or radiative heat transfer \cite{PhysRevLett.105.234301}.

We also studied the spatio-temporal distribution of the energy in a strongly-driven time-dependent open quantum system, shedding light on the internal dynamics of the system as well as on the evolution of the system-reservoir interface.  The interface contribution to the internal energy of the system was shown to be localized at the junctions between the system and the reservoirs.  For the
case of a metallic reservoir in the nearest-neighbor tight-binding model, with only nearest-neighbor bonds to the system, the
interface energy was found to be entirely localized on the system-reservoir bonds.  The sign of the interface energy can vary from negative to positive, corresponding to bonding or antibonding character of the interface, respectively.

We applied our formal results to a strongly driven quantum machine utilizing Rabi Oscillations between states of a double quantum dot coupled to metallic reservoirs. We presented a full thermodynamic analysis of the machine's operation as both an electrochemical pump and a heat engine, illustrating the utility of our theoretical framework and the power of our computational approach based on NEGF. 

It should be remarked that although the First Law clearly holds at the level of quantum statistical averages, provided the energetics are partitioned properly, one cannot expect it to hold at the level of individual quantum trajectories, since the operators for internal energy, heat current, and chemical power (among others) do not commute, and hence these are incompatible observables \cite{talknerColloquiumStatisticalMechanics2020,kerremansProbabilisticallyViolatingFirst2022}. 
In particular, in addition to thermal fluctuations, there exist also quantum fluctuations of the various energetic factors appearing in the First Law due to
the generalized uncertainty principle for non-commuting observables.

It is hoped that our very general derivation of the First Law of Thermodynamics in time-dependent open quantum systems, illustrated with applications of our theory to specific examples of model quantum thermal machines, will help to resolve once and for all the controversy over the validity of First Law of Thermodynamics in open quantum systems.

\section{Acknowledgments}\label{acknowledge_sec}
It is a pleasure to acknowledge several helpful discussions with Caleb M.\ Webb, Marco A.\ Jimenez-Valencia, and Carter S.\ Eckel during various stages of this work. We also thank J.\ M.\ van Ruitenbeek for suggesting the valuable generalization of incorporating phonons in our model. This work was partially supported by the U.S.\ Department of Energy (DOE), Office of Science under Award No. DE-SC0006699. %
After submission of the first version of this manuscript, we became aware of Ref.\ \cite{dannUnificationFirstLaw2022}, which addresses some issues similar to those considered in this article.

\bibliography{firstlaw_main}%

\begin{thebibliography}{89}%
\makeatletter
\providecommand \@ifxundefined [1]{%
 \@ifx{#1\undefined}
}%
\providecommand \@ifnum [1]{%
 \ifnum #1\expandafter \@firstoftwo
 \else \expandafter \@secondoftwo
 \fi
}%
\providecommand \@ifx [1]{%
 \ifx #1\expandafter \@firstoftwo
 \else \expandafter \@secondoftwo
 \fi
}%
\providecommand \natexlab [1]{#1}%
\providecommand \enquote  [1]{``#1''}%
\providecommand \bibnamefont  [1]{#1}%
\providecommand \bibfnamefont [1]{#1}%
\providecommand \citenamefont [1]{#1}%
\providecommand \href@noop [0]{\@secondoftwo}%
\providecommand \href [0]{\begingroup \@sanitize@url \@href}%
\providecommand \@href[1]{\@@startlink{#1}\@@href}%
\providecommand \@@href[1]{\endgroup#1\@@endlink}%
\providecommand \@sanitize@url [0]{\catcode `\\12\catcode `\$12\catcode
  `\&12\catcode `\#12\catcode `\^12\catcode `\_12\catcode `\%12\relax}%
\providecommand \@@startlink[1]{}%
\providecommand \@@endlink[0]{}%
\providecommand \url  [0]{\begingroup\@sanitize@url \@url }%
\providecommand \@url [1]{\endgroup\@href {#1}{\urlprefix }}%
\providecommand \urlprefix  [0]{URL }%
\providecommand \Eprint [0]{\href }%
\providecommand \doibase [0]{https://doi.org/}%
\providecommand \selectlanguage [0]{\@gobble}%
\providecommand \bibinfo  [0]{\@secondoftwo}%
\providecommand \bibfield  [0]{\@secondoftwo}%
\providecommand \translation [1]{[#1]}%
\providecommand \BibitemOpen [0]{}%
\providecommand \bibitemStop [0]{}%
\providecommand \bibitemNoStop [0]{.\EOS\space}%
\providecommand \EOS [0]{\spacefactor3000\relax}%
\providecommand \BibitemShut  [1]{\csname bibitem#1\endcsname}%
\let\auto@bib@innerbib\@empty
\bibitem [{\citenamefont {Binder}\ \emph {et~al.}(2018)\citenamefont {Binder},
  \citenamefont {Correa}, \citenamefont {Gogolin}, \citenamefont {Anders},\
  and\ \citenamefont {Adesso}}]{binderThermodynamicsQuantumRegime2018}%
  \BibitemOpen
  \bibinfo {editor} {\bibfnamefont {F.}~\bibnamefont {Binder}}, \bibinfo
  {editor} {\bibfnamefont {L.~A.}\ \bibnamefont {Correa}}, \bibinfo {editor}
  {\bibfnamefont {C.}~\bibnamefont {Gogolin}}, \bibinfo {editor} {\bibfnamefont
  {J.}~\bibnamefont {Anders}},\ and\ \bibinfo {editor} {\bibfnamefont
  {G.}~\bibnamefont {Adesso}},\ eds.,\ \href
  {https://doi.org/10.1007/978-3-319-99046-0} {\emph {\bibinfo {title}
  {Thermodynamics in the {{Quantum Regime}}: {{Fundamental Aspects}} and {{New
  Directions}}}}},\ \bibinfo {series} {Fundamental {{Theories}} of
  {{Physics}}}, Vol.\ \bibinfo {volume} {195}\ (\bibinfo  {publisher}
  {{Springer International Publishing}},\ \bibinfo {address} {{Cham}},\
  \bibinfo {year} {2018})\BibitemShut {NoStop}%
\bibitem [{\citenamefont {Shastry}\ \emph {et~al.}(2020)\citenamefont
  {Shastry}, \citenamefont {Inui},\ and\ \citenamefont
  {Stafford}}]{Shastry2020_STTh}%
  \BibitemOpen
  \bibfield  {author} {\bibinfo {author} {\bibfnamefont {A.}~\bibnamefont
  {Shastry}}, \bibinfo {author} {\bibfnamefont {S.}~\bibnamefont {Inui}},\ and\
  \bibinfo {author} {\bibfnamefont {C.~A.}\ \bibnamefont {Stafford}},\
  }\bibfield  {title} {\bibinfo {title} {Scanning tunneling thermometry},\
  }\href {https://doi.org/10.1103/PhysRevApplied.13.024065} {\bibfield
  {journal} {\bibinfo  {journal} {Physical Review Applied}\ }\textbf {\bibinfo
  {volume} {13}},\ \bibinfo {pages} {024065} (\bibinfo {year}
  {2020})}\BibitemShut {NoStop}%
\bibitem [{\citenamefont {Mecklenburg}\ \emph {et~al.}(2015)\citenamefont
  {Mecklenburg}, \citenamefont {Hubbard}, \citenamefont {White}, \citenamefont
  {Dhall}, \citenamefont {Cronin}, \citenamefont {Aloni},\ and\ \citenamefont
  {Regan}}]{mecklenburgNanoscaleTemperatureMapping2015}%
  \BibitemOpen
  \bibfield  {author} {\bibinfo {author} {\bibfnamefont {M.}~\bibnamefont
  {Mecklenburg}}, \bibinfo {author} {\bibfnamefont {W.~A.}\ \bibnamefont
  {Hubbard}}, \bibinfo {author} {\bibfnamefont {E.~R.}\ \bibnamefont {White}},
  \bibinfo {author} {\bibfnamefont {R.}~\bibnamefont {Dhall}}, \bibinfo
  {author} {\bibfnamefont {S.~B.}\ \bibnamefont {Cronin}}, \bibinfo {author}
  {\bibfnamefont {S.}~\bibnamefont {Aloni}},\ and\ \bibinfo {author}
  {\bibfnamefont {B.~C.}\ \bibnamefont {Regan}},\ }\bibfield  {title} {\bibinfo
  {title} {Nanoscale temperature mapping in operating microelectronic
  devices},\ }\href {https://doi.org/10.1126/science.aaa2433} {\bibfield
  {journal} {\bibinfo  {journal} {Science (New York, N.Y.)}\ }\textbf {\bibinfo
  {volume} {347}},\ \bibinfo {pages} {629} (\bibinfo {year}
  {2015})}\BibitemShut {NoStop}%
\bibitem [{\citenamefont {Neumann}\ \emph {et~al.}(2013)\citenamefont
  {Neumann}, \citenamefont {Jakobi}, \citenamefont {Dolde}, \citenamefont
  {Burk}, \citenamefont {Reuter}, \citenamefont {Waldherr}, \citenamefont
  {Honert}, \citenamefont {Wolf}, \citenamefont {Brunner}, \citenamefont
  {Shim}, \citenamefont {Suter}, \citenamefont {Sumiya}, \citenamefont
  {Isoya},\ and\ \citenamefont
  {Wrachtrup}}]{neumannHighprecisionNanoscaleTemperature2013}%
  \BibitemOpen
  \bibfield  {author} {\bibinfo {author} {\bibfnamefont {P.}~\bibnamefont
  {Neumann}}, \bibinfo {author} {\bibfnamefont {I.}~\bibnamefont {Jakobi}},
  \bibinfo {author} {\bibfnamefont {F.}~\bibnamefont {Dolde}}, \bibinfo
  {author} {\bibfnamefont {C.}~\bibnamefont {Burk}}, \bibinfo {author}
  {\bibfnamefont {R.}~\bibnamefont {Reuter}}, \bibinfo {author} {\bibfnamefont
  {G.}~\bibnamefont {Waldherr}}, \bibinfo {author} {\bibfnamefont
  {J.}~\bibnamefont {Honert}}, \bibinfo {author} {\bibfnamefont
  {T.}~\bibnamefont {Wolf}}, \bibinfo {author} {\bibfnamefont {A.}~\bibnamefont
  {Brunner}}, \bibinfo {author} {\bibfnamefont {J.~H.}\ \bibnamefont {Shim}},
  \bibinfo {author} {\bibfnamefont {D.}~\bibnamefont {Suter}}, \bibinfo
  {author} {\bibfnamefont {H.}~\bibnamefont {Sumiya}}, \bibinfo {author}
  {\bibfnamefont {J.}~\bibnamefont {Isoya}},\ and\ \bibinfo {author}
  {\bibfnamefont {J.}~\bibnamefont {Wrachtrup}},\ }\bibfield  {title} {\bibinfo
  {title} {High-precision nanoscale temperature sensing using single defects in
  diamond},\ }\href {https://doi.org/10.1021/nl401216y} {\bibfield  {journal}
  {\bibinfo  {journal} {Nano Letters}\ }\textbf {\bibinfo {volume} {13}},\
  \bibinfo {pages} {2738} (\bibinfo {year} {2013})},\ \bibinfo {note} {pMID:
  23721106}\BibitemShut {NoStop}%
\bibitem [{\citenamefont {Jeong}\ \emph {et~al.}(2015)\citenamefont {Jeong},
  \citenamefont {Hur}, \citenamefont {Meyhofer},\ and\ \citenamefont
  {Reddy}}]{jeongScanningProbeMicroscopy2015}%
  \BibitemOpen
  \bibfield  {author} {\bibinfo {author} {\bibfnamefont {W.}~\bibnamefont
  {Jeong}}, \bibinfo {author} {\bibfnamefont {S.}~\bibnamefont {Hur}}, \bibinfo
  {author} {\bibfnamefont {E.}~\bibnamefont {Meyhofer}},\ and\ \bibinfo
  {author} {\bibfnamefont {P.}~\bibnamefont {Reddy}},\ }\bibfield  {title}
  {\bibinfo {title} {Scanning probe microscopy for thermal transport
  measurements},\ }\href {https://doi.org/10.1080/15567265.2015.1109740}
  {\bibfield  {journal} {\bibinfo  {journal} {Nanoscale and Microscale
  Thermophysical Engineering}\ }\textbf {\bibinfo {volume} {19}},\ \bibinfo
  {pages} {279} (\bibinfo {year} {2015})}\BibitemShut {NoStop}%
\bibitem [{\citenamefont {Menges}\ \emph {et~al.}(2016)\citenamefont {Menges},
  \citenamefont {Mensch}, \citenamefont {Schmid}, \citenamefont {Riel},
  \citenamefont {Stemmer},\ and\ \citenamefont
  {Gotsmann}}]{mengesTemperatureMappingOperating2016}%
  \BibitemOpen
  \bibfield  {author} {\bibinfo {author} {\bibfnamefont {F.}~\bibnamefont
  {Menges}}, \bibinfo {author} {\bibfnamefont {P.}~\bibnamefont {Mensch}},
  \bibinfo {author} {\bibfnamefont {H.}~\bibnamefont {Schmid}}, \bibinfo
  {author} {\bibfnamefont {H.}~\bibnamefont {Riel}}, \bibinfo {author}
  {\bibfnamefont {A.}~\bibnamefont {Stemmer}},\ and\ \bibinfo {author}
  {\bibfnamefont {B.}~\bibnamefont {Gotsmann}},\ }\bibfield  {title} {\bibinfo
  {title} {Temperature mapping of operating nanoscale devices by scanning probe
  thermometry},\ }\href {http://dx.doi.org/10.1038/ncomms10874} {\bibfield
  {journal} {\bibinfo  {journal} {Nature Communications}\ }\textbf {\bibinfo
  {volume} {7}},\ \bibinfo {pages} {10874 EP } (\bibinfo {year} {2016})},\
  \bibinfo {note} {article}\BibitemShut {NoStop}%
\bibitem [{\citenamefont {Shi}\ \emph {et~al.}(2009)\citenamefont {Shi},
  \citenamefont {Zhou}, \citenamefont {Kim}, \citenamefont {Bachtold},
  \citenamefont {Majumdar},\ and\ \citenamefont
  {McEuen}}]{shiThermalProbingEnergy2009}%
  \BibitemOpen
  \bibfield  {author} {\bibinfo {author} {\bibfnamefont {L.}~\bibnamefont
  {Shi}}, \bibinfo {author} {\bibfnamefont {J.}~\bibnamefont {Zhou}}, \bibinfo
  {author} {\bibfnamefont {P.}~\bibnamefont {Kim}}, \bibinfo {author}
  {\bibfnamefont {A.}~\bibnamefont {Bachtold}}, \bibinfo {author}
  {\bibfnamefont {A.}~\bibnamefont {Majumdar}},\ and\ \bibinfo {author}
  {\bibfnamefont {P.~L.}\ \bibnamefont {McEuen}},\ }\bibfield  {title}
  {\bibinfo {title} {Thermal probing of energy dissipation in current-carrying
  carbon nanotubes},\ }\href {https://doi.org/10.1063/1.3126708} {\bibfield
  {journal} {\bibinfo  {journal} {Journal of Applied Physics}\ }\textbf
  {\bibinfo {volume} {105}},\ \bibinfo {pages} {104306} (\bibinfo {year}
  {2009})}\BibitemShut {NoStop}%
\bibitem [{\citenamefont {Kim}\ \emph {et~al.}(2012)\citenamefont {Kim},
  \citenamefont {Jeong}, \citenamefont {Lee},\ and\ \citenamefont
  {Reddy}}]{kimUltrahighVacuumScanning2012}%
  \BibitemOpen
  \bibfield  {author} {\bibinfo {author} {\bibfnamefont {K.}~\bibnamefont
  {Kim}}, \bibinfo {author} {\bibfnamefont {W.}~\bibnamefont {Jeong}}, \bibinfo
  {author} {\bibfnamefont {W.}~\bibnamefont {Lee}},\ and\ \bibinfo {author}
  {\bibfnamefont {P.}~\bibnamefont {Reddy}},\ }\bibfield  {title} {\bibinfo
  {title} {Ultra-high vacuum scanning thermal microscopy for nanometer
  resolution quantitative thermometry},\ }\href
  {https://doi.org/10.1021/nn300774n} {\bibfield  {journal} {\bibinfo
  {journal} {ACS Nano}\ }\textbf {\bibinfo {volume} {6}},\ \bibinfo {pages}
  {4248} (\bibinfo {year} {2012})},\ \bibinfo {note} {pMID:
  22530657}\BibitemShut {NoStop}%
\bibitem [{\citenamefont {Lee}\ \emph {et~al.}(2013)\citenamefont {Lee},
  \citenamefont {Kim}, \citenamefont {Jeong}, \citenamefont {Zotti},
  \citenamefont {Pauly}, \citenamefont {Cuevas},\ and\ \citenamefont
  {Reddy}}]{Lee2013}%
  \BibitemOpen
  \bibfield  {author} {\bibinfo {author} {\bibfnamefont {W.}~\bibnamefont
  {Lee}}, \bibinfo {author} {\bibfnamefont {K.}~\bibnamefont {Kim}}, \bibinfo
  {author} {\bibfnamefont {W.}~\bibnamefont {Jeong}}, \bibinfo {author}
  {\bibfnamefont {L.~A.}\ \bibnamefont {Zotti}}, \bibinfo {author}
  {\bibfnamefont {F.}~\bibnamefont {Pauly}}, \bibinfo {author} {\bibfnamefont
  {J.~C.}\ \bibnamefont {Cuevas}},\ and\ \bibinfo {author} {\bibfnamefont
  {P.}~\bibnamefont {Reddy}},\ }\bibfield  {title} {\bibinfo {title} {Heat
  dissipation in atomic-scale junctions},\ }\href
  {http://dx.doi.org/10.1038/nature12183} {\bibfield  {journal} {\bibinfo
  {journal} {Nature}\ }\textbf {\bibinfo {volume} {498}},\ \bibinfo {pages}
  {209} (\bibinfo {year} {2013})}\BibitemShut {NoStop}%
\bibitem [{\citenamefont {Gom{\`e}s}\ \emph {et~al.}(2015)\citenamefont
  {Gom{\`e}s}, \citenamefont {Assy},\ and\ \citenamefont
  {Chapuis}}]{Gomes2015}%
  \BibitemOpen
  \bibfield  {author} {\bibinfo {author} {\bibfnamefont {S.}~\bibnamefont
  {Gom{\`e}s}}, \bibinfo {author} {\bibfnamefont {A.}~\bibnamefont {Assy}},\
  and\ \bibinfo {author} {\bibfnamefont {P.-O.}\ \bibnamefont {Chapuis}},\
  }\bibfield  {title} {\bibinfo {title} {Scanning thermal microscopy: {{A}}
  review},\ }\href {https://doi.org/10.1002/pssa.201400360} {\bibfield
  {journal} {\bibinfo  {journal} {physica status solidi (a)}\ }\textbf
  {\bibinfo {volume} {212}},\ \bibinfo {pages} {477} (\bibinfo {year}
  {2015})}\BibitemShut {NoStop}%
\bibitem [{\citenamefont {Cui}\ \emph {et~al.}(2017)\citenamefont {Cui},
  \citenamefont {Jeong}, \citenamefont {Hur}, \citenamefont {Matt},
  \citenamefont {Kl{\"o}ckner}, \citenamefont {Pauly}, \citenamefont {Nielaba},
  \citenamefont {Cuevas}, \citenamefont {Meyhofer},\ and\ \citenamefont
  {Reddy}}]{cuiQuantizedThermalTransport2017}%
  \BibitemOpen
  \bibfield  {author} {\bibinfo {author} {\bibfnamefont {L.}~\bibnamefont
  {Cui}}, \bibinfo {author} {\bibfnamefont {W.}~\bibnamefont {Jeong}}, \bibinfo
  {author} {\bibfnamefont {S.}~\bibnamefont {Hur}}, \bibinfo {author}
  {\bibfnamefont {M.}~\bibnamefont {Matt}}, \bibinfo {author} {\bibfnamefont
  {J.~C.}\ \bibnamefont {Kl{\"o}ckner}}, \bibinfo {author} {\bibfnamefont
  {F.}~\bibnamefont {Pauly}}, \bibinfo {author} {\bibfnamefont
  {P.}~\bibnamefont {Nielaba}}, \bibinfo {author} {\bibfnamefont {J.~C.}\
  \bibnamefont {Cuevas}}, \bibinfo {author} {\bibfnamefont {E.}~\bibnamefont
  {Meyhofer}},\ and\ \bibinfo {author} {\bibfnamefont {P.}~\bibnamefont
  {Reddy}},\ }\bibfield  {title} {\bibinfo {title} {Quantized thermal transport
  in single-atom junctions},\ }\href {https://doi.org/10.1126/science.aam6622}
  {\bibfield  {journal} {\bibinfo  {journal} {Science (New York, N.Y.)}\
  }\textbf {\bibinfo {volume} {355}},\ \bibinfo {pages} {1192} (\bibinfo {year}
  {2017})}\BibitemShut {NoStop}%
\bibitem [{\citenamefont {Mosso}\ \emph {et~al.}(2017)\citenamefont {Mosso},
  \citenamefont {Drechsler}, \citenamefont {Menges}, \citenamefont {Nirmalraj},
  \citenamefont {Karg}, \citenamefont {Riel},\ and\ \citenamefont
  {Gotsmann}}]{Mosso2017}%
  \BibitemOpen
  \bibfield  {author} {\bibinfo {author} {\bibfnamefont {N.}~\bibnamefont
  {Mosso}}, \bibinfo {author} {\bibfnamefont {U.}~\bibnamefont {Drechsler}},
  \bibinfo {author} {\bibfnamefont {F.}~\bibnamefont {Menges}}, \bibinfo
  {author} {\bibfnamefont {P.}~\bibnamefont {Nirmalraj}}, \bibinfo {author}
  {\bibfnamefont {S.}~\bibnamefont {Karg}}, \bibinfo {author} {\bibfnamefont
  {H.}~\bibnamefont {Riel}},\ and\ \bibinfo {author} {\bibfnamefont
  {B.}~\bibnamefont {Gotsmann}},\ }\bibfield  {title} {\bibinfo {title} {Heat
  transport through atomic contacts},\ }\href
  {http://dx.doi.org/10.1038/nnano.2016.302} {\bibfield  {journal} {\bibinfo
  {journal} {Nat Nano}\ }\textbf {\bibinfo {volume} {12}},\ \bibinfo {pages}
  {430} (\bibinfo {year} {2017})},\ \bibinfo {note} {letter}\BibitemShut
  {NoStop}%
\bibitem [{\citenamefont {Toyabe}\ \emph {et~al.}(2010)\citenamefont {Toyabe},
  \citenamefont {Sagawa}, \citenamefont {Ueda}, \citenamefont {Muneyuki},\ and\
  \citenamefont {Sano}}]{Toyabi2010_Szilard_exp}%
  \BibitemOpen
  \bibfield  {author} {\bibinfo {author} {\bibfnamefont {S.}~\bibnamefont
  {Toyabe}}, \bibinfo {author} {\bibfnamefont {T.}~\bibnamefont {Sagawa}},
  \bibinfo {author} {\bibfnamefont {M.}~\bibnamefont {Ueda}}, \bibinfo {author}
  {\bibfnamefont {E.}~\bibnamefont {Muneyuki}},\ and\ \bibinfo {author}
  {\bibfnamefont {M.}~\bibnamefont {Sano}},\ }\bibfield  {title} {\bibinfo
  {title} {Experimental demonstration of information-to-energy conversion and
  validation of the generalized {{Jarzynski}} equality},\ }\href
  {https://doi.org/10.1038/nphys1821} {\bibfield  {journal} {\bibinfo
  {journal} {Nature Physics}\ }\textbf {\bibinfo {volume} {6}},\ \bibinfo
  {pages} {988} (\bibinfo {year} {2010})},\ \Eprint
  {https://arxiv.org/abs/1009.5287} {arxiv:1009.5287 [cond-mat.stat-mech]}
  \BibitemShut {NoStop}%
\bibitem [{\citenamefont {B{\'e}rut}\ \emph {et~al.}(2012)\citenamefont
  {B{\'e}rut}, \citenamefont {Arakelyan}, \citenamefont {Petrosyan},
  \citenamefont {Ciliberto}, \citenamefont {Dillenschneider},\ and\
  \citenamefont {Lutz}}]{berutExperimentalVerificationLandauer2012}%
  \BibitemOpen
  \bibfield  {author} {\bibinfo {author} {\bibfnamefont {A.}~\bibnamefont
  {B{\'e}rut}}, \bibinfo {author} {\bibfnamefont {A.}~\bibnamefont
  {Arakelyan}}, \bibinfo {author} {\bibfnamefont {A.}~\bibnamefont
  {Petrosyan}}, \bibinfo {author} {\bibfnamefont {S.}~\bibnamefont
  {Ciliberto}}, \bibinfo {author} {\bibfnamefont {R.}~\bibnamefont
  {Dillenschneider}},\ and\ \bibinfo {author} {\bibfnamefont {E.}~\bibnamefont
  {Lutz}},\ }\bibfield  {title} {\bibinfo {title} {Experimental verification of
  {{Landauer}}'s principle linking information and thermodynamics},\ }\href
  {https://doi.org/10.1038/nature10872} {\bibfield  {journal} {\bibinfo
  {journal} {Nature}\ }\textbf {\bibinfo {volume} {483}},\ \bibinfo {pages}
  {187} (\bibinfo {year} {2012})}\BibitemShut {NoStop}%
\bibitem [{\citenamefont {Koski}\ \emph {et~al.}(2014)\citenamefont {Koski},
  \citenamefont {Maisi}, \citenamefont {Pekola},\ and\ \citenamefont {{Dmitri
  V. Averin}}}]{koskiExperimentalRealizationSzilard2014}%
  \BibitemOpen
  \bibfield  {author} {\bibinfo {author} {\bibfnamefont {J.~V.}\ \bibnamefont
  {Koski}}, \bibinfo {author} {\bibfnamefont {V.~F.}\ \bibnamefont {Maisi}},
  \bibinfo {author} {\bibfnamefont {J.~P.}\ \bibnamefont {Pekola}},\ and\
  \bibinfo {author} {\bibnamefont {{Dmitri V. Averin}}},\ }\bibfield  {title}
  {\bibinfo {title} {Experimental realization of a {{Szilard}} engine with a
  single electron},\ }\href {https://doi.org/10.1073/pnas.1406966111}
  {\bibfield  {journal} {\bibinfo  {journal} {Proceedings of the National
  Academy of Sciences}\ }\textbf {\bibinfo {volume} {111}},\ \bibinfo {pages}
  {13786} (\bibinfo {year} {2014})}\BibitemShut {NoStop}%
\bibitem [{\citenamefont {Devoret}\ \emph {et~al.}(2014)\citenamefont
  {Devoret}, \citenamefont {Huard}, \citenamefont {Schoelkopf},\ and\
  \citenamefont {Cugliandolo}}]{devoretQuantumMachinesMeasurement2014a}%
  \BibitemOpen
  \bibinfo {editor} {\bibfnamefont {M.~H.}\ \bibnamefont {Devoret}}, \bibinfo
  {editor} {\bibfnamefont {B.}~\bibnamefont {Huard}}, \bibinfo {editor}
  {\bibfnamefont {R.}~\bibnamefont {Schoelkopf}},\ and\ \bibinfo {editor}
  {\bibfnamefont {L.~F.}\ \bibnamefont {Cugliandolo}},\ eds.,\ \href
  {10.1093/acprof:oso/9780199681181.001.0001} {\emph {\bibinfo {title} {Quantum
  Machines: Measurement and Control of Engineered Quantum Systems}}},\ \bibinfo
  {edition} {first edition}\ ed.\ (\bibinfo  {publisher} {{Oxford University
  Press}},\ \bibinfo {address} {{Oxford, United Kingdom}},\ \bibinfo {year}
  {2014})\BibitemShut {NoStop}%
\bibitem [{\citenamefont {Liu}\ \emph {et~al.}(2021)\citenamefont {Liu},
  \citenamefont {Jung},\ and\ \citenamefont
  {Segal}}]{liuPeriodicallyDrivenQuantum2021}%
  \BibitemOpen
  \bibfield  {author} {\bibinfo {author} {\bibfnamefont {J.}~\bibnamefont
  {Liu}}, \bibinfo {author} {\bibfnamefont {K.~A.}\ \bibnamefont {Jung}},\ and\
  \bibinfo {author} {\bibfnamefont {D.}~\bibnamefont {Segal}},\ }\bibfield
  {title} {\bibinfo {title} {Periodically {{Driven Quantum Thermal Machines}}
  from {{Warming}} up to {{Limit Cycle}}},\ }\href
  {https://doi.org/10.1103/PhysRevLett.127.200602} {\bibfield  {journal}
  {\bibinfo  {journal} {Physical Review Letters}\ }\textbf {\bibinfo {volume}
  {127}},\ \bibinfo {pages} {200602} (\bibinfo {year} {2021})}\BibitemShut
  {NoStop}%
\bibitem [{\citenamefont {Campisi}\ \emph {et~al.}(2011)\citenamefont
  {Campisi}, \citenamefont {H{\"a}nggi},\ and\ \citenamefont
  {Talkner}}]{campisiColloquiumQuantumFluctuation2011}%
  \BibitemOpen
  \bibfield  {author} {\bibinfo {author} {\bibfnamefont {M.}~\bibnamefont
  {Campisi}}, \bibinfo {author} {\bibfnamefont {P.}~\bibnamefont
  {H{\"a}nggi}},\ and\ \bibinfo {author} {\bibfnamefont {P.}~\bibnamefont
  {Talkner}},\ }\bibfield  {title} {\bibinfo {title} {{\emph{Colloquium}} :
  {{Quantum}} fluctuation relations: {{Foundations}} and applications},\ }\href
  {https://doi.org/10.1103/RevModPhys.83.771} {\bibfield  {journal} {\bibinfo
  {journal} {Reviews of Modern Physics}\ }\textbf {\bibinfo {volume} {83}},\
  \bibinfo {pages} {771} (\bibinfo {year} {2011})}\BibitemShut {NoStop}%
\bibitem [{\citenamefont
  {Jarzynski}(2011)}]{jarzynskiEqualitiesInequalitiesIrreversibility2011}%
  \BibitemOpen
  \bibfield  {author} {\bibinfo {author} {\bibfnamefont {C.}~\bibnamefont
  {Jarzynski}},\ }\bibfield  {title} {\bibinfo {title} {Equalities and
  {{Inequalities}}: {{Irreversibility}} and the {{Second Law}} of
  {{Thermodynamics}} at the {{Nanoscale}}},\ }\href
  {https://doi.org/10.1146/annurev-conmatphys-062910-140506} {\bibfield
  {journal} {\bibinfo  {journal} {Annual Review of Condensed Matter Physics}\
  }\textbf {\bibinfo {volume} {2}},\ \bibinfo {pages} {329} (\bibinfo {year}
  {2011})}\BibitemShut {NoStop}%
\bibitem [{\citenamefont
  {Seifert}(2012)}]{seifertStochasticThermodynamicsFluctuation2012}%
  \BibitemOpen
  \bibfield  {author} {\bibinfo {author} {\bibfnamefont {U.}~\bibnamefont
  {Seifert}},\ }\bibfield  {title} {\bibinfo {title} {Stochastic
  thermodynamics, fluctuation theorems, and molecular machines},\ }\href
  {https://doi.org/10.1088/0034-4885/75/12/126001} {\bibfield  {journal}
  {\bibinfo  {journal} {Reports on Progress in Physics}\ }\textbf {\bibinfo
  {volume} {75}},\ \bibinfo {pages} {126001} (\bibinfo {year} {2012})},\
  \Eprint {https://arxiv.org/abs/1205.4176} {arxiv:1205.4176} \BibitemShut
  {NoStop}%
\bibitem [{\citenamefont {Deffner}\ and\ \citenamefont
  {Campbell}(2019)}]{10.1088/2053-2571/ab21c6}%
  \BibitemOpen
  \bibfield  {author} {\bibinfo {author} {\bibfnamefont {S.}~\bibnamefont
  {Deffner}}\ and\ \bibinfo {author} {\bibfnamefont {S.}~\bibnamefont
  {Campbell}},\ }\href {https://doi.org/10.1088/2053-2571/ab21c6} {\emph
  {\bibinfo {title} {Quantum Thermodynamics}}},\ 2053-2571\ (\bibinfo
  {publisher} {{Morgan \& Claypool Publishers}},\ \bibinfo {year}
  {2019})\BibitemShut {NoStop}%
\bibitem [{\citenamefont {Talkner}\ and\ \citenamefont
  {H{\"a}nggi}(2020)}]{talknerColloquiumStatisticalMechanics2020}%
  \BibitemOpen
  \bibfield  {author} {\bibinfo {author} {\bibfnamefont {P.}~\bibnamefont
  {Talkner}}\ and\ \bibinfo {author} {\bibfnamefont {P.}~\bibnamefont
  {H{\"a}nggi}},\ }\bibfield  {title} {\bibinfo {title} {{\emph{Colloquium}} :
  {{Statistical}} mechanics and thermodynamics at strong coupling: {{Quantum}}
  and classical},\ }\href {https://doi.org/10.1103/RevModPhys.92.041002}
  {\bibfield  {journal} {\bibinfo  {journal} {Reviews of Modern Physics}\
  }\textbf {\bibinfo {volume} {92}},\ \bibinfo {pages} {041002} (\bibinfo
  {year} {2020})}\BibitemShut {NoStop}%
\bibitem [{\citenamefont {Ludovico}\ \emph {et~al.}(2014)\citenamefont
  {Ludovico}, \citenamefont {Lim}, \citenamefont {Moskalets}, \citenamefont
  {Arrachea},\ and\ \citenamefont
  {S{\'a}nchez}}]{ludovicoDynamicalEnergyTransfer2014}%
  \BibitemOpen
  \bibfield  {author} {\bibinfo {author} {\bibfnamefont {M.~F.}\ \bibnamefont
  {Ludovico}}, \bibinfo {author} {\bibfnamefont {J.~S.}\ \bibnamefont {Lim}},
  \bibinfo {author} {\bibfnamefont {M.}~\bibnamefont {Moskalets}}, \bibinfo
  {author} {\bibfnamefont {L.}~\bibnamefont {Arrachea}},\ and\ \bibinfo
  {author} {\bibfnamefont {D.}~\bibnamefont {S{\'a}nchez}},\ }\bibfield
  {title} {\bibinfo {title} {Dynamical energy transfer in ac-driven quantum
  systems},\ }\href {https://doi.org/10.1103/PhysRevB.89.161306} {\bibfield
  {journal} {\bibinfo  {journal} {Physical Review B}\ }\textbf {\bibinfo
  {volume} {89}},\ \bibinfo {pages} {161306} (\bibinfo {year}
  {2014})}\BibitemShut {NoStop}%
\bibitem [{\citenamefont {Bruch}\ \emph {et~al.}(2016)\citenamefont {Bruch},
  \citenamefont {Thomas}, \citenamefont {Viola~Kusminskiy}, \citenamefont {{von
  Oppen}},\ and\ \citenamefont
  {Nitzan}}]{bruchQuantumThermodynamicsDriven2016}%
  \BibitemOpen
  \bibfield  {author} {\bibinfo {author} {\bibfnamefont {A.}~\bibnamefont
  {Bruch}}, \bibinfo {author} {\bibfnamefont {M.}~\bibnamefont {Thomas}},
  \bibinfo {author} {\bibfnamefont {S.}~\bibnamefont {Viola~Kusminskiy}},
  \bibinfo {author} {\bibfnamefont {F.}~\bibnamefont {{von Oppen}}},\ and\
  \bibinfo {author} {\bibfnamefont {A.}~\bibnamefont {Nitzan}},\ }\bibfield
  {title} {\bibinfo {title} {Quantum thermodynamics of the driven resonant
  level model},\ }\href {https://doi.org/10.1103/PhysRevB.93.115318} {\bibfield
   {journal} {\bibinfo  {journal} {Physical Review B}\ }\textbf {\bibinfo
  {volume} {93}},\ \bibinfo {pages} {115318} (\bibinfo {year}
  {2016})}\BibitemShut {NoStop}%
\bibitem [{\citenamefont {Esposito}\ \emph {et~al.}(2010)\citenamefont
  {Esposito}, \citenamefont {Lindenberg},\ and\ \citenamefont {den
  Broeck}}]{espositoEntropyProductionCorrelation2010}%
  \BibitemOpen
  \bibfield  {author} {\bibinfo {author} {\bibfnamefont {M.}~\bibnamefont
  {Esposito}}, \bibinfo {author} {\bibfnamefont {K.}~\bibnamefont
  {Lindenberg}},\ and\ \bibinfo {author} {\bibfnamefont {C.~V.}\ \bibnamefont
  {den Broeck}},\ }\bibfield  {title} {\bibinfo {title} {Entropy production as
  correlation between system and reservoir},\ }\href
  {https://doi.org/10.1088/1367-2630/12/1/013013} {\bibfield  {journal}
  {\bibinfo  {journal} {New Journal of Physics}\ }\textbf {\bibinfo {volume}
  {12}},\ \bibinfo {pages} {013013} (\bibinfo {year} {2010})},\ \bibinfo {note}
  {comment: 1 figure, 4 pages},\ \Eprint {https://arxiv.org/abs/0908.1125}
  {arxiv:0908.1125} \BibitemShut {NoStop}%
\bibitem [{\citenamefont {Strasberg}\ and\ \citenamefont
  {Winter}(2021)}]{strasbergFirstSecondLaw2021a}%
  \BibitemOpen
  \bibfield  {author} {\bibinfo {author} {\bibfnamefont {P.}~\bibnamefont
  {Strasberg}}\ and\ \bibinfo {author} {\bibfnamefont {A.}~\bibnamefont
  {Winter}},\ }\bibfield  {title} {\bibinfo {title} {First and {{Second Law}}
  of {{Quantum Thermodynamics}}: {{A Consistent Derivation Based}} on a
  {{Microscopic Definition}} of {{Entropy}}},\ }\href
  {https://doi.org/10.1103/PRXQuantum.2.030202} {\bibfield  {journal} {\bibinfo
   {journal} {PRX Quantum}\ }\textbf {\bibinfo {volume} {2}},\ \bibinfo {pages}
  {030202} (\bibinfo {year} {2021})}\BibitemShut {NoStop}%
\bibitem [{\citenamefont {Lacerda}\ \emph {et~al.}(2023)\citenamefont
  {Lacerda}, \citenamefont {Purkayastha}, \citenamefont {Kewming},
  \citenamefont {Landi},\ and\ \citenamefont
  {Goold}}]{lacerdaQuantumThermodynamicsFast2023}%
  \BibitemOpen
  \bibfield  {author} {\bibinfo {author} {\bibfnamefont {A.~M.}\ \bibnamefont
  {Lacerda}}, \bibinfo {author} {\bibfnamefont {A.}~\bibnamefont
  {Purkayastha}}, \bibinfo {author} {\bibfnamefont {M.}~\bibnamefont
  {Kewming}}, \bibinfo {author} {\bibfnamefont {G.~T.}\ \bibnamefont {Landi}},\
  and\ \bibinfo {author} {\bibfnamefont {J.}~\bibnamefont {Goold}},\ }\bibfield
   {title} {\bibinfo {title} {Quantum thermodynamics with fast driving and
  strong coupling via the mesoscopic leads approach},\ }\href
  {https://doi.org/10.1103/PhysRevB.107.195117} {\bibfield  {journal} {\bibinfo
   {journal} {Physical Review B}\ }\textbf {\bibinfo {volume} {107}},\ \bibinfo
  {pages} {195117} (\bibinfo {year} {2023})}\BibitemShut {NoStop}%
\bibitem [{\citenamefont {Bergmann}\ and\ \citenamefont
  {Galperin}(2021)}]{bergmannGreenfunctionPerspective2021}%
  \BibitemOpen
  \bibfield  {author} {\bibinfo {author} {\bibfnamefont {N.}~\bibnamefont
  {Bergmann}}\ and\ \bibinfo {author} {\bibfnamefont {M.}~\bibnamefont
  {Galperin}},\ }\bibfield  {title} {\bibinfo {title} {A {{Green}}'s function
  perspective on the nonequilibrium thermodynamics of open quantum systems
  strongly coupled to baths: {{Nonequilibrium}} quantum thermodynamics},\
  }\href {https://doi.org/10.1140/epjs/s11734-021-00067-3} {\bibfield
  {journal} {\bibinfo  {journal} {The European Physical Journal Special
  Topics}\ }\textbf {\bibinfo {volume} {230}},\ \bibinfo {pages} {859}
  (\bibinfo {year} {2021})}\BibitemShut {NoStop}%
\bibitem [{\citenamefont {Ochoa}\ \emph {et~al.}(2016)\citenamefont {Ochoa},
  \citenamefont {Bruch},\ and\ \citenamefont
  {Nitzan}}]{ochoaEnergyDistributionLocal2016}%
  \BibitemOpen
  \bibfield  {author} {\bibinfo {author} {\bibfnamefont {M.~A.}\ \bibnamefont
  {Ochoa}}, \bibinfo {author} {\bibfnamefont {A.}~\bibnamefont {Bruch}},\ and\
  \bibinfo {author} {\bibfnamefont {A.}~\bibnamefont {Nitzan}},\ }\bibfield
  {title} {\bibinfo {title} {Energy distribution and local fluctuations in
  strongly coupled open quantum systems: {{The}} extended resonant level
  model},\ }\href {https://doi.org/10.1103/PhysRevB.94.035420} {\bibfield
  {journal} {\bibinfo  {journal} {Physical Review B}\ }\textbf {\bibinfo
  {volume} {94}},\ \bibinfo {pages} {035420} (\bibinfo {year}
  {2016})}\BibitemShut {NoStop}%
\bibitem [{\citenamefont {Haug}\ and\ \citenamefont
  {Jauho}(2007)}]{haugQuantumKineticsTransport2007}%
  \BibitemOpen
  \bibfield  {author} {\bibinfo {author} {\bibfnamefont {H.}~\bibnamefont
  {Haug}}\ and\ \bibinfo {author} {\bibfnamefont {A.-P.}\ \bibnamefont
  {Jauho}},\ }\href@noop {} {\emph {\bibinfo {title} {Quantum {{Kinetics}} in
  {{Transport}} and {{Optics}} of {{Semiconductors}}}}},\ \bibinfo {edition}
  {2nd}\ ed.\ (\bibinfo  {publisher} {{Springer}},\ \bibinfo {address} {{Berlin
  ; New York}},\ \bibinfo {year} {2007})\BibitemShut {NoStop}%
\bibitem [{\citenamefont {Stefanucci}\ and\ \citenamefont {{van
  Leeuwen}}(2013)}]{stefanucciNonequilibriumManyBodyTheory2013}%
  \BibitemOpen
  \bibfield  {author} {\bibinfo {author} {\bibfnamefont {G.}~\bibnamefont
  {Stefanucci}}\ and\ \bibinfo {author} {\bibfnamefont {R.}~\bibnamefont {{van
  Leeuwen}}},\ }\href {https://doi.org/10.1017/CBO9781139023979} {\emph
  {\bibinfo {title} {Nonequilibrium {{Many-Body Theory}} of {{Quantum
  Systems}}: {{A Modern Introduction}}}}}\ (\bibinfo  {publisher} {{Cambridge
  University Press}},\ \bibinfo {address} {{Cambridge}},\ \bibinfo {year}
  {2013})\BibitemShut {NoStop}%
\bibitem [{\citenamefont {Dittrich}\ \emph {et~al.}(1998)\citenamefont
  {Dittrich}, \citenamefont {P.}, \citenamefont {Ingold}, \citenamefont
  {Kramer}, \citenamefont {G.},\ and\ \citenamefont
  {Zwerger}}]{dittrichQuantumTransportDissipation1998a}%
  \BibitemOpen
  \bibfield  {author} {\bibinfo {author} {\bibfnamefont {T.}~\bibnamefont
  {Dittrich}}, \bibinfo {author} {\bibfnamefont {H.}~\bibnamefont {P.}},
  \bibinfo {author} {\bibfnamefont {G.-L.}\ \bibnamefont {Ingold}}, \bibinfo
  {author} {\bibfnamefont {B.}~\bibnamefont {Kramer}}, \bibinfo {author}
  {\bibfnamefont {S.}~\bibnamefont {G.}},\ and\ \bibinfo {author}
  {\bibfnamefont {W.}~\bibnamefont {Zwerger}},\ }\href@noop {} {\emph {\bibinfo
  {title} {Quantum {{Transport}} and {{Dissipation}}}}},\ \bibinfo {edition}
  {hardcover}\ ed.\ (\bibinfo  {publisher} {{Wiley-VCH}},\ \bibinfo {year}
  {1998})\BibitemShut {NoStop}%
\bibitem [{\citenamefont
  {B{\"u}ttiker}(2000)}]{buttikerTimeDependentTransportMesoscopic2000}%
  \BibitemOpen
  \bibfield  {author} {\bibinfo {author} {\bibfnamefont {M.}~\bibnamefont
  {B{\"u}ttiker}},\ }\bibfield  {title} {\bibinfo {title} {Time-{{Dependent
  Transport}} in {{Mesoscopic Structures}}},\ }\href
  {https://doi.org/10.1023/A:1004622924099} {\bibfield  {journal} {\bibinfo
  {journal} {Journal of Low Temperature Physics}\ }\textbf {\bibinfo {volume}
  {118}},\ \bibinfo {pages} {519} (\bibinfo {year} {2000})}\BibitemShut
  {NoStop}%
\bibitem [{\citenamefont {Shastry}\ \emph {et~al.}(2019)\citenamefont
  {Shastry}, \citenamefont {Xu},\ and\ \citenamefont
  {Stafford}}]{shastryThirdLawThermodynamics2019}%
  \BibitemOpen
  \bibfield  {author} {\bibinfo {author} {\bibfnamefont {A.}~\bibnamefont
  {Shastry}}, \bibinfo {author} {\bibfnamefont {Y.}~\bibnamefont {Xu}},\ and\
  \bibinfo {author} {\bibfnamefont {C.~A.}\ \bibnamefont {Stafford}},\
  }\bibfield  {title} {\bibinfo {title} {The third law of thermodynamics in
  open quantum systems},\ }\href {https://doi.org/10.1063/1.5100182} {\bibfield
   {journal} {\bibinfo  {journal} {The Journal of Chemical Physics}\ }\textbf
  {\bibinfo {volume} {151}},\ \bibinfo {pages} {064115} (\bibinfo {year}
  {2019})}\BibitemShut {NoStop}%
\bibitem [{\citenamefont
  {Shastry}(2019)}]{shastryTheoryThermodynamicMeasurements2019}%
  \BibitemOpen
  \bibfield  {author} {\bibinfo {author} {\bibfnamefont {A.}~\bibnamefont
  {Shastry}},\ }\href {https://doi.org/10.1007/978-3-030-33574-8} {\emph
  {\bibinfo {title} {Theory of {{Thermodynamic Measurements}} of {{Quantum
  Systems Far}} from {{Equilibrium}}}}},\ Springer {{Theses}}\ (\bibinfo
  {publisher} {{Springer International Publishing}},\ \bibinfo {address}
  {{Cham}},\ \bibinfo {year} {2019})\BibitemShut {NoStop}%
\bibitem [{\citenamefont {Webb}\ and\ \citenamefont
  {Stafford}(2023)}]{webb2023}%
  \BibitemOpen
  \bibfield  {author} {\bibinfo {author} {\bibfnamefont {C.~M.}\ \bibnamefont
  {Webb}}\ and\ \bibinfo {author} {\bibfnamefont {C.~A.}\ \bibnamefont
  {Stafford}},\ }\href@noop {} {\bibinfo {title} {How to partition a quantum
  observable}},\ \bibinfo {howpublished} {unpublished} (\bibinfo {year}
  {2023})\BibitemShut {NoStop}%
\bibitem [{Note1()}]{Note1}%
  \BibitemOpen
  \bibinfo {note} {\label {NDOfootnote} This form of the electronic Hamiltonian
  is often referred to as the \protect \textit {Complete Neglect of
  Differential Overlap} approximation and can also be formulated in terms of
  Wannier orbitals \cite {PhysRevB.86.115403}.}\BibitemShut {Stop}%
\bibitem [{Note2()}]{Note2}%
  \BibitemOpen
  \bibinfo {note} {\label {ph_ortho_transf_footnote} Details of the orthogonal
  transformation used in Eq.\ \protect \textup {\hbox {\mathsurround \z@
  \protect \normalfont (\ignorespaces \ref {sys_ph_harm_hamiltonian_eqn}\unskip
  \@@italiccorr )}}: \begin {equation} Q_i=\DOTSB \sum@ \slimits@
  _{r}C_{ir}Q_r\protect \,, \end {equation} where from orthogonality we have
  $\DOTSB \sum@ \slimits@ _{r}C^{\protect \mathrm {T}}_{ir}C_{rj}=\DOTSB \sum@
  \slimits@ _{r}C_{ri}C_{rj}=\delta _{ij}$, and \begin {equation} P_i=\DOTSB
  \sum@ \slimits@ _{r}C_{ir}P_r \protect \,, \end {equation}and we can write
  \begin {equation} \DOTSB \sum@ \slimits@ _{ij} C_{ir}K_{ij}C_{sj}={\omega
  }_r^2 \delta _{rs} \protect \,, \end {equation} and \begin {equation}
  F_{i}(t)=\DOTSB \sum@ \slimits@ _r C_{ir}F_r(t)\protect \,, \end {equation}
  where $F_r$ is the force acting on the $r^{th}$ oscillator normal
  mode.}\BibitemShut {Stop}%
\bibitem [{\citenamefont
  {Arnold}(1987)}]{arnoldSuperconductingTunnelingTunneling1987}%
  \BibitemOpen
  \bibfield  {author} {\bibinfo {author} {\bibfnamefont {G.~B.}\ \bibnamefont
  {Arnold}},\ }\bibfield  {title} {\bibinfo {title} {Superconducting tunneling
  without the tunneling {{Hamiltonian}}. {{II}}. {{Subgap}} harmonic
  structure},\ }\href {https://doi.org/10.1007/BF00682620} {\bibfield
  {journal} {\bibinfo  {journal} {Journal of Low Temperature Physics}\ }\textbf
  {\bibinfo {volume} {68}},\ \bibinfo {pages} {1} (\bibinfo {year}
  {1987})}\BibitemShut {NoStop}%
\bibitem [{\citenamefont {Averin}\ and\ \citenamefont
  {Bardas}(1995)}]{averinAcJosephsonEffect1995}%
  \BibitemOpen
  \bibfield  {author} {\bibinfo {author} {\bibfnamefont {D.}~\bibnamefont
  {Averin}}\ and\ \bibinfo {author} {\bibfnamefont {A.}~\bibnamefont
  {Bardas}},\ }\bibfield  {title} {\bibinfo {title} {Ac {{Josephson Effect}} in
  a {{Single Quantum Channel}}},\ }\href
  {https://doi.org/10.1103/PhysRevLett.75.1831} {\bibfield  {journal} {\bibinfo
   {journal} {Physical Review Letters}\ }\textbf {\bibinfo {volume} {75}},\
  \bibinfo {pages} {1831} (\bibinfo {year} {1995})}\BibitemShut {NoStop}%
\bibitem [{\citenamefont {Cuevas}\ \emph {et~al.}(1996)\citenamefont {Cuevas},
  \citenamefont {{Mart{\'i}n-Rodero}},\ and\ \citenamefont
  {Yeyati}}]{cuevasHamiltonianApproachTransport1996}%
  \BibitemOpen
  \bibfield  {author} {\bibinfo {author} {\bibfnamefont {J.~C.}\ \bibnamefont
  {Cuevas}}, \bibinfo {author} {\bibfnamefont {A.}~\bibnamefont
  {{Mart{\'i}n-Rodero}}},\ and\ \bibinfo {author} {\bibfnamefont {A.~L.}\
  \bibnamefont {Yeyati}},\ }\bibfield  {title} {\bibinfo {title} {Hamiltonian
  approach to the transport properties of superconducting quantum point
  contacts},\ }\href {https://doi.org/10.1103/PhysRevB.54.7366} {\bibfield
  {journal} {\bibinfo  {journal} {Physical Review B}\ }\textbf {\bibinfo
  {volume} {54}},\ \bibinfo {pages} {7366} (\bibinfo {year}
  {1996})}\BibitemShut {NoStop}%
\bibitem [{\citenamefont {Scheer}\ \emph {et~al.}(1997)\citenamefont {Scheer},
  \citenamefont {Joyez}, \citenamefont {Esteve}, \citenamefont {Urbina},\ and\
  \citenamefont {Devoret}}]{scheerConductionChannelTransmissions1997}%
  \BibitemOpen
  \bibfield  {author} {\bibinfo {author} {\bibfnamefont {E.}~\bibnamefont
  {Scheer}}, \bibinfo {author} {\bibfnamefont {P.}~\bibnamefont {Joyez}},
  \bibinfo {author} {\bibfnamefont {D.}~\bibnamefont {Esteve}}, \bibinfo
  {author} {\bibfnamefont {C.}~\bibnamefont {Urbina}},\ and\ \bibinfo {author}
  {\bibfnamefont {M.~H.}\ \bibnamefont {Devoret}},\ }\bibfield  {title}
  {\bibinfo {title} {Conduction {{Channel Transmissions}} of {{Atomic-Size
  Aluminum Contacts}}},\ }\href {https://doi.org/10.1103/PhysRevLett.78.3535}
  {\bibfield  {journal} {\bibinfo  {journal} {Physical Review Letters}\
  }\textbf {\bibinfo {volume} {78}},\ \bibinfo {pages} {3535} (\bibinfo {year}
  {1997})}\BibitemShut {NoStop}%
\bibitem [{\citenamefont {Barr}\ \emph {et~al.}(2012)\citenamefont {Barr},
  \citenamefont {Stafford},\ and\ \citenamefont
  {Bergfield}}]{PhysRevB.86.115403}%
  \BibitemOpen
  \bibfield  {author} {\bibinfo {author} {\bibfnamefont {J.~D.}\ \bibnamefont
  {Barr}}, \bibinfo {author} {\bibfnamefont {C.~A.}\ \bibnamefont {Stafford}},\
  and\ \bibinfo {author} {\bibfnamefont {J.~P.}\ \bibnamefont {Bergfield}},\
  }\bibfield  {title} {\bibinfo {title} {Effective field theory of interacting
  {{$\pi$}} electrons},\ }\href {https://doi.org/10.1103/PhysRevB.86.115403}
  {\bibfield  {journal} {\bibinfo  {journal} {Physical Review B}\ }\textbf
  {\bibinfo {volume} {86}},\ \bibinfo {pages} {115403} (\bibinfo {year}
  {2012})}\BibitemShut {NoStop}%
\bibitem [{\citenamefont {Bergfield}\ \emph {et~al.}(2011)\citenamefont
  {Bergfield}, \citenamefont {Barr},\ and\ \citenamefont
  {Stafford}}]{bergfieldNumberTransmissionChannels2011a}%
  \BibitemOpen
  \bibfield  {author} {\bibinfo {author} {\bibfnamefont {J.~P.}\ \bibnamefont
  {Bergfield}}, \bibinfo {author} {\bibfnamefont {J.~D.}\ \bibnamefont
  {Barr}},\ and\ \bibinfo {author} {\bibfnamefont {C.~A.}\ \bibnamefont
  {Stafford}},\ }\bibfield  {title} {\bibinfo {title} {The {{Number}} of
  {{Transmission Channels Through}} a {{Single-Molecule Junction}}},\ }\href
  {https://doi.org/10.1021/nn1030753} {\bibfield  {journal} {\bibinfo
  {journal} {ACS Nano}\ }\textbf {\bibinfo {volume} {5}},\ \bibinfo {pages}
  {2707} (\bibinfo {year} {2011})},\ \bibinfo {note} {doi:
  10.1021/nn1030753}\BibitemShut {NoStop}%
\bibitem [{\citenamefont {Bergfield}\ \emph {et~al.}(2012)\citenamefont
  {Bergfield}, \citenamefont {Barr},\ and\ \citenamefont
  {Stafford}}]{bergfieldTransmissionEigenvalueDistributions2012a}%
  \BibitemOpen
  \bibfield  {author} {\bibinfo {author} {\bibfnamefont {J.~P.}\ \bibnamefont
  {Bergfield}}, \bibinfo {author} {\bibfnamefont {J.~D.}\ \bibnamefont
  {Barr}},\ and\ \bibinfo {author} {\bibfnamefont {C.~A.}\ \bibnamefont
  {Stafford}},\ }\bibfield  {title} {\bibinfo {title} {Transmission eigenvalue
  distributions in highly conductive molecular junctions.},\ }\href
  {https://doi.org/10.3762/bjnano.3.5} {\bibfield  {journal} {\bibinfo
  {journal} {Beilstein journal of nanotechnology}\ }\textbf {\bibinfo {volume}
  {3}},\ \bibinfo {pages} {40} (\bibinfo {year} {2012})}\BibitemShut {NoStop}%
\bibitem [{Note3()}]{Note3}%
  \BibitemOpen
  \bibinfo {note} {The clarification about the spatial extent of the coupling
  is of direct relevance to the central result of this work and one that needs
  highlighting in view of the claim that inclusion of Reservoir degrees of
  freedom in the System Internal Energy is ``problematic from an operational
  perspective'' \cite
  {strasbergQuantumStochasticThermodynamics2022_ch3}.}\BibitemShut {Stop}%
\bibitem [{\citenamefont
  {Strasberg}(2022)}]{strasbergQuantumStochasticThermodynamics2022_ch3}%
  \BibitemOpen
  \bibfield  {author} {\bibinfo {author} {\bibfnamefont {P.}~\bibnamefont
  {Strasberg}},\ }\bibfield  {title} {\bibinfo {title} {Chapter 3 - {{Quantum
  Thermodynamics Without Measurements}}},\ }in\ \href@noop {} {\emph {\bibinfo
  {booktitle} {Quantum {{Stochastic Thermodynamics}}: {{Foundations}} and
  {{Selected Applications}}}}},\ \bibinfo {series and number} {Oxford
  {{Graduate Texts}}}\ (\bibinfo  {publisher} {{Oxford University Press}},\
  \bibinfo {address} {{Oxford, New York}},\ \bibinfo {year} {2022})\BibitemShut
  {NoStop}%
\bibitem [{\citenamefont {Ashcroft}\ and\ \citenamefont
  {Mermin}(1976)}]{ashcroft_solid_1976_ch20}%
  \BibitemOpen
  \bibfield  {author} {\bibinfo {author} {\bibfnamefont {N.~W.}\ \bibnamefont
  {Ashcroft}}\ and\ \bibinfo {author} {\bibfnamefont {N.~D.}\ \bibnamefont
  {Mermin}},\ }\bibfield  {title} {\bibinfo {title} {Chapter 20 - {{Cohesive
  Energy}}},\ }in\ \href@noop {} {\emph {\bibinfo {booktitle} {Solid State
  Physics}}}\ (\bibinfo  {publisher} {{Holt, Rinehart and Winston}},\ \bibinfo
  {address} {{New York}},\ \bibinfo {year} {1976})\ pp.\ \bibinfo {pages}
  {395--414}\BibitemShut {NoStop}%
\bibitem [{\citenamefont
  {Jarzynski}(2007)}]{jarzynskiComparisonFarfromequilibriumWork2007}%
  \BibitemOpen
  \bibfield  {author} {\bibinfo {author} {\bibfnamefont {C.}~\bibnamefont
  {Jarzynski}},\ }\bibfield  {title} {\bibinfo {title} {Comparison of
  far-from-equilibrium work relations},\ }\href
  {https://doi.org/10.1016/j.crhy.2007.04.010} {\bibfield  {journal} {\bibinfo
  {journal} {Comptes Rendus Physique}\ }\textbf {\bibinfo {volume} {8}},\
  \bibinfo {pages} {495} (\bibinfo {year} {2007})}\BibitemShut {NoStop}%
\bibitem [{\citenamefont {Bochkov}\ and\ \citenamefont
  {Kuzovlev}(1977)}]{1977ZhETF..72..238B}%
  \BibitemOpen
  \bibfield  {author} {\bibinfo {author} {\bibfnamefont {G.~N.}\ \bibnamefont
  {Bochkov}}\ and\ \bibinfo {author} {\bibfnamefont {{\relax Iu}.~E.}\
  \bibnamefont {Kuzovlev}},\ }\bibfield  {title} {\bibinfo {title} {General
  theory of thermal fluctuations in nonlinear systems},\ }\href@noop {}
  {\bibfield  {journal} {\bibinfo  {journal} {Zhurnal Eksperimentalnoi i
  Teoreticheskoi Fiziki}\ }\textbf {\bibinfo {volume} {72}},\ \bibinfo {pages}
  {238} (\bibinfo {year} {1977})}\BibitemShut {NoStop}%
\bibitem [{\citenamefont
  {Ballentine}(2014)}]{ballentineChapterKinematicsDynamics2014}%
  \BibitemOpen
  \bibfield  {author} {\bibinfo {author} {\bibfnamefont {L.~E.}\ \bibnamefont
  {Ballentine}},\ }\bibfield  {title} {\bibinfo {title} {Chapter 3 -
  {{Kinematics}} and {{Dynamics}}},\ }in\ \href@noop {} {\emph {\bibinfo
  {booktitle} {Quantum {{Mechanics}}: {{A Modern Development}}}}}\ (\bibinfo
  {publisher} {{World Scientific Publishing}},\ \bibinfo {year} {2014})\
  \bibinfo {edition} {2nd}\ ed.\BibitemShut {Stop}%
\bibitem [{\citenamefont {Landau}\ and\ \citenamefont
  {Lifshitz}(1980)}]{LANDAU198034}%
  \BibitemOpen
  \bibfield  {author} {\bibinfo {author} {\bibfnamefont {L.}~\bibnamefont
  {Landau}}\ and\ \bibinfo {author} {\bibfnamefont {E.}~\bibnamefont
  {Lifshitz}},\ }\bibfield  {title} {\bibinfo {title} {{{CHAPTER II}} -
  {{THERMODYNAMIC QUANTITIES}}},\ }in\ \href
  {https://doi.org/10.1016/B978-0-08-057046-4.50009-9} {\emph {\bibinfo
  {booktitle} {Statistical Physics (Third Edition)}}}\ (\bibinfo  {publisher}
  {{Butterworth-Heinemann}},\ \bibinfo {address} {{Oxford}},\ \bibinfo {year}
  {1980})\ \bibinfo {edition} {third edition}\ ed.,\ pp.\ \bibinfo {pages}
  {34--78}\BibitemShut {NoStop}%
\bibitem [{\citenamefont {Callen}(1985)}]{Callen1985}%
  \BibitemOpen
  \bibfield  {author} {\bibinfo {author} {\bibfnamefont {H.~B.}\ \bibnamefont
  {Callen}},\ }\bibfield  {title} {\bibinfo {title} {Thermodynamics and an
  introduction to thermostatistics; 2nd ed.}\ }(\bibinfo  {publisher}
  {{Wiley}},\ \bibinfo {address} {{New York, NY}},\ \bibinfo {year} {1985})\
  pp.\ \bibinfo {pages} {18--37}\BibitemShut {NoStop}%
\bibitem [{\citenamefont
  {B{\"u}ttiker}(1986)}]{buttikerRoleQuantumCoherence1986}%
  \BibitemOpen
  \bibfield  {author} {\bibinfo {author} {\bibfnamefont {M.}~\bibnamefont
  {B{\"u}ttiker}},\ }\bibfield  {title} {\bibinfo {title} {Role of quantum
  coherence in series resistors},\ }\href
  {https://doi.org/10.1103/PhysRevB.33.3020} {\bibfield  {journal} {\bibinfo
  {journal} {Physical Review B}\ }\textbf {\bibinfo {volume} {33}},\ \bibinfo
  {pages} {3020} (\bibinfo {year} {1986})}\BibitemShut {NoStop}%
\bibitem [{\citenamefont {Baranger}\ and\ \citenamefont
  {Stone}(1989)}]{barangerElectricalLinearresponseTheory1989}%
  \BibitemOpen
  \bibfield  {author} {\bibinfo {author} {\bibfnamefont {H.~U.}\ \bibnamefont
  {Baranger}}\ and\ \bibinfo {author} {\bibfnamefont {A.~D.}\ \bibnamefont
  {Stone}},\ }\bibfield  {title} {\bibinfo {title} {Electrical linear-response
  theory in an arbitrary magnetic field: {{A}} new {{Fermi-surface}}
  formation},\ }\href {https://doi.org/10.1103/PhysRevB.40.8169} {\bibfield
  {journal} {\bibinfo  {journal} {Physical Review B}\ }\textbf {\bibinfo
  {volume} {40}},\ \bibinfo {pages} {8169} (\bibinfo {year}
  {1989})}\BibitemShut {NoStop}%
\bibitem [{\citenamefont {N{\"o}ckel}\ \emph {et~al.}(1993)\citenamefont
  {N{\"o}ckel}, \citenamefont {Stone},\ and\ \citenamefont
  {Baranger}}]{nockelAdiabaticTurnonAsymptotic1993}%
  \BibitemOpen
  \bibfield  {author} {\bibinfo {author} {\bibfnamefont {J.~U.}\ \bibnamefont
  {N{\"o}ckel}}, \bibinfo {author} {\bibfnamefont {A.~D.}\ \bibnamefont
  {Stone}},\ and\ \bibinfo {author} {\bibfnamefont {H.~U.}\ \bibnamefont
  {Baranger}},\ }\bibfield  {title} {\bibinfo {title} {Adiabatic turn-on and
  the asymptotic limit in linear-response theory for open systems},\ }\href
  {https://doi.org/10.1103/PhysRevB.48.17569} {\bibfield  {journal} {\bibinfo
  {journal} {Physical Review B}\ }\textbf {\bibinfo {volume} {48}},\ \bibinfo
  {pages} {17569} (\bibinfo {year} {1993})}\BibitemShut {NoStop}%
\bibitem [{\citenamefont {{von Neumann}}\ \emph {et~al.}(2018)\citenamefont
  {{von Neumann}}, \citenamefont {Wheeler},\ and\ \citenamefont
  {Beyer}}]{vonneumannMathematicalFoundationsQuantum2018}%
  \BibitemOpen
  \bibfield  {author} {\bibinfo {author} {\bibfnamefont {J.}~\bibnamefont {{von
  Neumann}}}, \bibinfo {author} {\bibfnamefont {N.~A.}\ \bibnamefont
  {Wheeler}},\ and\ \bibinfo {author} {\bibfnamefont {R.~T.}\ \bibnamefont
  {Beyer}},\ }\href {https://doi.org/10.1515/9781400889921} {\emph {\bibinfo
  {title} {Mathematical Foundations of Quantum Mechanics, New Edition}}},\
  \bibinfo {edition} {ned - new edition}\ ed.\ (\bibinfo  {publisher}
  {{Princeton: Princeton University Press}},\ \bibinfo {address}
  {{Princeton}},\ \bibinfo {year} {2018})\BibitemShut {NoStop}%
\bibitem [{\citenamefont {Esposito}\ \emph {et~al.}(2015)\citenamefont
  {Esposito}, \citenamefont {Ochoa},\ and\ \citenamefont
  {Galperin}}]{espositoNatureHeatStrongly2015}%
  \BibitemOpen
  \bibfield  {author} {\bibinfo {author} {\bibfnamefont {M.}~\bibnamefont
  {Esposito}}, \bibinfo {author} {\bibfnamefont {M.~A.}\ \bibnamefont
  {Ochoa}},\ and\ \bibinfo {author} {\bibfnamefont {M.}~\bibnamefont
  {Galperin}},\ }\bibfield  {title} {\bibinfo {title} {Nature of heat in
  strongly coupled open quantum systems},\ }\href
  {https://doi.org/10.1103/PhysRevB.92.235440} {\bibfield  {journal} {\bibinfo
  {journal} {Physical Review B}\ }\textbf {\bibinfo {volume} {92}},\ \bibinfo
  {pages} {235440} (\bibinfo {year} {2015})}\BibitemShut {NoStop}%
\bibitem [{\citenamefont {Bergfield}\ and\ \citenamefont
  {Stafford}(2009{\natexlab{a}})}]{bergfieldThermoelectricSignaturesCoherent2009}%
  \BibitemOpen
  \bibfield  {author} {\bibinfo {author} {\bibfnamefont {J.~P.}\ \bibnamefont
  {Bergfield}}\ and\ \bibinfo {author} {\bibfnamefont {C.~A.}\ \bibnamefont
  {Stafford}},\ }\href {https://doi.org/10.1021/nl901554s} {\bibinfo {title}
  {Thermoelectric {{Signatures}} of {{Coherent Transport}} in {{Single-Molecule
  Heterojunctions}}}} (\bibinfo {year} {2009}{\natexlab{a}})\BibitemShut
  {NoStop}%
\bibitem [{\citenamefont {Bergfield}\ and\ \citenamefont
  {Stafford}(2009{\natexlab{b}})}]{bergfieldThermoelectricSignaturesCoherent2009d}%
  \BibitemOpen
  \bibfield  {author} {\bibinfo {author} {\bibfnamefont {J.~P.}\ \bibnamefont
  {Bergfield}}\ and\ \bibinfo {author} {\bibfnamefont {C.~A.}\ \bibnamefont
  {Stafford}},\ }\bibfield  {title} {\bibinfo {title} {Thermoelectric
  signatures of coherent transport in single-molecule heterojunctions},\ }\href
  {https://doi.org/10.1021/nl901554s} {\bibfield  {journal} {\bibinfo
  {journal} {Nano Letters}\ }\textbf {\bibinfo {volume} {9}},\ \bibinfo {pages}
  {3072} (\bibinfo {year} {2009}{\natexlab{b}})}\BibitemShut {NoStop}%
\bibitem [{\citenamefont {Galperin}\ \emph {et~al.}(2007)\citenamefont
  {Galperin}, \citenamefont {Nitzan},\ and\ \citenamefont
  {Ratner}}]{galperinHeatConductionMolecular2007}%
  \BibitemOpen
  \bibfield  {author} {\bibinfo {author} {\bibfnamefont {M.}~\bibnamefont
  {Galperin}}, \bibinfo {author} {\bibfnamefont {A.}~\bibnamefont {Nitzan}},\
  and\ \bibinfo {author} {\bibfnamefont {M.~A.}\ \bibnamefont {Ratner}},\
  }\bibfield  {title} {\bibinfo {title} {Heat conduction in molecular transport
  junctions},\ }\href {https://doi.org/10.1103/PhysRevB.75.155312} {\bibfield
  {journal} {\bibinfo  {journal} {Physical Review B}\ }\textbf {\bibinfo
  {volume} {75}},\ \bibinfo {pages} {155312} (\bibinfo {year}
  {2007})}\BibitemShut {NoStop}%
\bibitem [{\citenamefont
  {Leek}(2012)}]{leekMathematicalDetailsApplication2012}%
  \BibitemOpen
  \bibfield  {author} {\bibinfo {author} {\bibfnamefont {M.~L.}\ \bibnamefont
  {Leek}},\ }\href {http://arxiv.org/abs/1207.6204} {\bibinfo {title}
  {Mathematical {{Details}} in the application of {{Non-equilibrium Green}}'s
  {{Functions}} ({{NEGF}}) and {{Quantum Kinetic Equations}} ({{QKE}}) to
  {{Thermal Transport}}}} (\bibinfo {year} {2012}),\ \Eprint
  {https://arxiv.org/abs/1207.6204} {arxiv:1207.6204 [cond-mat]} \BibitemShut
  {NoStop}%
\bibitem [{\citenamefont {Wang}\ \emph {et~al.}(2014)\citenamefont {Wang},
  \citenamefont {Agarwalla}, \citenamefont {Li},\ and\ \citenamefont
  {Thingna}}]{wangNonequilibriumGreenFunction2014}%
  \BibitemOpen
  \bibfield  {author} {\bibinfo {author} {\bibfnamefont {J.-S.}\ \bibnamefont
  {Wang}}, \bibinfo {author} {\bibfnamefont {B.~K.}\ \bibnamefont {Agarwalla}},
  \bibinfo {author} {\bibfnamefont {H.}~\bibnamefont {Li}},\ and\ \bibinfo
  {author} {\bibfnamefont {J.}~\bibnamefont {Thingna}},\ }\bibfield  {title}
  {\bibinfo {title} {Nonequilibrium {{Green}}'s function method for quantum
  thermal transport},\ }\href {https://doi.org/10.1007/s11467-013-0340-x}
  {\bibfield  {journal} {\bibinfo  {journal} {Frontiers of Physics}\ }\textbf
  {\bibinfo {volume} {9}},\ \bibinfo {pages} {673} (\bibinfo {year}
  {2014})}\BibitemShut {NoStop}%
\bibitem [{\citenamefont {Stafford}\ and\ \citenamefont
  {Shastry}(2017)}]{staffordLocalEntropyNonequilibrium2017}%
  \BibitemOpen
  \bibfield  {author} {\bibinfo {author} {\bibfnamefont {C.~A.}\ \bibnamefont
  {Stafford}}\ and\ \bibinfo {author} {\bibfnamefont {A.}~\bibnamefont
  {Shastry}},\ }\bibfield  {title} {\bibinfo {title} {Local entropy of a
  nonequilibrium fermion system},\ }\href {https://doi.org/10.1063/1.4975810}
  {\bibfield  {journal} {\bibinfo  {journal} {The Journal of Chemical Physics}\
  }\textbf {\bibinfo {volume} {146}},\ \bibinfo {pages} {092324} (\bibinfo
  {year} {2017})}\BibitemShut {NoStop}%
\bibitem [{Note4()}]{Note4}%
  \BibitemOpen
  \bibinfo {note} {\label {prev_ver_footnote}We note that in the previous
  version of this preprint \cite {kumarFirstLawThermodynamics2022} we presented
  arguments which lead to an agreement with the internal energy identification
  of Refs. \cite
  {espositoEntropyProductionCorrelation2010,strasbergFirstSecondLaw2021a,lacerdaQuantumThermodynamicsFast2023}
  and a disagreement with that of Refs. \cite
  {ludovicoDynamicalEnergyTransfer2014,bruchQuantumThermodynamicsDriven2016}--a
  view that has been reversed in light of the arguments presented in the
  preceding sections.}\BibitemShut {Stop}%
\bibitem [{\citenamefont {Talkner}\ \emph {et~al.}(2007)\citenamefont
  {Talkner}, \citenamefont {Lutz},\ and\ \citenamefont
  {H{\"a}nggi}}]{talknerFluctuationTheoremsWork2007}%
  \BibitemOpen
  \bibfield  {author} {\bibinfo {author} {\bibfnamefont {P.}~\bibnamefont
  {Talkner}}, \bibinfo {author} {\bibfnamefont {E.}~\bibnamefont {Lutz}},\ and\
  \bibinfo {author} {\bibfnamefont {P.}~\bibnamefont {H{\"a}nggi}},\ }\bibfield
   {title} {\bibinfo {title} {Fluctuation theorems: {{Work}} is not an
  observable},\ }\href {https://doi.org/10.1103/PhysRevE.75.050102} {\bibfield
  {journal} {\bibinfo  {journal} {Physical Review E}\ }\textbf {\bibinfo
  {volume} {75}},\ \bibinfo {pages} {050102} (\bibinfo {year}
  {2007})}\BibitemShut {NoStop}%
\bibitem [{\citenamefont {Talkner}\ and\ \citenamefont
  {H{\"a}nggi}(2016)}]{talknerAspectsQuantumWork2016}%
  \BibitemOpen
  \bibfield  {author} {\bibinfo {author} {\bibfnamefont {P.}~\bibnamefont
  {Talkner}}\ and\ \bibinfo {author} {\bibfnamefont {P.}~\bibnamefont
  {H{\"a}nggi}},\ }\bibfield  {title} {\bibinfo {title} {Aspects of quantum
  work},\ }\href {https://doi.org/10.1103/PhysRevE.93.022131} {\bibfield
  {journal} {\bibinfo  {journal} {Physical Review E}\ }\textbf {\bibinfo
  {volume} {93}},\ \bibinfo {pages} {022131} (\bibinfo {year}
  {2016})}\BibitemShut {NoStop}%
\bibitem [{\citenamefont {Stafford}\ \emph {et~al.}(2022)\citenamefont
  {Stafford}, \citenamefont {Jimenez~Valencia}, \citenamefont {Webb},\ and\
  \citenamefont {Evers}}]{2022APS_MM_poster_densityandtopology}%
  \BibitemOpen
  \bibfield  {author} {\bibinfo {author} {\bibfnamefont {C.~A.}\ \bibnamefont
  {Stafford}}, \bibinfo {author} {\bibfnamefont {M.~A.}\ \bibnamefont
  {Jimenez~Valencia}}, \bibinfo {author} {\bibfnamefont {C.~M.}\ \bibnamefont
  {Webb}},\ and\ \bibinfo {author} {\bibfnamefont {F.}~\bibnamefont {Evers}},\
  }\bibfield  {title} {\bibinfo {title} {Entropy density and flux in
  topological and nonequilibrium quantum systems},\ }in\ \href
  {https://meetings.aps.org/Meeting/MAR22/Session/K50.5} {\emph {\bibinfo
  {booktitle} {{{APS}} March Meeting Abstracts}}},\ \bibinfo {series} {{{APS}}
  Meeting Abstracts}, Vol.\ \bibinfo {volume} {2022}\ (\bibinfo {year} {2022})\
  p.\ \bibinfo {pages} {K50.00005}\BibitemShut {NoStop}%
\bibitem [{\citenamefont {Kadanoff}\ and\ \citenamefont
  {Baym}(1962)}]{kadanoffQuantumStatisticalMechanics1962}%
  \BibitemOpen
  \bibfield  {author} {\bibinfo {author} {\bibfnamefont {L.~P.}\ \bibnamefont
  {Kadanoff}}\ and\ \bibinfo {author} {\bibfnamefont {G.}~\bibnamefont
  {Baym}},\ }\href@noop {} {\emph {\bibinfo {title} {Quantum Statistical
  Mechanics}}}\ (\bibinfo  {publisher} {{New York, W.A. Benjamin}},\ \bibinfo
  {address} {{New York}},\ \bibinfo {year} {1962})\BibitemShut {NoStop}%
\bibitem [{\citenamefont
  {Keldysh}(1964)}]{keldyshDiagramTechniqueNonequilibrium1964}%
  \BibitemOpen
  \bibfield  {author} {\bibinfo {author} {\bibfnamefont {L.~V.}\ \bibnamefont
  {Keldysh}},\ }\bibfield  {title} {\bibinfo {title} {Diagram technique for
  nonequilibrium processes},\ }\href@noop {} {\bibfield  {journal} {\bibinfo
  {journal} {Zh. Eksp. Teor. Fiz.}\ }\textbf {\bibinfo {volume} {47}},\
  \bibinfo {pages} {1515} (\bibinfo {year} {1964})}\BibitemShut {NoStop}%
\bibitem [{\citenamefont {Rammer}(2007)}]{rammerQuantumFieldTheory2007}%
  \BibitemOpen
  \bibfield  {author} {\bibinfo {author} {\bibfnamefont {J.}~\bibnamefont
  {Rammer}},\ }\href {https://doi.org/10.1017/CBO9780511618956} {\emph
  {\bibinfo {title} {Quantum {{Field Theory}} of {{Non-equilibrium States}}}}}\
  (\bibinfo  {publisher} {{Cambridge: Cambridge University Press}},\ \bibinfo
  {address} {{Cambridge}},\ \bibinfo {year} {2007})\BibitemShut {NoStop}%
\bibitem [{\citenamefont {Gasparian}\ \emph {et~al.}(1996)\citenamefont
  {Gasparian}, \citenamefont {Christen},\ and\ \citenamefont
  {B{\"u}ttiker}}]{gasparianPartialDensitiesStates1996}%
  \BibitemOpen
  \bibfield  {author} {\bibinfo {author} {\bibfnamefont {V.}~\bibnamefont
  {Gasparian}}, \bibinfo {author} {\bibfnamefont {T.}~\bibnamefont
  {Christen}},\ and\ \bibinfo {author} {\bibfnamefont {M.}~\bibnamefont
  {B{\"u}ttiker}},\ }\bibfield  {title} {\bibinfo {title} {Partial densities of
  states, scattering matrices, and {{Green}}'s functions},\ }\href
  {https://doi.org/10.1103/PhysRevA.54.4022} {\bibfield  {journal} {\bibinfo
  {journal} {Physical Review A}\ }\textbf {\bibinfo {volume} {54}},\ \bibinfo
  {pages} {4022} (\bibinfo {year} {1996})}\BibitemShut {NoStop}%
\bibitem [{\citenamefont {Breuer}\ and\ \citenamefont
  {Petruccione}(2002)}]{breuerTheoryOpenQuantum2002a}%
  \BibitemOpen
  \bibfield  {author} {\bibinfo {author} {\bibfnamefont {H.-P.}\ \bibnamefont
  {Breuer}}\ and\ \bibinfo {author} {\bibfnamefont {F.}~\bibnamefont
  {Petruccione}},\ }\href@noop {} {\emph {\bibinfo {title} {The Theory of Open
  Quantum Systems}}}\ (\bibinfo  {publisher} {{Oxford University Press}},\
  \bibinfo {address} {{Oxford ; New York}},\ \bibinfo {year}
  {2002})\BibitemShut {NoStop}%
\bibitem [{\citenamefont {Wingreen}\ \emph {et~al.}(1993)\citenamefont
  {Wingreen}, \citenamefont {Jauho},\ and\ \citenamefont
  {Meir}}]{wingreenTimedependentTransportMesoscopic1993}%
  \BibitemOpen
  \bibfield  {author} {\bibinfo {author} {\bibfnamefont {N.~S.}\ \bibnamefont
  {Wingreen}}, \bibinfo {author} {\bibfnamefont {A.-P.}\ \bibnamefont
  {Jauho}},\ and\ \bibinfo {author} {\bibfnamefont {Y.}~\bibnamefont {Meir}},\
  }\bibfield  {title} {\bibinfo {title} {Time-dependent transport through a
  mesoscopic structure},\ }\href {https://doi.org/10.1103/PhysRevB.48.8487}
  {\bibfield  {journal} {\bibinfo  {journal} {Physical Review B}\ }\textbf
  {\bibinfo {volume} {48}},\ \bibinfo {pages} {8487} (\bibinfo {year}
  {1993})}\BibitemShut {NoStop}%
\bibitem [{\citenamefont {Jauho}\ \emph {et~al.}(1994)\citenamefont {Jauho},
  \citenamefont {Wingreen},\ and\ \citenamefont
  {Meir}}]{jauhoTimedependentTransportInteracting1994}%
  \BibitemOpen
  \bibfield  {author} {\bibinfo {author} {\bibfnamefont {A.-P.}\ \bibnamefont
  {Jauho}}, \bibinfo {author} {\bibfnamefont {N.~S.}\ \bibnamefont
  {Wingreen}},\ and\ \bibinfo {author} {\bibfnamefont {Y.}~\bibnamefont
  {Meir}},\ }\bibfield  {title} {\bibinfo {title} {Time-dependent transport in
  interacting and noninteracting resonant-tunneling systems},\ }\href
  {https://doi.org/10.1103/PhysRevB.50.5528} {\bibfield  {journal} {\bibinfo
  {journal} {Physical Review B}\ }\textbf {\bibinfo {volume} {50}},\ \bibinfo
  {pages} {5528} (\bibinfo {year} {1994})}\BibitemShut {NoStop}%
\bibitem [{\citenamefont {Meir}\ and\ \citenamefont
  {Wingreen}(1992)}]{meirLandauerFormulaCurrent1992}%
  \BibitemOpen
  \bibfield  {author} {\bibinfo {author} {\bibfnamefont {Y.}~\bibnamefont
  {Meir}}\ and\ \bibinfo {author} {\bibfnamefont {N.~S.}\ \bibnamefont
  {Wingreen}},\ }\bibfield  {title} {\bibinfo {title} {Landauer formula for the
  current through an interacting electron region},\ }\href
  {https://doi.org/10.1103/PhysRevLett.68.2512} {\bibfield  {journal} {\bibinfo
   {journal} {Physical Review Letters}\ }\textbf {\bibinfo {volume} {68}},\
  \bibinfo {pages} {2512} (\bibinfo {year} {1992})}\BibitemShut {NoStop}%
\bibitem [{\citenamefont {Lehmann}\ \emph {et~al.}(2018)\citenamefont
  {Lehmann}, \citenamefont {Croy}, \citenamefont {Guti{\'e}rrez},\ and\
  \citenamefont {Cuniberti}}]{lehmannTimedependentFrameworkEnergy2018}%
  \BibitemOpen
  \bibfield  {author} {\bibinfo {author} {\bibfnamefont {T.}~\bibnamefont
  {Lehmann}}, \bibinfo {author} {\bibfnamefont {A.}~\bibnamefont {Croy}},
  \bibinfo {author} {\bibfnamefont {R.}~\bibnamefont {Guti{\'e}rrez}},\ and\
  \bibinfo {author} {\bibfnamefont {G.}~\bibnamefont {Cuniberti}},\ }\bibfield
  {title} {\bibinfo {title} {Time-dependent framework for energy and charge
  currents in nanoscale systems},\ }\href
  {https://doi.org/10.1016/j.chemphys.2018.01.011} {\bibfield  {journal}
  {\bibinfo  {journal} {Chemical Physics}\ }\textbf {\bibinfo {volume} {514}},\
  \bibinfo {pages} {176} (\bibinfo {year} {2018})}\BibitemShut {NoStop}%
\bibitem [{\citenamefont {Stafford}\ and\ \citenamefont
  {Wingreen}(1996)}]{staffordResonantPhotonAssistedTunneling1996}%
  \BibitemOpen
  \bibfield  {author} {\bibinfo {author} {\bibfnamefont {C.~A.}\ \bibnamefont
  {Stafford}}\ and\ \bibinfo {author} {\bibfnamefont {N.~S.}\ \bibnamefont
  {Wingreen}},\ }\bibfield  {title} {\bibinfo {title} {Resonant
  {{Photon-Assisted Tunneling}} through a {{Double Quantum Dot}}: {{An Electron
  Pump}} from {{Spatial Rabi Oscillations}}},\ }\href
  {https://doi.org/10.1103/PhysRevLett.76.1916} {\bibfield  {journal} {\bibinfo
   {journal} {Physical Review Letters}\ }\textbf {\bibinfo {volume} {76}},\
  \bibinfo {pages} {1916} (\bibinfo {year} {1996})}\BibitemShut {NoStop}%
\bibitem [{\citenamefont
  {Datta}(1995)}]{dattaElectronicTransportMesoscopic1995}%
  \BibitemOpen
  \bibfield  {author} {\bibinfo {author} {\bibfnamefont {S.}~\bibnamefont
  {Datta}},\ }\href {https://doi.org/10.1017/CBO9780511805776} {\emph {\bibinfo
  {title} {Electronic {{Transport}} in {{Mesoscopic Systems}}}}},\ Cambridge
  {{Studies}} in {{Semiconductor Physics}} and {{Microelectronic Engineering}}\
  (\bibinfo  {publisher} {{Cambridge University Press}},\ \bibinfo {address}
  {{Cambridge}},\ \bibinfo {year} {1995})\BibitemShut {NoStop}%
\bibitem [{\citenamefont {Datta}(2005)}]{dattaChapterLevelBroadening2005}%
  \BibitemOpen
  \bibfield  {author} {\bibinfo {author} {\bibfnamefont {S.}~\bibnamefont
  {Datta}},\ }\bibfield  {title} {\bibinfo {title} {Chapter 8 - {{Level}}
  broadening},\ }in\ \href {https://doi.org/10.1017/CBO9781139164313} {\emph
  {\bibinfo {booktitle} {Quantum {{Transport}}: {{Atom}} to {{Transistor}}}}}\
  (\bibinfo  {publisher} {{Cambridge University Press}},\ \bibinfo {address}
  {{Cambridge}},\ \bibinfo {year} {2005})\BibitemShut {NoStop}%
\bibitem [{\citenamefont
  {Bergfield}(2010)}]{bergfieldManyBodyTheoryElectrical2010}%
  \BibitemOpen
  \bibfield  {author} {\bibinfo {author} {\bibfnamefont {J.}~\bibnamefont
  {Bergfield}},\ }\emph {\bibinfo {title} {Many-{{Body Theory}} of
  {{Electrical}}, {{Thermal}} and {{Optical Response}} of {{Molecular
  Heterojunctions}}}},\ \href {http://hdl.handle.net/10150/194386} {Ph.D.
  thesis},\ \bibinfo  {school} {The University of Arizona.} (\bibinfo {year}
  {2010})\BibitemShut {NoStop}%
\bibitem [{Note5()}]{Note5}%
  \BibitemOpen
  \bibinfo {note} {\label {numerics_bb_footnote}In the computational solution,
  the broad-band limit was used, which consists in taking the limit
  $t_0\rightarrow \infty $ while keeping $t^2/t_0$ fixed.}\BibitemShut {Stop}%
\bibitem [{\citenamefont {Biehs}\ \emph {et~al.}(2010)\citenamefont {Biehs},
  \citenamefont {Rousseau},\ and\ \citenamefont
  {Greffet}}]{PhysRevLett.105.234301}%
  \BibitemOpen
  \bibfield  {author} {\bibinfo {author} {\bibfnamefont {S.-A.}\ \bibnamefont
  {Biehs}}, \bibinfo {author} {\bibfnamefont {E.}~\bibnamefont {Rousseau}},\
  and\ \bibinfo {author} {\bibfnamefont {J.-J.}\ \bibnamefont {Greffet}},\
  }\bibfield  {title} {\bibinfo {title} {Mesoscopic description of radiative
  heat transfer at the nanoscale},\ }\href
  {https://doi.org/10.1103/PhysRevLett.105.234301} {\bibfield  {journal}
  {\bibinfo  {journal} {Physical Review Letters}\ }\textbf {\bibinfo {volume}
  {105}},\ \bibinfo {pages} {234301} (\bibinfo {year} {2010})}\BibitemShut
  {NoStop}%
\bibitem [{\citenamefont {Kerremans}\ \emph {et~al.}(2022)\citenamefont
  {Kerremans}, \citenamefont {Samuelsson},\ and\ \citenamefont
  {Potts}}]{kerremansProbabilisticallyViolatingFirst2022}%
  \BibitemOpen
  \bibfield  {author} {\bibinfo {author} {\bibfnamefont {T.}~\bibnamefont
  {Kerremans}}, \bibinfo {author} {\bibfnamefont {P.}~\bibnamefont
  {Samuelsson}},\ and\ \bibinfo {author} {\bibfnamefont {P.}~\bibnamefont
  {Potts}},\ }\bibfield  {title} {\bibinfo {title} {Probabilistically violating
  the first law of thermodynamics in a quantum heat engine},\ }\href
  {https://doi.org/10.21468/SciPostPhys.12.5.168} {\bibfield  {journal}
  {\bibinfo  {journal} {SciPost Physics}\ }\textbf {\bibinfo {volume} {12}},\
  \bibinfo {pages} {168} (\bibinfo {year} {2022})}\BibitemShut {NoStop}%
\bibitem [{\citenamefont {Dann}\ and\ \citenamefont
  {Kosloff}(2022)}]{dannUnificationFirstLaw2022}%
  \BibitemOpen
  \bibfield  {author} {\bibinfo {author} {\bibfnamefont {R.}~\bibnamefont
  {Dann}}\ and\ \bibinfo {author} {\bibfnamefont {R.}~\bibnamefont {Kosloff}},\
  }\href {http://arxiv.org/abs/2208.10561} {\bibinfo {title} {Unification of
  the first law of quantum thermodynamics}} (\bibinfo {year} {2022}),\ \Eprint
  {https://arxiv.org/abs/2208.10561} {arxiv:2208.10561 [quant-ph]} \BibitemShut
  {NoStop}%
\bibitem [{\citenamefont {Kumar}\ and\ \citenamefont
  {Stafford}(2022)}]{kumarFirstLawThermodynamics2022}%
  \BibitemOpen
  \bibfield  {author} {\bibinfo {author} {\bibfnamefont {P.}~\bibnamefont
  {Kumar}}\ and\ \bibinfo {author} {\bibfnamefont {C.~A.}\ \bibnamefont
  {Stafford}},\ }\href {http://arxiv.org/abs/2208.06544} {\bibinfo {title} {On
  the {{First Law}} of {{Thermodynamics}} in {{Time-Dependent Open Quantum
  Systems}}}} (\bibinfo {year} {2022}),\ \bibinfo {note} {comment: Version 1,
  19 pages, 7 figures},\ \Eprint {https://arxiv.org/abs/2208.06544v1}
  {arxiv:2208.06544v1 [cond-mat]} \BibitemShut {NoStop}%
\bibitem [{\citenamefont {Hedin}(1965)}]{hedinNewMethodCalculating1965}%
  \BibitemOpen
  \bibfield  {author} {\bibinfo {author} {\bibfnamefont {L.}~\bibnamefont
  {Hedin}},\ }\bibfield  {title} {\bibinfo {title} {New {{Method}} for
  {{Calculating}} the {{One-Particle Green}}'s {{Function}} with
  {{Application}} to the {{Electron-Gas Problem}}},\ }\href
  {https://doi.org/10.1103/PhysRev.139.A796} {\bibfield  {journal} {\bibinfo
  {journal} {Physical Review}\ }\textbf {\bibinfo {volume} {139}},\ \bibinfo
  {pages} {A796} (\bibinfo {year} {1965})}\BibitemShut {NoStop}%
\bibitem [{\citenamefont {Mahan}(2000)}]{mahanManyParticlePhysics2000}%
  \BibitemOpen
  \bibfield  {author} {\bibinfo {author} {\bibfnamefont {G.~D.}\ \bibnamefont
  {Mahan}},\ }\href@noop {} {\emph {\bibinfo {title} {Many-{{Particle
  Physics}}}}},\ \bibinfo {edition} {3rd}\ ed.\ (\bibinfo  {publisher} {{New
  York : Kluwer Academic/Plenum Publishers}},\ \bibinfo {address} {{New
  York}},\ \bibinfo {year} {2000})\BibitemShut {NoStop}%
\bibitem [{\citenamefont {Ziman}(2001)}]{zimanElectronsPhononsTheory2001}%
  \BibitemOpen
  \bibfield  {author} {\bibinfo {author} {\bibfnamefont {J.~M.}\ \bibnamefont
  {Ziman}},\ }\href {https://doi.org/10.1093/acprof:oso/9780198507796.001.0001}
  {\emph {\bibinfo {title} {Electrons and {{Phonons}}: {{The Theory}} of
  {{Transport Phenomena}} in {{Solids}}}}}\ (\bibinfo  {publisher} {{Oxford
  University Press}},\ \bibinfo {year} {2001})\BibitemShut {NoStop}%
\end{thebibliography}%

\newpage
\onecolumngrid  
	
\appendix
\section{Some details about Entropy and the Hilbert Space partition}
\label{Hspacediv_app}

This appendix provides some additional details of the analysis of entropy and Hilbert space partition given in Sec.\ \ref{heat_roleofres_subsec}. 

\subsection{Derivation of Eq.\ \eqref{lineardev_entropy_eqn}}

Eq.\ \eqref{lineardev_entropy_eqn} follows straightforwardly from Eq.\ (\ref{eq:vonNeumann}) using the properties of the density matrix and perturbation theory.  In its eigenbasis, the density matrix may be written
\begin{equation}
    \rho=\sum_\nu \lambda_\nu |\lambda_\nu\rangle \langle \lambda_\nu|,
\end{equation}
with $\sum_\nu \lambda_\nu=1$.  A small deviation of $\rho$ from an initial density matrix $\rho_0$ is given to linear order by
\begin{equation}
    d\rho =\sum_\nu \left[d\lambda_\nu |\lambda_\nu^0\rangle\langle\lambda_\nu^0|
    +\lambda_\nu^0 |d\lambda_\nu\rangle\langle\lambda_\nu^0|
    +\lambda_\nu^0 |\lambda_\nu^0\rangle\langle d\lambda_\nu|
    \right],
    \label{eq:drho}
\end{equation}
where $\langle d\lambda_\nu|\lambda_\nu^0\rangle=0$ without loss of generality.
The linear deviation of the entropy is
\begin{equation}
    dS=-\sum_\nu d\lambda_\nu \ln \lambda_\nu^0 -\sum_\nu d\lambda_\nu,
\end{equation}
where the second term on the rhs vanishes due to the invariance of the trace of $\rho$.
Finally, 
\begin{eqnarray}
    {\rm Tr}\{d\rho \ln \rho_0\} & = & \sum_\nu \langle \lambda_\nu^0 |d\rho|\lambda_\nu^0\rangle \ln \lambda_\nu^0
    \nonumber \\
    & = & \sum_\nu d\lambda_\nu \ln \lambda_\nu^0,
\end{eqnarray}
where the second line follows from Eq.\ (\ref{eq:drho}) because $\langle d\lambda_\nu|\lambda_\nu^0\rangle=0$.
This completes the proof of Eq.\ \eqref{lineardev_entropy_eqn}

\subsection{Derivation of Eq.\ \eqref{eq:dS_gamma}}

Using Eq.\ \eqref{eq:Tr_gamma} and $\omega A(\omega)=h^{(1)}A(\omega)$, Eq.\ \eqref{eq:dS_gamma} may be written
\begin{equation}
    T dS_\gamma = \int d\omega \mathbb{Tr}^{(1)}\{\Pi_\gamma (h^{(1)}-\mu)A(\omega)\} df(\omega).
\end{equation}
Because $[h^{(1)},A]=0$, we can rewrite the trace
\begin{eqnarray}
    \mathbb{Tr}^{(1)}\{\Pi_\gamma h^{(1)}A(\omega)\} & = & \frac{1}{2} \mathbb{Tr}^{(1)}\{\Pi_\gamma \{h^{(1)},A(\omega)\}\}
    \nonumber \\
    & = & \mathbb{Tr}^{(1)}\{\left.h^{(1)}\right|_\gamma A(\omega)\},
\end{eqnarray}
where the second line used Eq.\ \eqref{eq:H_gamma} and the cyclic property of the trace.
The expectation value of the Hamiltonian partitioned on subsystem $\gamma$ in thermal equilibrium is
\begin{eqnarray}
    \langle \left.H^{(1)}\right|_\gamma\rangle & = & \int d\omega \mathbb{Tr}^{(1)}\{\left.h^{(1)}\right|_\gamma A(\omega)\} f(\omega)
    \nonumber \\
    & = & -i \mathbb{Tr}^{(1)}_\gamma \left\{h^{(1)}(t)G^<(t,t)\right\},
    \label{eq:meanH_gamma}
\end{eqnarray}
where the second line holds also out of equilibrium.
Finally, including only the change in the statistical factors but not the change in the Hamiltonian itself,
\begin{equation}
    \int d\omega \mathbb{Tr}^{(1)}\{\left.h^{(1)}\right|_\gamma A(\omega)\} df(\omega)
    = d\langle \left.H^{(1)}\right|_\gamma \rangle -\langle \left.dH^{(1)}\right|_\gamma \rangle.
\end{equation}
This completes the derivation of Eq.\ \eqref{eq:dS_gamma}.

\subsubsection{Phonon analysis}

For the case of a harmonic phonon system, Eq.\ (\ref{eq:TdS}) takes the form
\begin{equation}
    T dS_\gamma^{ph} = \int d\omega \mathbb{Tr}^{(1)}_\gamma \{A^{ph}(\omega)\} \hbar \omega \, df^{Planck}(\omega),
    \label{eq:TdS_ph}
\end{equation}
where $A^{ph}(\omega)=\delta(\omega- \sqrt{K})$ and the harmonic interaction matrix $K$ was defined in Eq.\ (\ref{sys_ph_harm_hamiltonian_1_eqn}).  Eq.\ (\ref{eq:TdS_ph}) may be integrated to obtain the phonon subsystem entropy in equilibrium
\begin{equation}
    S^{ph}_\gamma = \int d\omega \mathbb{Tr}^{(1)}_\gamma \{A^{ph}(\omega)\} s_-(\omega),
\end{equation}
which is non-singular %
and respects the 3rd Law of Thermodynamics as $T\rightarrow 0$, where
\begin{eqnarray}
    s_-(\omega) = \beta \hbar \omega f^{Planck}(\omega) -\ln (1-e^{-\beta \hbar \omega}).
\end{eqnarray}
By arguments similar to those given above, one can readily show that the phonon contributions to the entropy, work, and internal energy changes
also satisfy Eq.\ \eqref{eq:dS_gamma} in the harmonic approximation (with $\mu=0$ for phonons). 

\subsection{Partition of the two-body interactions}

The electron-electron interaction term may be treated along the lines of Eq.\ \eqref{eq:meanH_gamma} with the help of the Coulomb self-energy $\Sigma_C$.  The expectation value of the electronic Hamiltonian, including electron-electron interactions is \cite{hedinNewMethodCalculating1965}
\begin{equation}
    \langle H(t)\rangle = -i \mathbb{Tr}^{(1)}\left\{h^{(1)}(t)G^<(t,t) + \frac{1}{2}\int dt^\prime  \left[\Sigma_C(t,t^\prime)G(t^\prime,t)\right]^< \right\}.
\end{equation}
Using the Langreth rules \cite{stefanucciNonequilibriumManyBodyTheory2013}, this may be expressed as
\begin{equation}
    \langle H(t)\rangle = -i \mathbb{Tr}^{(1)}\left\{h^{(1)}(t)G^<(t,t) + \frac{1}{2}\int dt^\prime  
    \left[\Sigma_C^R(t,t^\prime)G^<(t^\prime,t)
    +\Sigma_C^<(t,t^\prime)G^A(t^\prime,t)
    \right]\right\}.
    \label{eq:H1and2}
\end{equation}
The mean energy of subsystem $\gamma$ is then
\begin{equation}
    \langle \left.H\right|_\gamma(t)\rangle = -\frac{i}{2} \mathbb{Tr}^{(1)}_\gamma \left\{\{h^{(1)},(t)G^<(t,t)\} + \frac{1}{2}\int dt^\prime  
    \left[\{\Sigma_C^R(t,t^\prime),G^<(t^\prime,t)\}
    +\{\Sigma_C^<(t,t^\prime),G^A(t^\prime,t)\}
    \right]\right\}.
    \label{eq:H1and2_gamma}
\end{equation}
Because two-body interactions occur only within the system (by construction), the Coulomb self-energy is a system-diagonal matrix in the
one-particle Hilbert space.  The two-body interactions therefore do not contribute to the partition of the Hamiltonian onto the reservoirs,
confirming Eq.\ \eqref{eq:H|R}.  This result may also be obtained directly by partitioning $H^{(2)}$, as discussed in Ref.\
\cite{webb2023}.

In a similar way, it can be shown that phonon anharmonic terms and electron-phonon interactions, occuring only within the system, do not contribute to the partition of the Hamiltonian onto the reservoirs.

\section{Derivation of the NEGF formulas for all First Law quantities}
\label{NEGF_details_app}

This appendix gives details of the derivations of the key NEGF results presented in Sec.\ \ref{NEGFresults_sec}.

\subsection{Computations in the NEGF formalism}
Before giving the details of specific results we briefly review here the general procedure employed in computing the Non-Equilibrium Green's functions. Within the framework of NEGF, the Retarded Green's function $G^{R}(t,t')$, which describes the dynamics of the system, can be computed using the Dyson Equation
\begin{IEEEeqnarray}{lcr}
	G^{R}(t,t')= G^{R}_{0}(t,t')+\int_{-\infty}^{\infty} \,dt_1\int_{-\infty}^{\infty} \,dt_2  G^{R}_{0}(t,t_1)\Sigma^{R}(t_1,t_2)G^{R}(t_2,t') \,,
\end{IEEEeqnarray}
where $G^{R}_{0}(t,t')$ is the Green's function for the non-interacting and uncoupled system and the Self-Energy function which encodes the interactions and tunneling is given by
\begin{equation}
	\Sigma^{R}(t,t')=\Sigma^{R}_{int}+\Sigma^{R}_{tun}(t,t')\,,
\end{equation}
where $\Sigma^{R}_{int}$ is the retarded interaction self-energy of the system (for example, the Coulomb self-energy) and the retarded tunneling self-energy is given by
\begin{equation}
  \Sigma^{R}_{tun}(t,t')= \sum_{\alpha}\int_{-\infty}^{\infty} \,\frac{d\omega}{2\pi} e^{-i\omega(t-t')}[\Lambda_{\alpha}(\omega)-\frac{i}{2}\Gamma_{\alpha}(\omega)] \,,
\end{equation}
with $\Lambda_{\alpha}(\omega)=\mathcal{P}\int_{-\infty}^{\infty} \,\frac{d\omega'}{2\pi}\frac{\Gamma_{\alpha}(\omega')}{\omega-\omega'}$, where $\mathcal{P}$ denotes the Principal Value [see, for instance, Eq.\ ($2.26$) in \cite{stefanucciNonequilibriumManyBodyTheory2013}]. The Lesser Green's function $G^{<}(t,t')$, which describes the occupation of the system, can be computed using the Keldysh Equation 
\begin{equation}
	G^{<}(t,t')=\int_{-\infty}^{\infty} \,dt_1\int_{-\infty}^{\infty} \,dt_2  G^{R}(t,t_1)\Sigma^{<}(t_1,t_2) G^{A}(t_2,t') \,,
\end{equation}
where we have omitted the so-called memory term [see, for instance, Eq.\ (5.11) in \cite{haugQuantumKineticsTransport2007}] since it is not relevant for the protocols we consider in this article, and the lesser Self-Energy function is 
\begin{equation}
	\Sigma^{<}(t,t')=\Sigma^{<}_{int}(t,t')+i\sum_{\alpha}\int_{-\infty}^{\infty} \,\frac{d\omega}{2\pi} e^{-i\omega(t-t')} f_{\alpha}(\omega)\Gamma_{\alpha}(\omega) \,.
\end{equation}

The two-body Green's function $G^{(2)}(t,t)$ appearing in Eq.\ \eqref{HS_el_negf_eqn} can be evaluated directly using the Bethe-Salpeter equation \cite{stefanucciNonequilibriumManyBodyTheory2013}, or the contribution from electron-electron interactions to Eq.\ \eqref{HS_negf_eqn} can be evaluated using the Coulomb self-energy and the one-body Green's function, as in Eqs.\ \eqref{eq:H1and2}--\eqref{eq:H1and2_gamma}. The electron-phonon Green's function $G^{(2),el-ph}(t,t)$ and the higher-order phonon Green's functions such as $D^{ahm}(t,t,t)$, appearing in Eqs.\ \eqref{HS_ph_negf_eqn} and \eqref{HS_elph_negf_eqn}, can be evaluated using the well-developed theory of electron-phonon and phonon-phonon interactions \cite{mahanManyParticlePhysics2000,zimanElectronsPhononsTheory2001}.

\subsection{Electronic Heat current} 
Here we give the details of the derivation of the electronic heat current formula in terms of the System Green's functions using the Nonequilibrum Green's function (NEGF) formalism \cite{kadanoffQuantumStatisticalMechanics1962,keldyshDiagramTechniqueNonequilibrium1964,rammerQuantumFieldTheory2007,haugQuantumKineticsTransport2007,stefanucciNonequilibriumManyBodyTheory2013}. (The calculation of the interface internal energy [Eq.\ \eqref{HSR_negf_eqn}] proceeds in 
a nearly parallel fashion.)

To derive the conventional component of heat current formula [Eq.\ \eqref{el_heat_curr_negf_eqn}], we begin with the electronic heat current given in terms of expectation values in Eq.\ \eqref{conv_el_heatcurr_expval_eqn}. We then use the the lesser Tunneling Green's functions (defined in \ref{eqy_den_negf_subsec}) $G^{<}_{nk}(t,t')= i\langle c_{k}^{\dagger}(t')d_{n}(t)\rangle$ and $G^{<}_{kn}(t,t')=i\langle d_{n}^{\dagger}(t')c_{k}(t)\rangle$, where, to lighten notation we write $G^{<}_{tun,nk}(t,t')$ simply as $G^{<}_{nk}(t,t')$, and noting that at equal times $G^{<}_{nk}(t,t)=-[G^{<}_{kn}(t,t)]^{*}$, to write the heat current as
 \begin{equation}
	I_{\alpha,el}^{Q,conv}(t)=\frac{-2}{\hbar}\mathbb{Re}\Bigg[\sum_{k\in\alpha,n}(\epsilon_k-\mu_\alpha)V_{kn}^{el}G^{<}_{nk}(t,t)\Bigg] \,. 
	\label{heat_curr_Glesser_eqn}
\end{equation}
The NEGF prescription tells us that $G^{<}_{nk}(t,t')$ will be a component of the Contour-Ordered Tunneling Green's function
\begin{equation}
	G^{\mathcal{T}}_{nk}(\tau,\tau')=-i\langle\mathcal{T}(d_{n}(\tau)c_{k}^{\dagger}(\tau'))\rangle	\,,
\end{equation}
where $\tau$ is the contour time. The   equation of motion of the Contour-Ordered Tunneling Green's function is 
\begin{equation}
	-i\hbar\frac{\partial}{\partial \tau'}G^{\mathcal{T}}_{nk}(\tau,\tau')=\epsilon_{k}G^{\mathcal{T}}_{nk}(\tau,\tau') + \sum_{m}G^{\mathcal{T}}_{nm}(\tau,\tau') (V^{el}_{km})^{*} \,,
	\label{tun_contour_GF_eom_1_eqn}
\end{equation}
where \(G^{\mathcal{T}}_{nm}(\tau,\tau')=-i\langle\mathcal{T}(d_{n}(\tau)d_{m}^{\dagger}(\tau'))\rangle\) is the Contour-Ordered \emph{System} Green's function and we get 
\begin{equation}
	G^{\mathcal{T}}_{nk}(\tau,\tau')=\frac{1}{\hbar}\sum_{m} \int \,d\tau_1 G^{\mathcal{T}}_{nm}(\tau,\tau_1)(V^{el}_{km})^{*}g^{\mathcal{T}}_{kk}(\tau_1,\tau') \,,
	\label{tun_contour_GF_eom_2_eqn}
\end{equation}
using the equation of motion of the uncoupled Contour-Ordered reservoir  Green's function \(g_{kk}^{\mathcal{T}}(\tau,\tau')=-i\langle c_{k}(\tau)c^{\dagger}_{k}(\tau')\rangle\). The physical lesser component of the Contour-Ordered Tunneling Green's function [Eq.\ \eqref{tun_contour_GF_eom_2_eqn}] can be found using the so-called Langreth rules \cite{haugQuantumKineticsTransport2007,stefanucciNonequilibriumManyBodyTheory2013}, giving 
\begin{IEEEeqnarray}{lcr}
	G^{<}_{nk}(t,t')=\frac{1}{\hbar}\sum_{m} \int \,dt_1 (V^{el}_{km})^{*}[G^{R}_{nm}(t,t_1)g^{<}_{kk}(t_1,t')
	+G^{<}_{nm}(t,t_1)g^{A}_{kk}(t_1,t')] \,.\nonumber\\&&
\end{IEEEeqnarray}
The lesser and advanced uncoupled Reservoir Green's functions are given by 
\begin{equation}
	g^{<}_{kk}(t,t')=i\langle c^{\dagger}_{k}(t')c_{k}(t)\rangle= if(\epsilon_{k}^{0})\exp(\frac{-i}{\hbar}\epsilon_{k}(t-t')) 
\end{equation}
and 
\begin{IEEEeqnarray}{lcr}
	g^{A}_{kk}(t,t')=i\theta (t'-t)\langle \{c_{k}(t),c^{\dagger}_{k}(t')\}\rangle  =i\theta(t'-t)\exp(\frac{-i}{\hbar}\epsilon_{k}(t-t')) \,,
\end{IEEEeqnarray} 
respectively, where
\begin{equation}
	f(\epsilon_{k}^{0})=\frac{1}{e^{\beta_{\alpha}(\epsilon_{k}^{0}-\mu)}+1} 
\end{equation}
is the Fermi-Dirac distribution function. Putting this all together, we can write  
\begin{IEEEeqnarray}{lcr}
	I_{\alpha,el}^{Q,conv}(t)=
	\frac{2}{\hbar}\mathbb{Im}\Bigg[\sum_{k\in\alpha,n,m}(\epsilon_k-\mu_\alpha)\int_{-\infty}^{t} \,dt_1\{e^{\frac{-i}{\hbar}\epsilon_{k}(t_1-t)}V^{el}_{kn}(V^{el}_{km})^{*}[G^{<}_{nm}(t,t_1) + f_{\alpha}(\epsilon_{k})G^{R}_{nm}(t,t_1)]\}\Bigg] \,, \nonumber\\
\end{IEEEeqnarray}
which, upon defining the tunneling width-matrix as in Eq.\ \eqref{t_indpt_Gamma_eqn}, allows us to turn the momentum sum into an energy integral and can be written, by going over to the matrix notation, as 
\begin{IEEEeqnarray}{lcr}
	I_{\alpha,el}^{Q,conv}(t)=\frac{2}{\hbar}\int_{-\infty}^{\infty} \,\frac{d\omega}{2\pi}\int_{-\infty}^{t} \,dt_1(\hbar\omega-\mu_{\alpha})\mathbb{ImTr}\Bigg\{e^{-i\omega(t_1-t)}\Gamma_{\alpha}^{el}(\omega)[G^{<}(t,t_1) + f_{\alpha}(\omega)G^{R}(t,t_1)]\Bigg\} \,,
	\label{el_heat_curr_negf_app_eqn}
\end{IEEEeqnarray}
which is exactly Eq.\ \eqref{el_heat_curr_negf_eqn}. %
The interfacial electronic contribution to the internal energy, $\langle H_{SR,el,\alpha} \rangle$ [Eq.\ \eqref{HSR_el_negf_eqn}], can be derived in a similar fashion. The time derivative of this interfacial energy (divided by two) gives the transient component of the Heat Current 
[Eq.\ \eqref{tr_el_heatcurr_defn_eqn}]. 

\subsection{Energy Density}
\label{egyden_app_subsec}
The energy density $\rho_H(x,t)$ [Eq.\ \eqref{eq:rho_H}] can also be evaluated in terms of the system Green's function. Since $\rho_{H_{S}}(x,t)$ is already given in terms of the system Green's function in  Eq.\ \eqref{egy_den_HS_el_negf_eqn}, we need to evaluate only $\rho_{H_{SR}}(x,t)$ [Eq.\ \eqref{egy_den_HSR_el_negf_eqn}]. Again, by employing the same procedure used for deriving the heat current equation [Eq.\ \eqref{el_heat_curr_negf_app_eqn}] just discussed, we get
\begin{equation}
	\rho_{H_{SR}}(x,t)= %
	\int_{-\infty}^{\infty} \,\frac{d\omega}{2\pi} \int_{-\infty}^{t} \,dt_1 \mathbb{Im}\Bigg\{\langle x|e^{-i\omega(t_1-t)}[G^{<}(t,t_1)\Sigma^{A}_{\alpha}(\omega)+ G^{R}(t,t_1)\Sigma^{<}_{\alpha}(\omega)]|x\rangle\Bigg\}\,.
\end{equation}

\subsection{Phononic heat current}
\label{phonon_heat_current_derv_app}
To derive the conventional component of the phonon heat current formula, Eq.\ \eqref{ph_heat_curr_negf_eqn}, we note that the heat current formula written in terms of the expectation values, Eq.\ \eqref{conv_ph_heatcurr_expval_eqn}, upon defining the tunneling Green's function $D^{<}_{tun,rq}(t,t')=-i\langle U_q(t')Q_r(t)\rangle$ and using its properties, can be written as %
\begin{equation}
    I_{\alpha,ph}^{Q,conv}(t)= i\sum_{q\in\alpha,r} \Bigg[V_{qr}\frac{\partial}{\partial t'}D^{<}_{tun,rq}(t,t')\Bigg]_{t=t'}\,.
    \label{ph_heat_curr_derv_app_eqn}
\end{equation}

A closed equation of motion for the tunneling phonon green's function results in second-order time derivatives. It can be shown that 
\begin{equation}
  \Bigg[\frac{{\partial}^2}{\partial {\tau'}^2}+\omega_{q}^2\Bigg]D^{\mathcal{T}}_{tun,rq}(\tau,\tau')=-\sum_{s}V_{sq}D^{\mathcal{T}}_{rs}(\tau,\tau')
\end{equation}
where we have moved to the contour variables $(\tau,\tau')$, defining the Contour-Ordered tunneling Green's function as $D^{\mathcal{T}}_{tun,rq}(\tau,\tau')=-i\langle \mathcal{T}(Q_r(\tau)U_q(\tau'))\rangle$, and the Contour-Ordered system Green's function is as $D^{\mathcal{T}}_{rs}(\tau,\tau')=-i\langle \mathcal{T}(Q_r(\tau)Q_s(\tau'))\rangle$. It can be also be shown that the Contour-Ordered uncoupled reservoir Green's function $d_{qq}^{\mathcal{T}}(\tau,\tau')=-i\langle\mathcal{T}(U_q(\tau)U_q(\tau'))\rangle$ obeys the closed second-order equation-of-motion 
\begin{equation}
    \Bigg[\frac{{\partial}^2}{\partial {\tau'}^2}+\omega_{q}^2\Bigg]d_{qq}^{\mathcal{T}}(\tau,\tau')=-\hbar\delta(\tau,\tau')\,.
\end{equation}
Putting the last two equations together it is easy to see that the tunneling Green's function can be written in terms of the reservoir and system Green's functions  as
\begin{equation}
 D^{\mathcal{T}}_{tun,rq}(\tau,\tau') =\frac{1}{\hbar} \sum_{q\in\alpha,s} \int \,d\tau_1 D^{\mathcal{T}}_{rs}(\tau,\tau_1)V_{qs}d^{\mathcal{T}}_{qq}(\tau_1,\tau)\,.
\end{equation}
Substituting the lesser component of this contour-ordered Green's function (obtained using the Langreth rules) into Eq.\ \eqref{ph_heat_curr_derv_app_eqn}, and using the fact that the reservoirs are time-independent [$d^{\mathcal{T}}_{qq}(\tau_1,\tau)=d^{\mathcal{T}}_{qq}(\tau_1-\tau)$], we arrive at the expression 
\begin{equation}
    I_{\alpha,ph}^{Q,conv}(t)= \frac{1}{\hbar}\int_{-\infty}^{+\infty}\,\frac{d\omega}{2\pi} \omega \int_{-\infty}^{+\infty}\,dt_1 \mathbb{Tr}\Bigg\{e^{-i\omega(t_1-t)}[D^{<}(t,t_1)\Pi^{tun,A}_{\alpha}(\omega)+D^{R}(t,t_1)\Pi^{tun,<}_{\alpha}(\omega)]\Bigg\}\,,
\end{equation}
which is exactly the result stated in Eq.\ \eqref{ph_heat_curr_negf_eqn}. The interfacial phononic contribution to the internal energy $\langle H_{SR,ph,\alpha} \rangle$ [Eq.\ \eqref{HSR_ph_negf_eqn}] can be derived in a similar fashion.

\subsection{External work done on Phonons by mechanical drive [Eq.\ \eqref{wext_ph_negf_eqn}]} \label{phonon_wext_NEGF_app}

Here we give some details for the derivation of Eq.\ \eqref{wext_ph_negf_eqn}. We begin by noting that the equation of motion of the displacement operator in the system is given by 
\begin{equation} \label{disp_op_eom_1_eqn}
    \frac{\partial^2}{\partial {t}^2}\langle Q_r(t)\rangle=\frac{-i}{\hbar}\langle [P_r(t),H_{S,ph}(t)+H_{S,el-ph}+H_{SR,ph}]\rangle\,.
\end{equation}
Proceeding similarly to the calculation in Appendix \ref{phonon_heat_current_derv_app}, we can write this as 
\begin{equation}\label{disp_op_eom_2_eqn}
  \Bigg[\frac{\partial^2}{\partial {t}^2}+\omega_{r}^2\Bigg]\langle Q_r(t)\rangle -\sum_{s}\int \,dt'\Pi^{tun}_{rs}(t,t')\langle Q_s(t)\rangle=F_r(t)+\frac{-i}{\hbar}\Bigg(\langle[P_r(t),H_{S,ph}^{ahm}]\rangle+
 \langle[P_r(t),H_{S,el-ph}]\rangle\Bigg) \,,
\end{equation}
where the commutators on the RHS can be evaluated as 
\begin{equation} \label{p_Hahm_com_eqn}
  [P_r(t),H_{S,ph}^{ahm}]=-3i\hbar\sum_{pu}[H_{S,ph}^{ahm}]_{rpu}Q_p(t)Q_u(t) 
\end{equation}
and 
\begin{equation} \label{p_Helph_com_eqn}
   [P_r(t),H_{S,el-ph}]=-i\hbar\sum_{nm}[H_{S,el-ph}]_{rnm} d_n^{\dagger}(t)d_m(t) \,.
\end{equation}
To cast Eq.\ \eqref{disp_op_eom_2_eqn} in a closed form we evaluate the equation of motion for the time-ordered System Phonon Green's function $\mathcal{D}_{rs}(t,t')=-i\langle \mathrm{T}(Q_r(t)Q_s(t'))\rangle$, in the absence of the time-dependent mechanical drive, giving
\begin{equation} \label{Gph_op_eom_1_eqn}
  \frac{-1}{i} \frac{\partial^2}{\partial{t}^2}\mathcal{D}_{rs}(t,t')=-\delta(t-t')\langle[P_r(t),Q_s(t')]\rangle+
  \frac{-i}{\hbar} \langle \mathrm{T}([P_r(t),H_{S,ph}(t)+H_{S,el-ph}+H_{SR,ph}]Q_r(t'))\rangle \,,
\end{equation}
which, similarly to Eq.\ \eqref{disp_op_eom_2_eqn}, can also be written as
\begin{eqnarray} \label{Gph_op_eom_2_eqn}
 \Bigg[\frac{\partial^2}{\partial {t}^2}+\omega_{r}^2\Bigg]\mathcal{D}_{rs}(t,t') - \sum_{p} \int \,dt_1\Pi^{tun}_{rp}(t,t_1)\mathcal{D}_{ps}(t,t')=  -\hbar\delta_{rs}\delta(t-t')+\frac{-i}{\hbar}\Bigg(\langle\mathrm{T}([P_r(t),H_{S,ph}^{ahm}]Q_s(t'))\rangle &+& \nonumber \\
 \langle\mathrm{T}([P_r(t),H_{S,el-ph}]Q_s(t'))\rangle\Bigg) \,.
\end{eqnarray}
The two commutator terms on the RHS of the above equation can be evaluated as
\begin{equation}
 \frac{-i}{\hbar}\langle\mathrm{T}([P_r(t),H_{S,ph}^{(3))}]Q_s(t'))\rangle=  -3\sum_{pu}[H_{S,ph}^{(3)}]_{rup}\langle\mathrm{T}(Q_u(t)Q_p(t)Q_s(t') )\rangle = -3i\sum_{pu}[H_{S,ph}^{(3)}]_{rup}D_{ups}^{ahm}(t,t,t')\,,
\end{equation}
and 
\begin{equation}
 \frac{-i}{\hbar}\langle\mathrm{T}([P_r(t),H_{S,el-ph}]Q_s(t'))\rangle=  -\sum_{nm}[H_{S,el-ph}]_{rnm}\langle\mathrm{T}(d_n^{\dagger}(t)d_m(t)Q_s(t'))\rangle = i\sum_{nm}[H_{S,el-ph}]_{rnm}G_{nms}^{el-ph}(t,t,t')\,.
\end{equation}
We now implicitly define the phonon self-energy due to the anharmonic term $\Pi^{ahm}(t,t')$ and the phonon self-energy due to the electron-phonon interaction $\Pi^{el-ph}(t,t')$ as 
\begin{equation}
    -i\sum_{pu}[H_{S,ph}^{(3)}]_{rup}D_{ups}^{ahm}(t,t,t')=\sum_{u}\int \,dt_1 \Pi^{ahm}_{ru}(t,t_1)\mathcal{D}_{us}(t_1,t')\,,
\end{equation}
and
\begin{equation}
   i\sum_{nm}[H_{S,el-ph}]_{rnm}G_{nms}^{el-ph}(t,t,t')=\sum_{l}\int \,dt_1 \Pi_{rl}^{el-ph}(t,t_1)\mathcal{D}_{ls}(t_1,t')\,,
\end{equation}
using which we can write the commutators \eqref{p_Hahm_com_eqn} and \eqref{p_Helph_com_eqn} as
\begin{equation}
  \frac{-i}{\hbar}\langle[P_r(t),H_{S,ph}^{ahm}]\rangle=3\sum_{u}\int \,dt_1\Pi^{ahm}_{ru}(t,t_1)
\langle Q_u(t_1)\rangle
\end{equation}
and
\begin{equation}
 \frac{-i}{\hbar} \langle [P_r(t),H_{S,el-ph}]\rangle=\sum_{l}\int \,dt_1\Pi_{rl}^{el-ph}(t,t_1) \langle Q_l(t_1)\rangle\,.
\end{equation}
Using the last four equations, we can write the equations of motion of the displacement operator [Eq.\ \eqref{disp_op_eom_2_eqn}] and the system Phonon Green's function [Eq.\ \eqref{Gph_op_eom_2_eqn}] as
\begin{eqnarray}
     \Bigg[\frac{\partial^2}{\partial {t}^2}+\omega_{r}^2\Bigg]\langle Q_r(t)\rangle -\sum_{s}\int \,dt'\Pi^{tun}_{rs}(t,t')\langle Q_s(t)\rangle -3\sum_{u}\int \,dt_1\Pi^{ahm}_{ru}(t,t_1)&-&\nonumber \\ \sum_{l}\int \,dt_1\Pi_{rl}^{el-ph}(t,t_1) \langle Q_l(t_1)\rangle=F_r(t)\,,
\end{eqnarray}
and
\begin{eqnarray}
 \Bigg[\frac{\partial^2}{\partial {t}^2}+\omega_{r}^2\Bigg]\mathcal{D}_{rs}(t,t') - \sum_{p} \int \,dt_1\Pi^{tun}_{rp}(t,t_1)\mathcal{D}_{ps}(t,t')-3\sum_{u}\int \,dt_1 \Pi^{ahm}_{ru}(t,t_1)\mathcal{D}_{us}(t_1,t')&-&\nonumber \\ \sum_{l}\int \,dt_1 \Pi_{rl}^{el-ph}(t,t_1)\mathcal{D}_{ls}(t_1,t')= -\hbar\delta_{rs}\delta(t-t')  \,.
\end{eqnarray}
From the last two equations it follows that (assuming homogeneous boundary conditions) we can write the displacement operator in terms of the Phonon Green's function as
\begin{equation}
  \langle Q_r(t)\rangle=\frac{-1}{\hbar}\sum_{s}\int_{-\infty}^{t}\,dt' F_s(t')\mathcal{D}_{rs}(t,t') \,,
\end{equation}
which upon substituting into Eq.\ \eqref{wext_ph_1_eqn} (and rearranging appropriately to recognize matrix multiplication) gives the stated result of Eq.\ \eqref{wext_ph_negf_eqn}.

\section{Generalization to Time-Dependent Reservoirs and Coupling}
\label{full_tdpt_generalize_app}

We generalize our results from Sec.\ \ref{NEGFresults_sec} to the case of a fully time-dependent Hamiltonian $H(t)$ where, in addition to the System, the Coupling and Reservoirs also acquire explicit time dependence, i.e.\ $H_{SR} \rightarrow H_{SR}(t)$ and $H_{R} \rightarrow H_{R}(t)$ so that
	
\begin{equation}
		H(t)= H_{S}(t) + H_{SR}(t) + H_{R}(t)\,.
\end{equation}
For notational simplicity, we will treat only electronic degrees of freedom here, but our results hold with full generality for phononic degrees of freedom as well. The time-dependent System Hamiltonian is then given by \begin{equation}
 H_{S}(t)=\sum_{n,m}\Bigg([H^{(1)}_{S}(t)]_{nm}d^{\dagger}_{n}d_{m} + [H^{(2)}_{S}]_{nm}d^{\dagger}_{n}d^{\dagger}_{m}d_{m}d_{n}\Bigg) \,.
\end{equation} 
The Reservoirs are modeled with
\begin{equation}
H_{R}(t)=	\sum_{k\in\alpha,\alpha}\epsilon_{k}(t)c_{k}^{\dagger}c_{k}\,,
\end{equation}
where, following \cite{wingreenTimedependentTransportMesoscopic1993}, the time dependence of the Reservoir energy levels is assumed to take the form $\epsilon_{k}(t)=\epsilon_{k}^{0}+\Delta_{\alpha}(t)$, where $\epsilon_{k}^{0}$ is the unperturbed reservoir energy and $\Delta_{\alpha}(t)$ is a rigid time-dependent energy shift. %
The System-Reservoir Coupling is given by 
\begin{equation}
	 H_{SR}(t)=\sum_{k\in\alpha,\alpha,n}(V_{kn}(t)c_{k}^{\dagger}d_{n}+h.c.)\,.
\end{equation}

The aim again is to compute the particle and heat currents flowing into the reservoirs and see what the First Law of thermodynamics looks like for this setup. The particle current was worked out as a generalization of the Meir-Wingreen formula in \cite{wingreenTimedependentTransportMesoscopic1993} and is given by
\begin{IEEEeqnarray}{lcr}
		I_{\alpha}^{N}(t)=
		\frac{2}{\hbar}\int_{-\infty}^{\infty} \,\frac{d\omega}{2\pi} \int_{-\infty}^{t} \,dt_1\mathbb{ImTr}\{e^{-i\omega(t_1-t)}\Gamma_{\alpha}(\omega;t_1,t)[G^{<}(t,t_1) + f_{\alpha}(\omega)G^{R}(t,t_1)]\}\,, %
\end{IEEEeqnarray}
where a generalized \emph{time-dependent} tunneling width matrix $\Gamma_{\alpha}(\omega;t_1,t)$ is identified as
\begin{IEEEeqnarray}{lcr}
		[\Gamma_{\alpha}(\omega;t_1,t)]_{nm}&=& \sum_{k\in\alpha}2\pi\delta(\hbar\omega-\epsilon_{k})V_{n}(\omega,t)V^{*}_{m}(\omega,t_1)e^{\frac{i}{\hbar}\int^{t_1}_{t}\,dt_2\Delta_{\alpha}(t_2)}\,,\nonumber\\
		&&
		\label{tdpt_Gamma}
\end{IEEEeqnarray}
with $V_{n}(\omega,t)=V_{kn}(t)$ when $ \hbar\omega=\epsilon_{k}$.

From Eq.\ \eqref{el_heat_curr_defn_eqn} it follows that the heat current for the case of time-dependent reservoirs can be written again as a sum of a conventional and a transient part
\begin{equation}
    I_{\alpha}^{Q}(t)=I_{\alpha}^{Q,conv}(t)+I_{\alpha}^{Q,tr}(t)
\end{equation}
where the conventional part is now given by 
\begin{equation}\label{fulltdpt_conv_heatcuur_def_eqn}
    I_{\alpha}^{Q,conv}(t)=\frac{d}{dt}\langle H_{R,\alpha}(t)\rangle -\langle \frac{\partial}{\partial t}H_{R,\alpha}(t)\rangle -\mu_{\alpha}(t)I_{\alpha}^{N}(t)\,,
\end{equation}
where $\mu_{\alpha}(t)=\mu_{\alpha}^0+\Delta_{\alpha}(t)$ (with $\mu_{\alpha}^0$ denoting the chemical potential of the reservoir in steady state), and the transient part is given by 
\begin{equation}\label{fulltdpt_tr_heatcuur_def_eqn}
   I_{\alpha}^{Q,tr}(t)= \frac{1}{2}\Bigg(\frac{d}{dt}\langle H_{SR,\alpha}(t)\rangle -\langle \frac{\partial}{\partial t}H_{SR,\alpha}(t)\rangle\Bigg)\,.
\end{equation}
We first compute the conventional part of the heat current. To evaluate the first term in \eqref{fulltdpt_conv_heatcuur_def_eqn}, we use the equation of motion of the expectation value of time-dependent operators 
\begin{equation}
  \frac{d}{dt}\langle H_{R,\alpha}(t)\rangle=\frac{-i}{\hbar}\langle[H_{R,\alpha}(t),H(t)]\rangle +\langle \frac{\partial}{\partial t}H_{R,\alpha}(t)\rangle\,. 
\end{equation}
The commutator in the above equation is evaluated as $[H_{R,\alpha}(t),H(t)]=[H_{R,\alpha}(t),H_{SR}(t)]$, and we get
\begin{equation}
  \frac{d}{dt}\langle H_{R,\alpha}(t)\rangle=\frac{-i}{\hbar} \langle[\sum_{k\in\alpha}\epsilon_{k}(t)c_{k}^{\dagger}c_{k},H_{SR}(t)]\rangle +\langle \frac{\partial}{\partial t}H_{SR,\alpha}(t)\rangle \,, 
\end{equation}
which upon substituting into Eq.\ \eqref{fulltdpt_conv_heatcuur_def_eqn} and simplifying gives
\begin{equation}
	I_{\alpha}^{Q,conv}(t)=\frac{-2}{\hbar}\mathbb{Re}\Bigg[\sum_{k\in\alpha,n}(\epsilon^{0}_k-\mu_{\alpha}^{0})V_{kn}(t)G^{<}_{nk}(t,t)\Bigg] \,.
\end{equation}
Evaluating $G^{<}_{nk}(t,t')$ exactly as in appendix \ref{NEGF_details_app}, gives
\begin{IEEEeqnarray}{lcr}
		I_{\alpha}^{Q,conv}(t)&=&
		\frac{2}{\hbar}\mathbb{Im}\Bigg[\sum_{k\in\alpha,n,m}(\epsilon^{0}_k-\mu_{\alpha}^0)\int_{-\infty}^{t} \,dt_1\{e^{\frac{-i}{\hbar}\epsilon_{k}^{0}(t_1-t)}V_{kn}(t)V^{*}_{km}(t_1)e^{\frac{i}{\hbar}\int^{t_1}_{t}\,dt_2\Delta_{\alpha}(t_2)}[G^{<}_{nm}(t,t_1) + f_{\alpha}(\omega)G^{R}_{nm}(t,t_1)]\}\Bigg] \,.
  \nonumber\\
		&&
\end{IEEEeqnarray}
where we have used the fact that the uncoupled (but driven) reservoir Green's functions are now given by
\begin{equation}
		g^{<}_{kk}(t,t')=if(\epsilon_{k}^{0})\exp(\frac{-i}{\hbar}\int_{t'}^{t}\,dt_1\epsilon_{k}(t_1))
\end{equation}
and 
\begin{equation} 
		g^{A}_{kk}(t,t')=i\theta(t'-t)\exp(\frac{-i}{\hbar}\int_{t'}^{t}\,dt_1\epsilon_{k}(t_1)) \,.
\end{equation} 
The conventional part of the heat current is thus given, upon going over to the matrix notation, by 
\begin{IEEEeqnarray}{lcr}
		I_{\alpha}^{Q,conv}(t)=\frac{2}{\hbar}\int_{-\infty}^{\infty} \,\frac{d\omega}{2\pi}\int_{-\infty}^{t} \,dt_1(\hbar\omega-\mu_{\alpha})\mathbb{ImTr}\{e^{-i\omega(t_1-t)}\Gamma_{\alpha}(\omega;t_1,t)[G^{<}(t,t_1) + f_{\alpha}(\omega)G^{R}(t,t_1)]\}\,,
\end{IEEEeqnarray}
where we have relabeled $\mu_{\alpha}^0$ as $\mu_{\alpha}$.

Next, the first term in the transient part of the heat current [Eq.\ \eqref{fulltdpt_tr_heatcuur_def_eqn}] can be computed by taking the derivative of 
\begin{equation}
    \langle H_{SR,\alpha}(t)\rangle = 2\int_{-\infty}^{+\infty}\,dt_1\mathbb{ImTr}\Bigg\{e^{-i\omega(t_1-t)}[G^{<}(t,t_1)\Sigma^{A}_{\alpha}(t_1,t)+ G^{R}(t,t_1)\Sigma^{<}_{\alpha}(t_1,t)]\Bigg\}\,,
\end{equation}
which generalizes Eq.\ \eqref{HSR_el_negf_eqn} to the case of time-dependent reservoirs and coupling, where the time-dependent self-energy is given by  $[\Sigma^{A/<}_{\alpha}(t,t')]_{nm}(t,t')=\sum_{k\in\alpha}V_{kn}(t')g^{A/<}_{kk}(t,t')V_{km}^{*}(t)$. The second term in \eqref{fulltdpt_tr_heatcuur_def_eqn} can be evaluated simply as
\begin{equation}
 \langle \frac{\partial}{\partial t}H_{SR,\alpha}(t)\rangle  
    =\frac{-i}{\hbar}\mathbb{Tr}\Bigg\{
 \dot{H}_{SR}(t)G_{tun}^{<}(t,t)\Bigg\}\,.
\end{equation}

The rate of external work done $\dot{W}_{ext}$ is now given by
\begin{IEEEeqnarray}{lcr}
 \dot{W}_{ext}(t)=\frac{-i}{\hbar}\mathbb{Tr}\Bigg\{\dot{H}_S(t)G^{<}(t,t)+\dot{H}_{R}(t)g^{<}(t,t)+
 \dot{H}_{SR}(t)G_{tun}^{<}(t,t)\Bigg\}\,.       
\end{IEEEeqnarray}

We can then see that the Internal Energy operator is still identified as in Eq.\ \eqref{internal_energy_op_eqn} (with $H_{SR}$ replaced with $H_{SR}(t)$) and the First Law Equation [Eq.\ \eqref{firstlaw1_eqn}] still holds exactly, even when the Reservoirs and Coupling are explicitly time-dependent.

\section{\texorpdfstring{$3-\tau$}{3 tau} plot for the Electrochemical pump}\label{3_tau_fig_app}
\begin{figure}
    \centering
    \includegraphics[height=15cm,width=7cm]{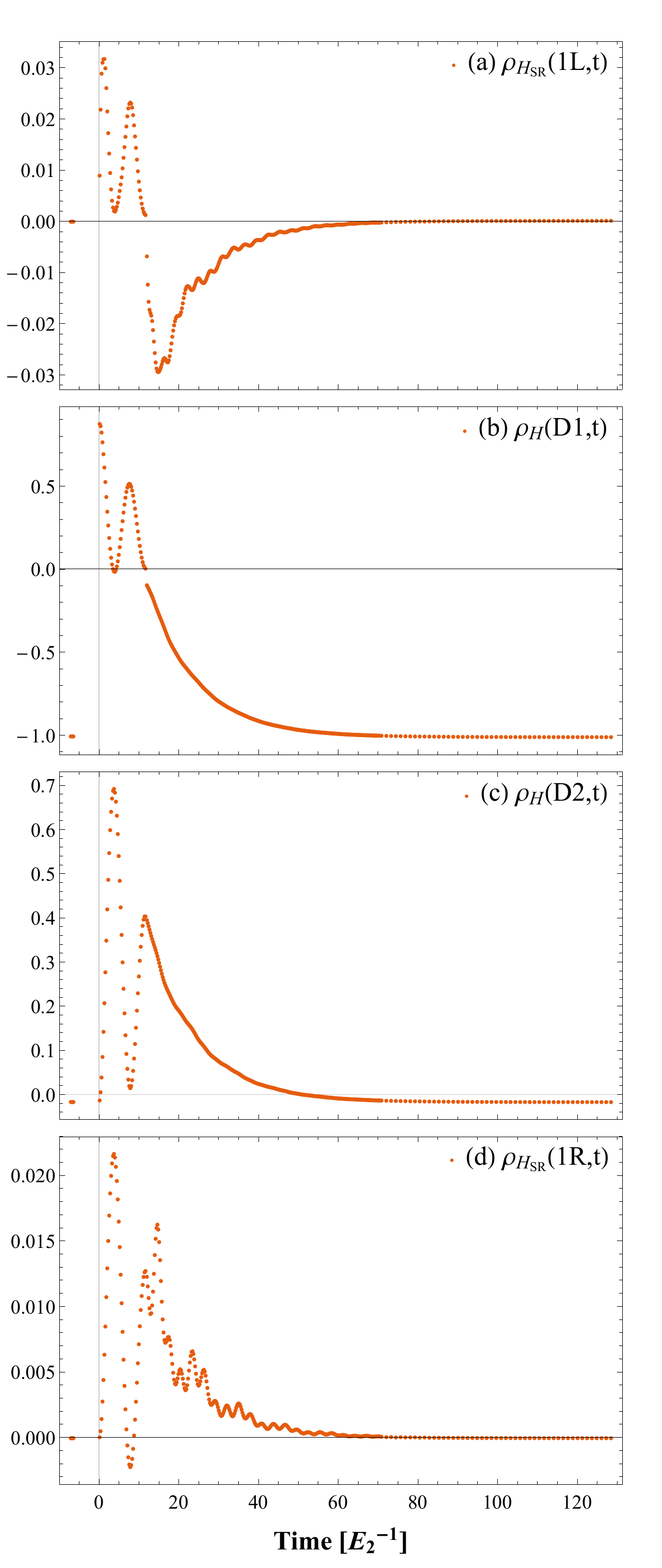}
	\captionsetup{justification=raggedright, singlelinecheck=false}
	\caption{%
 Plots of spatially resolved energy as a function
of time for the system depicted in Fig.\ \ref{chain_schematic_fig}, in the Electrochemcical pump configuration, with pulse duration set as a $3\pi$-pulse instead of a $\pi$-pulse (all other parameters are the same as those used for Fig.\ \ref{echempump_energy_density_plots_fig}). The $1.5$ periods of Rabi oscillations during the pulse (which is now active up to a time of $11.78$ in units of $E_2^{-1}$ ) are quite evident in this case.  Note that the energy scale for $\rho_{H_{SR}}(1L,t)$ and $\rho_{H_{SR}}(1R,t)$ (panels (a) and (d)) is different from that for $\rho_{H}(D1,t)$ and $\rho_{H}(D2,t)$ (panels (b) and (c)) in the figure.}
	
    \label{fig:Usys_den_3tau}
\end{figure}

\end{document}